\documentclass[pre,onecolumn,showpacs]{revtex4}

\usepackage{graphicx}
\usepackage{amsmath}
\usepackage{mathtools}

\newcommand{\bnabla}{\mbox{\boldmath $\nabla$}}

\begin{document}

\title{Mean-field electrostatics beyond the point-charge description}

\author{Derek Frydel}
\affiliation{
School of Chemistry and Chemical Engineering and Institute of Natural Sciences, 
Shanghai Jiao Tong University, Shanghai 200240, China\\}
\affiliation{Laboratoire de Physico-Chime Theorique, ESPCI, CNRS Gulliver, 
10 Rue Vauquelin, 75005 Paris, France}

\date{\today}

\begin{abstract}
This review explores the number of mean-field constructions for ions whose 
structure goes beyond the point-charge description, a representation used in the standard
Poisson-Boltzmann equation.  The exploration is motivated by a body of experimental 
work which indicates that ion-specific effects play a significant role, where ions of the same
valence charge but different size, polarizability, or shape yield quite different, and sometimes
surprising results.  
Furthermore, there are many large ions encountered in soft-matter and biophysics that
do not fit into a point-charge description, and their extension in space and shape 
must be taken into account of any reasonable representation. 
\end{abstract}

\pacs{
}

\maketitle

 \section[Introduction]{Introduction}
The present review explores the number of mean-field constructions for ions whose 
structure goes beyond the point-charge description, the representation used in the standard 
Poisson-Boltzman equation.  
The structural details neglected by a point-charge picture can be related to electrostatic 
structure of an ion, and they lead to polarizability, asymmetric interactions, or softening of Coulomb 
interactions if a distribution of an ion charge is extended in space, or they can be linked 
to the Pauli exclusion principle, and lead to the excluded volume effects or other softer types 
of repulsive interactions.  

The interest in incorporating these details came with the observation that charge is not the 
sole parameter that describes an ion \cite{David09}, as 
various ion-specific effects emerged in experimental work, beginning with the now famous work 
of Hofmeister \cite{Hofmeister}.  In the Hofmeister series ions of the same valance number can 
be arranged into a sequence according to what effect they have on protein solubility.  
The larger and more polarizable ions weaken hydrophobicity of a protein, while the smaller ones
strengthen it \cite{Zhang06}.  
Fluid interfaces is another place where ion specificity plays an important role in regulating a 
surface tension \cite{Yan10}.  
Dielectric decrement of an electrolyte upon addition of a salt is yet another example where 
specificity of an ion can be measured \cite{Collie48,David11} and that is associated with
how a given ion affects a water structure.  

Another motivation for a model with structured particles is to come up with 
a more realistic description for water, 
which within the standard Poisson-Boltzmann equation is represented as a background 
dielectric constants \cite{David07}.  A polar solvent dissociates 
salts and dissolves ions, but electrostatic interactions between dissolved ions are not the same 
as in vacuum.  The solvent background screens interactions as dipoles of a solvent 
molecules align themselves along field lines coming from ion centers.  This type of screening can 
be represented as an increased dielectric constant of a background, which is about eighty times
larger than that in vacuum.  This description is accurate if alignment of solvent molecules
is proportional to electrostatic field.  But the proportionality relation breaks down when 
alignment becomes complete and solvent no longer responds to an external field, thus the 
screening effect slowly goes away.  

Finally, as a number of problems in soft-matter and biophysics increases and involves
larger ions with significant size and definite shape, a point-charge representation no longer 
serves as an accurate representation.  Charge of an ion has to be represented as extended in 
space in order to capture novel results that such ions give rise to \cite{Pincus08}.  

The mean-field approximation is a powerful method and yet very simple.  
It represents pair interactions as an effective one-body external potential, 
\begin{equation}
w({\bf r}) = \int d{\bf r}'\rho({\bf r}')u({\bf r},{\bf r}'),
\end{equation}
which is a mean-potential 
a particle feels due to other particles in a system.  $u({\bf r},{\bf r}')$ is the true pair
interaction and $\rho({\bf r})$ is the equilibrium density of all particles.  
To construct a number density we write 
\begin{equation}
\rho({\bf r}) = \rho_be^{\beta w_b} e^{-\beta [V({\bf r})+w({\bf r})]},
\end{equation}
where $\rho_b$ and $w_b$ is the value of $\rho({\bf r})$ and $w({\bf r})$ in a bulk, and 
$V({\bf r})$ is the external potential.  The density now needs to be obtained self-consistently.  

The mean-field approximation includes particle interactions, and this is what sets it apart from 
the ideal-gas 
model, but it neglects correlations since it assumes that each particle interacts with a frozen
distribution of particles that does not change with the location of a test particle at ${\bf r}$.  
The test particle in this sense is invisible:  it measures but does not disturb the system.  
How accurate the mean-field is will depend on how negligible
correlations are \cite{Netz00a}.  
Among theoretical systems for which the mean-field yields exact density profile are the 
hard-sphere system in the infinite dimension limit \cite{Klein86}, and the system of particles 
interacting via potential of the form $U({r}) = \lambda^3u(\lambda r)$ in the limit $\lambda\to 0$, 
where $u(r)$ is bounded and with finite range \cite{Klein77b}.  
For physically relevant systems $\lambda$ remains finite, and so the corrections to the 
mean-field are always present and real.  An example of a system that is regarded as 
weakly-correlated liquid is the Gaussian core model at room temperature and so 
it is very accurately represented by the mean-field approximation \cite{Hansen00b} .

Correcting the mean-field with consecutive perturbative terms makes sense up to some value of 
an expansion parameter.  At some point this no longer makes sense and electrostatics enters into 
the strong-coupling limit.  
This splits electrostatics into two "worlds" \cite{Rudi10}.  In the strong-coupling "wonder world" 
things stand up on their head:  the same-charged particles attract each other 
\cite{Linse99}, for a counterion only system the distribution of counterions near a charged wall 
falls off exponentially as in the ideal-gas model 
\cite{Shklovskii99a,Shklovskii99b,Netz00b,Trizac11}.  The mean-field approximation 
has nothing to say in this regime \cite{Bloomfield96,Yan02,Shklovskii02}.

\section[The mean-field approximation]{The mean-field approximation}
\label{sec:MF}

In this section we go over the mean-field approximation by starting from a complete partition
function.  
To proceed, we postulate the system with scaled particle interactions, 
$u_{\lambda}({\bf r},{\bf r}')=\lambda u({\bf r},{\bf r}')$, where $\lambda=0$ corresponds to 
the ideal-gas limit, and $\lambda=1$ recovers the true system.  The partition function for this 
system reads
\begin{equation}
Z_{\lambda} = 
\frac{1}{N!\Lambda^{3N}}\int \prod_{i=1}^N d{\bf r}_i
e^{-\beta\sum_{i<j}\lambda u({\bf r}_i,{\bf r}_j)}e^{-\beta \sum_i V_{\lambda}({\bf r}_i)},
\end{equation}
where $V_{\lambda}$ is the $\lambda$-dependent external potential.  Its precise form is not 
required as it will not appear in the final expression.  It suffices to know that the sole function 
of $V_{\lambda}$ is to maintain the equilibrium density 
$\lambda$-independent, $\rho_{\lambda}=\rho$,  so that the density always corresponds to 
a physical system.  $\Lambda$ in the partition function is a length scale, and $\beta=1/(k_BT)$.  

The free energy is obtained from thermodynamic integration, 
\begin{eqnarray}
F &=& F_{\lambda=0} + \int_0^1 d\lambda\,\frac{\partial F_{\lambda}}{\partial \lambda},
\end{eqnarray}
where $\beta F_{\lambda}=-\log Z_{\lambda}$, 
$F_{\lambda=0}=F_{\rm id}+\int d{\bf r}\,\rho({\bf r})V_{\lambda=0}({\bf r})$, and the ideal-gas
contribution to the free energy is
\begin{equation}
\beta F_{\rm id} = \int d{\bf r}\,\rho\Big(\log\rho\Lambda^3-1\Big). 
\end{equation}  
The integrand in the free energy expression is
\begin{eqnarray}
\frac{\partial F_{\lambda}}{\partial \lambda} &=& 
\int d{\bf r}\,\rho({\bf r})\frac{\partial V_{\lambda}({\bf r})}{\partial\lambda}\nonumber\\
&+&\frac{1}{2}\int d{\bf r}\int d{\bf r}'\,\rho({\bf r})\rho({\bf r}')u({\bf r},{\bf r}')
\nonumber\\
&+&\frac{1}{2}\int d{\bf r}\int d{\bf r}'\,\rho({\bf r})\rho({\bf r}')u({\bf r},{\bf r}')h_{\lambda}({\bf r},{\bf r}'),
\end{eqnarray}
where $h_{\lambda}$ is the $\lambda$-dependent correlation function.  
After insertion into the free energy expression we find
\begin{eqnarray}
F &=& F_{\rm id} +\int d{\bf r}\,\rho({\bf r})V({\bf r}) 
+\frac{1}{2}\int d{\bf r}\int d{\bf r}'\,\rho({\bf r})\rho({\bf r}')u({\bf r},{\bf r}')\nonumber\\
&+&\frac{1}{2}\int d{\bf r}\int d{\bf r}'\,\rho({\bf r})\rho({\bf r}')u({\bf r},{\bf r}')
\int_0^1 d\lambda\,h_{\lambda}({\bf r},{\bf r}'),\nonumber\\
\end{eqnarray}
where the $\lambda$-dependent external potential disappears from the final expression.  

The mean-field approximation is obtained by setting correlations to zero, $h_{\lambda}=0$,
\begin{eqnarray}
F_{\rm mf} &=& F_{\rm id}+\int d{\bf r}\,V({\bf r})\rho({\bf r}) 
+\frac{1}{2}\int d{\bf r}\int d{\bf r}'\,\rho({\bf r})\rho({\bf r}')u({\bf r},{\bf r}'),
\label{eq:F_mf}
\end{eqnarray}
where the neglected correlation term is
\begin{equation}
F_{c} = \frac{1}{2}
\int d{\bf r}\int d{\bf r}'\,\rho({\bf r})\rho({\bf r}')u({\bf r},{\bf r}')\int_0^1 d\lambda\,
h_{\lambda}({\bf r},{\bf r}').
\label{eq:F_c}
\end{equation}
How accurate the mean-field is depends on the value of $F_c$.  

The mean-field free energy is written as a functional of a density that is not known {\sl a priori}.  
It has to be obtained variationally, knowing that it minimizes the free energy, a condition
expressed as functional derivative, 
\begin{equation}
\frac{\delta F_{\rm }}{\delta\rho({\bf r})} = 0,
\end{equation}
which yields
\begin{equation}
\rho({\bf r}) = \rho_be^{\beta w_b} e^{-\beta V({\bf r})-\beta\int d{\bf r}'\,\rho({\bf r}')u({\bf r},{\bf r}')},
\label{eq:rho_MF}
\end{equation}
where $w_b=\rho_b\int d{\bf r}'\,u({\bf r},{\bf r}')$ and ensures that a correct bulk limit is 
recovered.

Something more needs to be said about correlations.  We have seen how correlations are
eliminated from free energy.  Explicitly, thus, correlations are not part of the description.  
However, implicitly they are there.  If we take the exact relation,
\begin{equation}
-\frac{\delta\rho({\bf r})}{\delta \beta V({\bf r}')} = \rho({\bf r}')\rho({\bf r}')h({\bf r},{\bf r}') 
+ \rho({\bf r})\delta({\bf r}-{\bf r}'),
\label{eq:delta_rho}
\end{equation}
which can be verified from complete partition function, 
we should expect that the mean-field density in Eq. (\ref{eq:rho_MF}) would yield
\begin{equation}
-\frac{\delta\rho_{\rm }({\bf r})}{\delta \beta V({\bf r}')} = \rho({\bf r})\delta({\bf r}-{\bf r}'),
\end{equation}
since correlations were set to zero.
Indeed, this is what one gets for an ideal-gas.  But the mean-field density leads instead to a 
self-consistent equation indicating the presence of correlations,
\begin{equation}
-\frac{\delta\rho({\bf r})}{\delta \beta V({\bf r}')} = \rho({\bf r})\delta({\bf r}-{\bf r}')
+\int d{\bf r}''\,\beta u({\bf r},{\bf r}'')\frac{\delta\rho({\bf r}'')}{\delta \beta V({\bf r}')}.
\label{eq:delta_rho2}
\end{equation}
In combination with Eq. (\ref{eq:delta_rho}) it can be put into the Ornstein-Zernike equation
format,
\begin{equation}
h_{\rm }({\bf r},{\bf r}') = -\beta u({\bf r},{\bf r}') 
- \int d{\bf r}''\,\rho_{\rm}({\bf r}'')h({\bf r}',{\bf r}'')\beta u({\bf r},{\bf r}''),
\label{eq:h_mf}
\end{equation}
where the direct correlation function in the mean-field is approximated as
\begin{equation}
c_{\rm }({\bf r},{\bf r}') = -\beta u({\bf r},{\bf r}').
\label{eq:c_mf}
\end{equation}

\section{Point-ion with a structure}

\subsection{The standard Poisson-Boltzmann equation}

Within the standard Poisson-Boltzmann equation ions are represented 
as point-charges.  The model is constructed from the Poisson equation
\begin{equation}
\epsilon\nabla^2\psi({\bf r}) = -\rho_c({\bf r}),
\label{eq:poisson}
\end{equation}
which expresses the relation between the electrostatic potential $\psi$ and the charge 
density
\begin{equation}
\rho_c({\bf r}) = \sum_{i=1}^Kq_i\rho_i({\bf r}),
\end{equation}
where the subscript $i$ indicates an ion species, $K$ is the number of species, and $q_i$ is 
the charge of an ion species $i$.  In the Poisson equation $\epsilon$ is a background dielectric
constant representing a solvent medium.  The relation makes more sense if we transform it into 
somewhat different format,
\begin{equation}
\psi({\bf r}) = \int d{\bf r}'\,\rho_c({\bf r}')C({\bf r},{\bf r}'),
\label{eq:psi}
\end{equation}
where the Green's function $C({\bf r},{\bf r}')$ denotes the functional form of Coulomb 
interactions,
\begin{equation}
C({\bf r}-{\bf r}') = \frac{1}{4\pi\epsilon |{\bf r}-{\bf r}'|}, 
\end{equation}
and satisfies the fundamental equation 
\begin{equation}
\epsilon\nabla^2C({\bf r}-{\bf r}') = -\delta({\bf r}-{\bf r}').
\label{eq:C}
\end{equation}
The Poisson equation seen through the format in Eq. (\ref{eq:psi}) is nothing more than a 
restated definition of an electrostatic potential, due to some distribution in space of Coulomb 
charges.  

Within this description different ionic species are distinguished by their 
valence number alone.  Another approximation lies in the treatment of a solvent as a 
background dielectric constant $\epsilon$.   This description assumes that a polarization 
density of a polar solvent responds linearly to an electrostatic field, 
${\bf P}=\gamma{\bf E}$, where ${\bf E}=-\bnabla\psi$ is the local electrostatic field. 
The total dielectric constant then is $\epsilon=\epsilon_0+\gamma$, where $\epsilon_0$ 
is the dielectric constant in vacuum, and the action of a solvent is to screen electrostatic 
interactions by rising dielectric constant.

To construct the Poisson-Boltzmann equation we need an expression for the charge density
in terms of an electrostatic potential.  This can be obtained from the number densities, $\rho_i$, 
obtained, in turn, from the mean-potential that an ion of a species $i$ feels due to other 
ions in a system,
\begin{equation}
w_{i}({\bf r}) = q_i\int d{\bf r}'\,\rho_c({\bf r}')C({\bf r},{\bf r}') = q_i\psi({\bf r}),
\end{equation}
which leads to the mean-field density, 
\begin{equation}
\rho_i({\bf r}) = c_ie^{-\beta q_i\psi({\bf r})}, 
\end{equation}
where $c_i$ is the bulk concentration.  Inserting this into the Poisson equation in 
Eq. (\ref{eq:poisson}) we arrive at the Poisson-Boltzman equation,
\begin{equation}
\epsilon\nabla^2\psi = -\sum_{i=1}^Kc_iq_ie^{-\beta q_i\psi}.
\end{equation}
Later in the work, we refer to the Poisson-Boltzmann equation as the PB equation.

For testing different mean-field models in this work we use the wall system where electrolyte 
is confined by a charged wall with a surface charge $\sigma_c$ to a half-space $x>0$.
The PB equation reduces to 1D, 
\begin{equation}
\epsilon\psi'' = -\sum_{i=1}^Kc_iq_ie^{-\beta q_i\psi}.
\end{equation}
The neutrality condition,
\begin{equation}
\int_0^{\infty} dx\,\rho_c = -\sigma_c,
\end{equation}
fixes the boundary conditions at the location of a surface charge,
\begin{equation}
-\epsilon\psi'_w = \sigma_c,
\end{equation}
where the subscript $w$ indicates the value of a function at a contact with a wall.  

Finally, we derive the contact value theorem for the PB equation, which relates the value of a 
density at a wall contact to bulk properties.  To proceed we multiply the Poisson-Boltzmann 
equation by $\psi'$,
\begin{equation}
\epsilon\psi'\psi'' = -\psi'\sum_{i=1}^Kc_iq_ie^{-\beta q_i\psi},
\end{equation}
and rewrite it as
\begin{equation}
\frac{\partial }{\partial x}\bigg(\frac{\epsilon}{2}\psi'^2\bigg) = 
\frac{\partial }{\partial x}\bigg(k_BT\sum_{i=1}^Kc_ie^{-\beta q_i\psi}\bigg).
\label{eq:PB_E}
\end{equation}
The right-hand side term in parentheses is $k_BT\rho$ where  
$\rho = \sum_{i=1}^K\rho_i$
is the total number density.  Integrating Eq. (\ref{eq:PB_E}) from zero to infinity and using 
the boundary conditions, we find
\begin{equation}
\rho_w = \rho_b + \frac{\beta \sigma_c^2}{2\epsilon}, 
\label{eq:CVT_PB}
\end{equation}
where $\rho_b=\sum_{i=1}^Kc_i$.  In the exact contact value theorem $\rho_b\to P$, where
$P$ is a bulk pressure.   The present result reflects the ideal-gas entropy of the Poisson-Boltzmann
model where $P=P_{\rm id}$.

\subsection{Dipolar Poisson-Boltzmann equation}
Another possible point-particle is a point-dipole.  
The charge distribution of a dipole is comprised of two opposite charges brought 
infinitesimally close to each other,
\begin{equation}
\lim_{\substack{q\to\infty \\ \varepsilon\to 0}}
\Big[q\delta({\bf r}-{\bf r}')-q\delta({\bf r}-{\bf r}'+\varepsilon{\bf n})\Big]
= -(\varepsilon q)\big[{\bf n}\cdot\bnabla\delta({\bf r}-{\bf r}')\big],
\label{eq:dipole}
\end{equation}   
where ${\bf n}$ is the unit vector, and $q\varepsilon=p$ is the strength of a dipole moment.
The limit $q\to\infty$ is necessary to prevent the two charges from annihilation.  Because of 
the limits, the distribution of a dipole can be represented as a gradient of a delta function.  

For the system of dipoles an additional stochastic degree of freedom comes out due to 
dipole orientation.  A complete one particle distribution is, therefore, a function of a position 
and orientation, $\varrho({\bf r},{\bf n})$, which reduces to the number density 
\begin{equation}
\rho_i({\bf r}) = \int d{\bf p}_i\,\varrho_i({\bf r},{\bf n}). 
\end{equation}
A full distribution, however, is required to obtain a polarization density, 
\begin{equation}
{\bf P}({\bf r}) =  \sum_{i=1}^Kp_i\int d{\bf n}\,\,{\bf n}\,\varrho_i({\bf r},{\bf n}).
\end{equation}
To derive the Poisson equation for point-dipoles what is still needed is the formal expression 
of a charge density.   
Recalling that a dipole consists of two opposite charges 
"glued" together, the charge density can be written as
\begin{eqnarray}
\rho_c({\bf r}) &=& \sum_{i=1}^K p_i\int d{\bf n}\,\lim _{\varepsilon\to 0}
\Bigg[\frac{\varrho_i({\bf r},{\bf n})-\varrho_i({\bf r}+\varepsilon{\bf n},{\bf n})}{\varepsilon}\Bigg]
\nonumber\\
&=&-\sum_{i=1}^Kp_i\int d{\bf n}\,\Big[{\bf n}\cdot\bnabla\varrho_i({\bf r},{\bf n})\Big]
\nonumber\\
&=&-\bnabla\cdot \Bigg[\sum_{i=1}^Kp_i\int d{\bf n}\,\,{\bf n}\,\varrho_i({\bf r},{\bf n})\Bigg]
\nonumber\\
&=&-\bnabla\cdot{\bf P}({\bf r}).  
\end{eqnarray}
The local charge density expressed as divergence of the polarization density can be understood 
as a charge transfer from one volume element to another, and the non-zero $(\bnabla\cdot{\bf P})$ 
implies that the charge that enters a given volume element is unbalanced by the charge that 
leaves it.  The Poisson equation for the distribution of dipoles becomes
\begin{equation}
\epsilon\nabla^2\psi = \bnabla\cdot{\bf P}.
\end{equation}

To obtain the mean-field description of the present system, it is required to have an appropriate 
expression for ${\bf P}$.  
As before, we start with the expression for a mean-potential.  Knowing that an energy 
of a dipole in an external field is $(-{\bf p}\cdot{\bf E})$, we write 
\begin{equation}
w_i({\bf r},\theta) = {\bf p}_i\cdot\bnabla\psi = -p_i |\bnabla\psi|\cos\theta,
\label{eq:w_D}
\end{equation}
where $\theta$ is the angle between a local field ${\bf E}$ and a dipole ${\bf p}_i$.  
The corresponding mean-field distribution is
\begin{equation}
\varrho_i({\bf r},\theta) \sim c_i e^{\beta p_i |\bnabla\psi|\cos\theta},
\end{equation}
and it reduces to the mean-field number density
\begin{eqnarray}
\rho_i({\bf r}) 
&\sim& c_i\int_0^{\pi} d\theta \,\sin\!\theta\,e^{\beta p_i|\bnabla\psi|\cos\!\theta}.
\end{eqnarray}
The properly normalized number density is
\begin{equation}
\rho_i({\bf r})= \frac{c_i\sinh\beta p_i|\bnabla\psi|}{\beta p_i|\bnabla\psi|},
\label{eq:rho_dipole}
\end{equation}
which reduces to a bulk density $c_i$ as a field vanishes.  If dipoles were always aligned with a 
local field, the number density would simply be 
$\rho_i\to c_ie^{\beta p_i |\bnabla\psi|}$.  The different functional form in Eq. (\ref{eq:rho_dipole}) 
reflects the fact that dipoles fluctuate around their preferred orientation.  

It remains now to obtain an expression for the polarization density,
\begin{eqnarray}
{\bf P} &\sim& 
\sum_{i=1}^Kp_ic_i\int d{\bf n}\,{\bf n}\,e^{-\beta p_i ({\bf n}\cdot\bnabla\psi)}.
\end{eqnarray}
As the polarization vector is aligned with the field,
\begin{equation}
{\bf P}=P\Bigg(\frac{\bf E}{E}\Bigg),
\end{equation}
we write $P$ as
\begin{eqnarray}
P &=& 
\sum_{i=1}^K
c_ip_i\int_0^{\pi} d\theta\,\sin\theta\cos\theta e^{\beta p_i |\bnabla\psi|\cos\theta}\nonumber\\
&=& \sum_{i=1}^Kp_i\Bigg(\frac{c_i\sinh\beta p_i|\bnabla\psi|}{\beta p_i|\bnabla\psi|}\Bigg)
\Bigg[\coth\beta p_i|\bnabla\psi|-\frac{1}{\beta p_i|\bnabla\psi|}\Bigg],
\label{eq:P}
\end{eqnarray}
or,
\begin{equation}
{\bf P} = -\bigg(\frac{\bnabla\psi}{|\bnabla\psi|}\bigg)
\sum_{i=1}^Kp_i\Bigg(\frac{c_i\sinh\beta p_i|\bnabla\psi|}{\beta p_i|\bnabla\psi|}\Bigg)
{\mathcal L}(\beta p_i|\bnabla\psi|),
\end{equation}
where 
\begin{equation}
{\mathcal L}(x)=\coth x-\frac{1}{x}
\end{equation}
is the Langevin function which describes the degree of
alignment of a dipole in a uniform electrostatic field.  An average dipole moment of a 
particle of a species $i$ is given as
\begin{equation}
\langle p\rangle = p_i{\mathcal L(\beta p_i|\bnabla\psi|)}.
\end{equation}
In the limit $\beta p_i|\bnabla\psi|\to 0$, ${\mathcal L}\approx \beta p_i |\bnabla\psi|/3$.  This is 
reasonable as 
an average dipole moment should be proportional to an electrostatic field.  But an alignment
eventually must reach saturation and ${\mathcal L}$ cannot exceed $1$, where 
${\mathcal L}= 1$ signals perfect alignment.  Note, however, that the limit ${\mathcal L}\to 1$
is approached slowly, in algebraic manner like ${\mathcal L}\approx 1-1/(\beta p_i |\bnabla\psi|)$.

We have now everything that is needed for writing down a modified PB equation 
for a system of dipoles,
\begin{equation}
\epsilon\nabla^2\psi = -\bnabla\cdot\Bigg[\bnabla\psi 
\sum_{i=1}^Kp_ic_i
\frac{\sinh(\beta p_i|\bnabla\psi|)}{\beta p_i|\bnabla\psi|^2}{\mathcal L}(\beta p_i |\bnabla\psi|)\Bigg].
\end{equation}
The dipolar PB equation was derived in \cite{David07}, using the field-theory 
formalism, and was further explored in \cite{David13}.  The motivation was to 
treat water solvent more explicitly than as a background dielectric constant, 
the way it is done in the standard PB model.  The standard PB 
equation assumes a linear and local relation between the polarization density and the 
electrostatic field, ${\bf P}= \gamma {\bf E}$, so that the contributions of a polar solvent are 
subsumed into a dielectric constant $\epsilon\to \epsilon_0 + \gamma$.  
The dipolar PB equation allows us to treat a polar solvent explicitly, 
\begin{equation}
\bnabla\cdot\Bigg[\Bigg(\epsilon_0 + 
\frac{p_0c_d\,\sinh(\beta p_0 |\bnabla\psi|)}{\beta p_0 |\bnabla\psi|^2}
{\mathcal L}(\beta p_0 |\bnabla\psi|)\Bigg)\bnabla\psi\Bigg] 
= -\sum_{i=1}^Kq_ic_ie^{-\beta q_i\psi},
\label{eq:DPB}
\end{equation}
where $p_0$ is the dipole moment of a solvent molecule.  The linear relation between the 
polarization density and the field are recovered in the limit $\beta p_0 E\to 0$,
\begin{equation}
P \to \bigg(\frac{\beta c_d p_0^2}{3}\bigg)|\bnabla\psi|,
\end{equation}
that recovers a space-independent dielectric constant, 
\begin{equation}
\epsilon_{\rm eff}\to\epsilon_0 + \frac{\beta c_d p_0^2}{3} = \epsilon_{\rm sol},
\end{equation}
where $\epsilon_{\rm sol}$ denotes the dielectric constant of water, 
$\epsilon_{\rm sol}/\epsilon_0=80$, 
and the parameters $c_d$ and $p_0$ are tuned to accurately recover this limit.  Within this 
linear regime the dipolar PB equation behaves like its standard counterpart.  For larger values 
of $\beta p_0 |\bnabla\psi|$ the linearity breaks down.  Within the present model there are two 
sources of nonlinearity.  The first one lies within the Langevin function and captures the 
saturation of a polarization when a dipole aligns itself along a field.   The second source of 
nonlinearity comes from the fact that point-dipoles are incompressible,
\begin{equation}
\rho_d({\bf r}) = \frac{c_d\,\sinh(\beta p_0 |\bnabla\psi|)}{\beta p_0 |\bnabla\psi|},
\end{equation}
and a local concentration can become arbitrarily large --- a description somewhat unrealistic 
for water which is much better represented as an incompressible fluid.  

For the wall geometry and symmetric electrolyte $1:1$ the dipolar PB equation becomes,
\begin{equation}
\epsilon_0\psi'' + \frac{\partial }{\partial x}
\bigg[\frac{p_0c_d\sinh(\beta p_0\psi')}{\beta p_0\psi'}{\mathcal L}(\beta p_0\psi')\bigg]
= 2ec_s\sinh\beta e\psi. 
\end{equation}
The boundary conditions at a wall are obtained, as they were for the standard PB equation, 
from the neutrality condition,
\begin{equation}
-\epsilon\psi' = \sigma_c + \sigma_p,
\end{equation}
where we introduce the polarization surface charge,
\begin{equation}
\sigma_p = \frac{p_0c_d\sinh(\beta p_0\psi'_w)}{\beta p_0\psi'_w}{\mathcal L}(\beta p_0\psi'_w),
\end{equation}  
a polarization charge that accumulates at a wall due to charge transfer for a nonuniform  
polarization density.  The polarization surface charge has always opposite sign to the bare 
surface charge $\sigma_c$, and the surface charge can be said to be screened.

\begin{figure}[h] 
 \begin{center}%
 \begin{tabular}{rr}
  \includegraphics[width=0.5\textwidth]{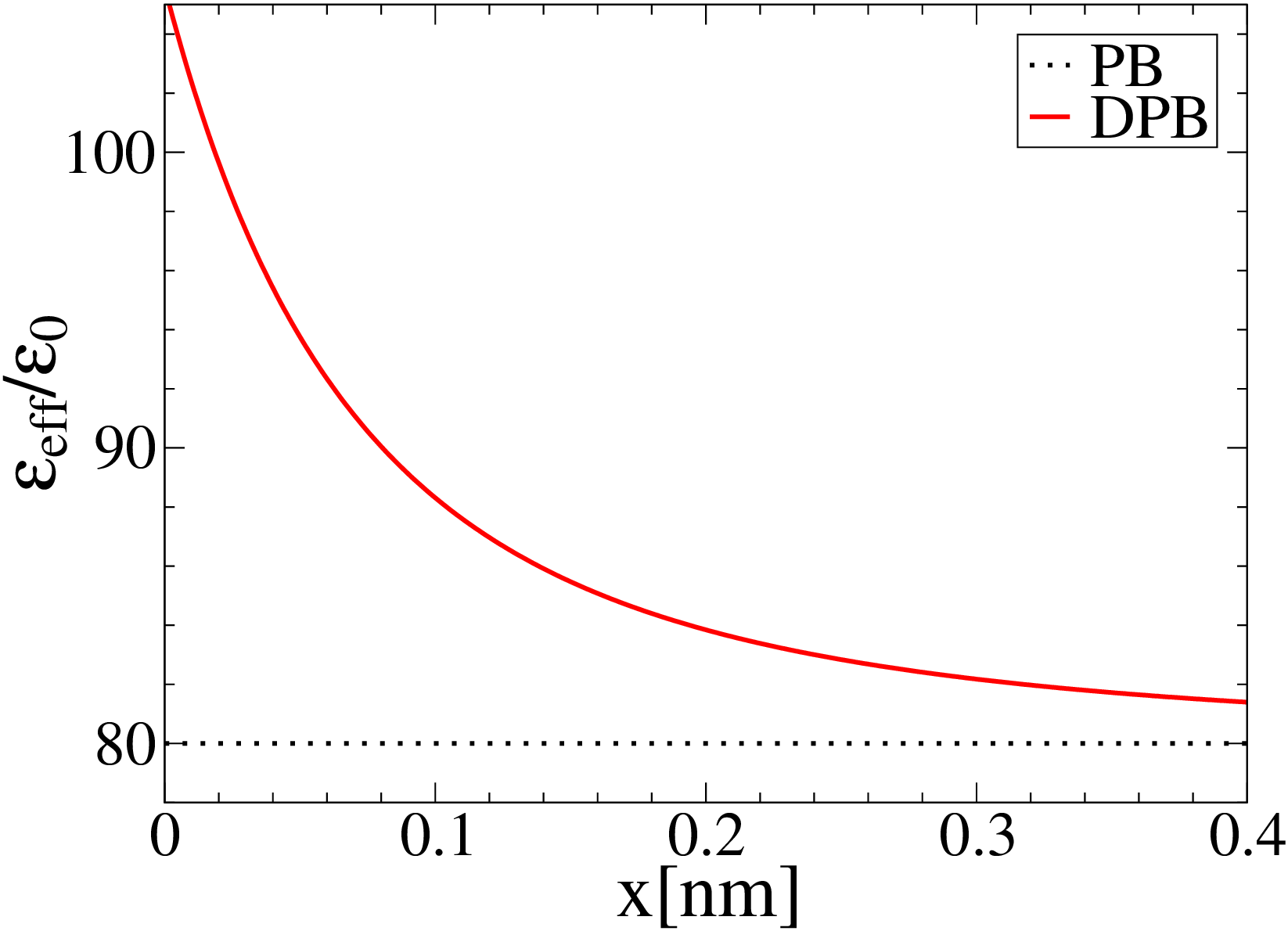}
  \includegraphics[width=0.5\textwidth]{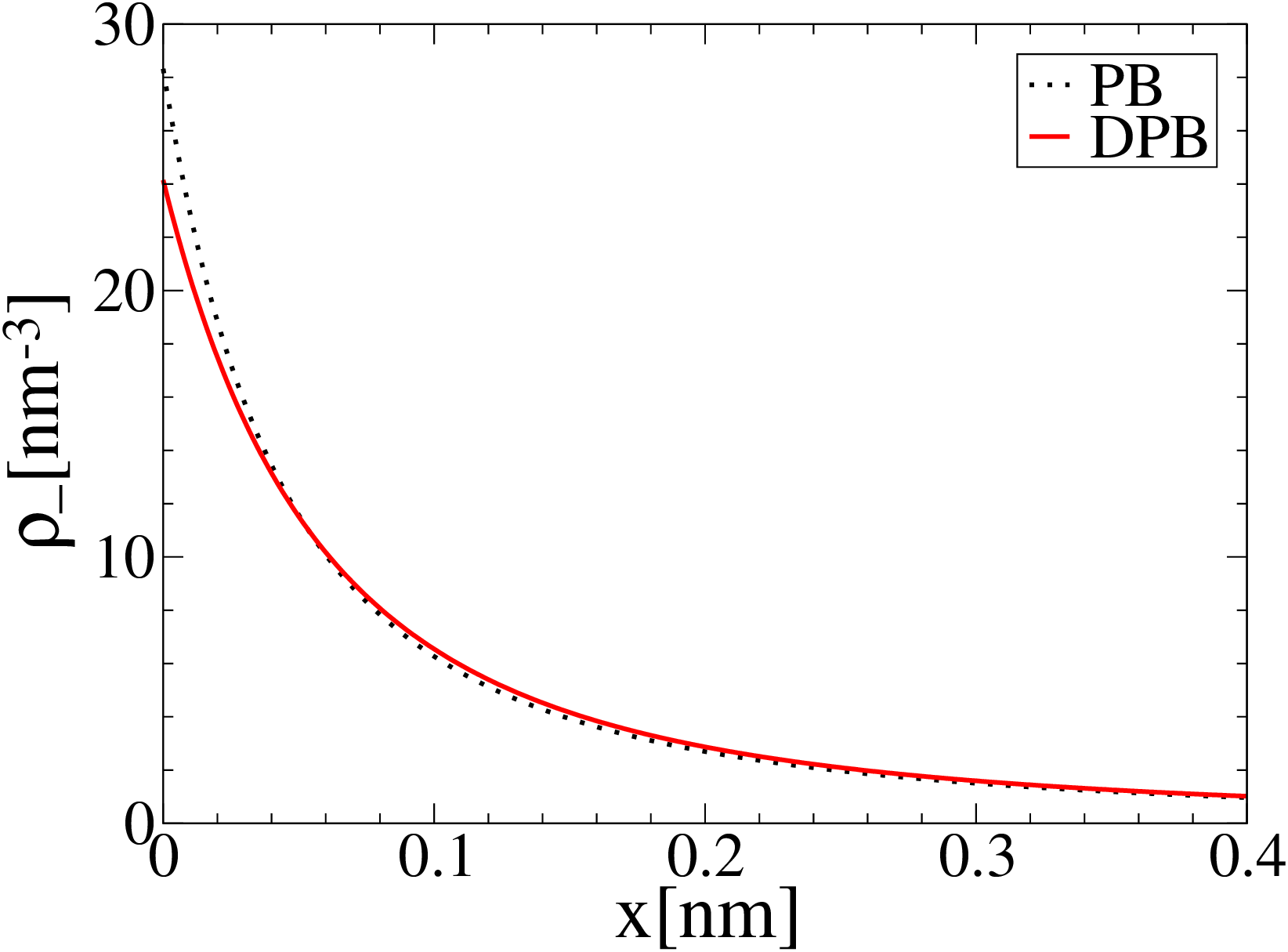}\\
\end{tabular}
 \end{center}
\caption{The effective dielectric constant, 
$\epsilon_{\rm eff}=\epsilon_0+p_0\rho_d{\mathcal L}/\psi'$, and the counterion density, $\rho_-$
(the density of coions is denoted as $\rho_+$), 
for a wall model with surface charge $\sigma_c=0.4\,{\rm Cm^{-2}}$.   The solvent parameters 
are:  $c_d=55\,{\rm M}$ and $p_0=4.78\,{\rm D}$, such that in the linear polarization regime the
dielectric constant of water is recovered, 
$\epsilon_{\rm sol}/\epsilon_0=1+\beta c_dp_0^2/3\epsilon_0=80$.  
The remaining parameters are 
$\lambda_B=\beta e^2/4\pi\epsilon_{\rm sol}=0.72\,{\rm nm}$ and $c_s=0.1\,{\rm M}$.}
\label{fig:DPB}
\end{figure}

In Fig. (\ref{fig:DPB}) we plot results for the wall model.  The dielectric constant is no longer
uniform for the dipolar PB equation as the screening increases in the wall vicinity.  Increased 
electrostatic screening reflects the excess of solvent molecules near a wall.  
\graphicspath{{figures/}}
The saturation effect of the Langevin function is, therefore, not dominant.  As a consequence, 
the counterion density is depleted from the wall region.  
(A dipole in a uniform electrostatic field adjusts its orientation but not is position since there can 
be no gain in energy.  In order to transport dipoles, a nonuniform field is required.  This type of
transport is referred to as dielectrophoresis.  The reason for the concentration gradient of dipoles 
near a wall in Fig. (\ref{fig:DPB}) reflects the fact that a field is nonuniform due to nonuniform
distribution of counterions and coions).

\begin{figure}[h] 
 \begin{center}%
 \begin{tabular}{rr}
  \includegraphics[width=0.5\textwidth]{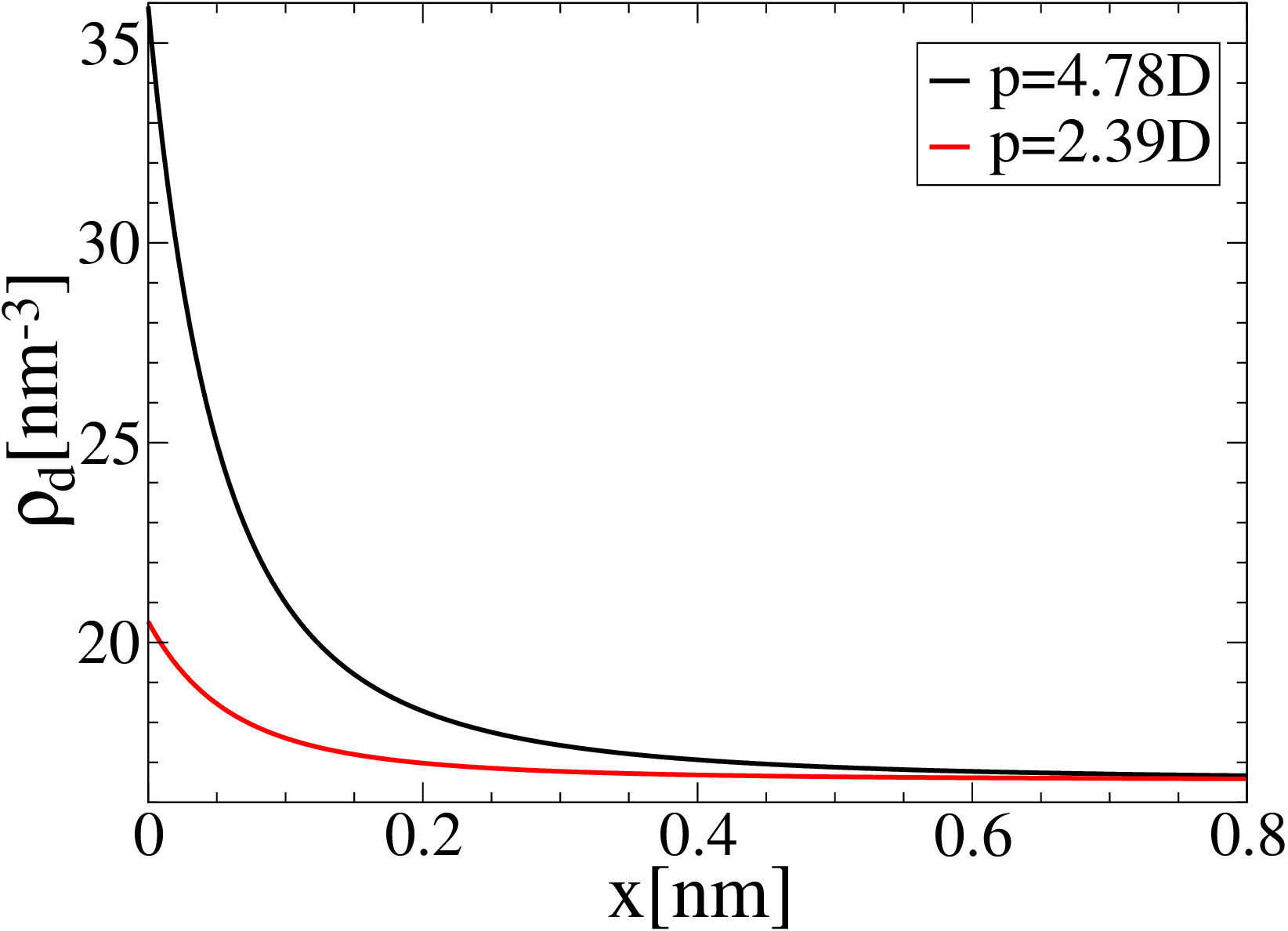}
\end{tabular}
 \end{center}
\caption{Densities of two dipolar species in a solvent mixture.  The same 
parameters as in Fig. (\ref{fig:DPB}), except the solvent parameters are:
 $c_{d1}=c_{d2}=27.5\,{\rm M}$, and $p_1=2p_2=4.78\,{\rm D}$.}
\label{fig:DPB1}
\end{figure}

The dipolar PB model can be employed for studies of solvent mixtures.  Given a mixture
of two solvents with different dipole moment, $p_1\neq p_2$, the solvent with higher polarity 
will prefer the vicinity of a charged surface, as a more efficient screening medium 
\cite{David09,David11,David13}.  The dipolar PB model captures this behavior as 
seen in Fig. (\ref{fig:DPB1}), where the hydration shell at a charged surface is comprised
primarily of solvent of higher polarity.  A heterogenous hydration shell formed around ions dissolved 
in a solvent mixture can induce additional, ion-hydration interactions, leading still to other 
ion-specific effects.  
\graphicspath{{figures/}}

\subsection{Langevin Poisson-Boltzmann equation}
\label{sec:LPB}

In the dipolar PB equation solvent obeys ideal-gas entropy.  But as water is not very compressible, 
a more realistic representation of a solvent should take into account excluded volume interactions.  
Such interactions have been implemented through a local scheme based on the lattice-gas 
entropy \cite{Birkman42,David97,Henri08,Henri09a,Henri09b,Bohinc10}.  Here, we make a simple 
assumption that
water is incompressible, and the solvent density is uniform everywhere, $\rho_d({\bf r})\to c_d$.  
This reduction leads to the Langevin PB equation, where the polarization density is determined 
solely by the Langevin function, $P=c_dp_0{\mathcal L}(\beta p_0 |\bnabla\psi|)$ 
\cite{Frydel11a,Frydel14}, 
\begin{equation}
\bnabla\cdot\Bigg[\Bigg(\epsilon_0
+\frac{c_dp_0\,{\mathcal L}(\beta p_0 |\bnabla\psi|)}{|\bnabla\psi|}\Bigg)\bnabla\psi\Bigg] = 
 -\sum_{i=1}^Kq_ic_ie^{-\beta q_i\psi}.
\end{equation}
The effective dielectric constant in parentheses has two limiting behaviors. 
In the limit $|\bnabla\psi|\to 0$, it is the same as for the dipolar PB equation,
\begin{equation}
\epsilon_{\rm }\to \epsilon_0 + \frac{\beta c_d p_0^2}{3} = \epsilon_{\rm sol},
\label{eq:eps_dilute}
\end{equation}
but now as $|\bnabla\psi|\to\infty$ and ${\mathcal L}\to 1$, the contributions of a solvent to 
dielectric response vanish, 
\begin{equation}
\epsilon_{\rm}\to \epsilon_0 + \frac{c_dp_0}{|\bnabla\psi|} 
\end{equation}
and the nonlinear contributions of the Langevin model lead to dielectric decrement.  

Dielectric decrement has been observed for bulk electrolytes, and it reflects structural 
rearrangement of water due to introduction of salt.  For salt concentrations ranging between 
zero and $1.5\,{\rm M}$, the dielectric constant was found to depend linearly on the salt
concentration, $\epsilon_{\rm eff}(c_s)=\epsilon + \alpha c_s$ \cite{Collie48,David11,David12}.
The rearrangement of water structure occurs around dissolved ions within the hydration shell.   
The orientation of these water dipoles is fixed by field lines originating from ion centers,
and they cannot respond to an external source of field.   
This behavior can be quantified with a crude model.  As dipoles within the hydration shell
are excluded from screening an external electrostatic field, the effective density of free water 
dipoles becomes reduced, $c_d\to c_d - (M_+c_++M_-c_-)$, where $M_{\pm}$ is the solvation 
number of water molecules in a hydration shell around either a cation or anion.  In the linear
regime the dielectric constant of water is $\epsilon = \epsilon_0 + \beta c_d p_0^2/3$.  
After addition of salt the effective concentration of salt is $c_d\to c_d-(M_+c_++M_-c_-)$ 
and the dielectric constant becomes
\begin{equation}
\epsilon\to \epsilon-(c_+M_++c_-M_-)\frac{(\epsilon-\epsilon_0)}{c_d}.
\label{eq:eps_cs}
\end{equation}
In this simple picture $M_{\pm}$ is salt specific.  

For the wall model and $1:1$ symmetric electrolyte, the Langevin PB equation becomes,
\begin{equation}
\epsilon_0\psi'' + c_dp_0{\mathcal L}'(\beta p_0\psi') = 2ec_s\sinh\beta e\psi.
\label{eq:LPB_1D}
\end{equation}
The boundary conditions obtained from the neutrality condition is
\begin{equation}
-\epsilon_0\psi'_w = \sigma_c +\sigma_p,
\end{equation}
where 
\begin{equation}
\sigma_p = c_dp_0{\mathcal L}(\beta p_0 \psi'_w),
\end{equation}
is the polarization surface charge.

To obtain the contact value theorem, the Langevin PB equation is multiplied by $\psi'$, 
\begin{equation}
\epsilon_0\psi'\psi'' = 2ec_s\sinh(\beta e\psi)\psi' + c_dp_0{\mathcal L}'(\beta p_0\psi')\psi',
\label{eq:CVT1}
\end{equation}
and after some manipulation we get,
\begin{eqnarray}
\frac{\partial}{\partial x}\big(\epsilon_0\psi'^2\big) &=&  
\frac{\partial}{\partial x}\big(2c_s\beta^{-1}\cosh\beta e\psi\big)
+\frac{\partial}{\partial x}\bigg(c_dp_0{\mathcal L}(\beta p_0\psi')\psi' \bigg) \nonumber\\
&-&\frac{\partial}{\partial x}\bigg(c_d\beta^{-1}
\log\bigg[\frac{\sinh\beta p_0\psi'}{\beta p_0\psi'}\bigg]\bigg).
\label{eq:CVT_LPB}
\end{eqnarray}
After integration the contact value relation becomes
\begin{equation}
\rho_w = \rho_b + \frac{\beta}{2\epsilon_0}\bigg(\sigma_c^2-\sigma_p^2\bigg)
-c_d\log\bigg[\frac{\sinh[\beta p_0(\sigma_c+\sigma_p)/\epsilon_0]}
{\beta p_0(\sigma_c+\sigma_p)/\epsilon_0}\bigg].
\end{equation}

\begin{figure}[h] 
 \begin{center}%
 \begin{tabular}{rr}
  \includegraphics[width=0.5\textwidth]{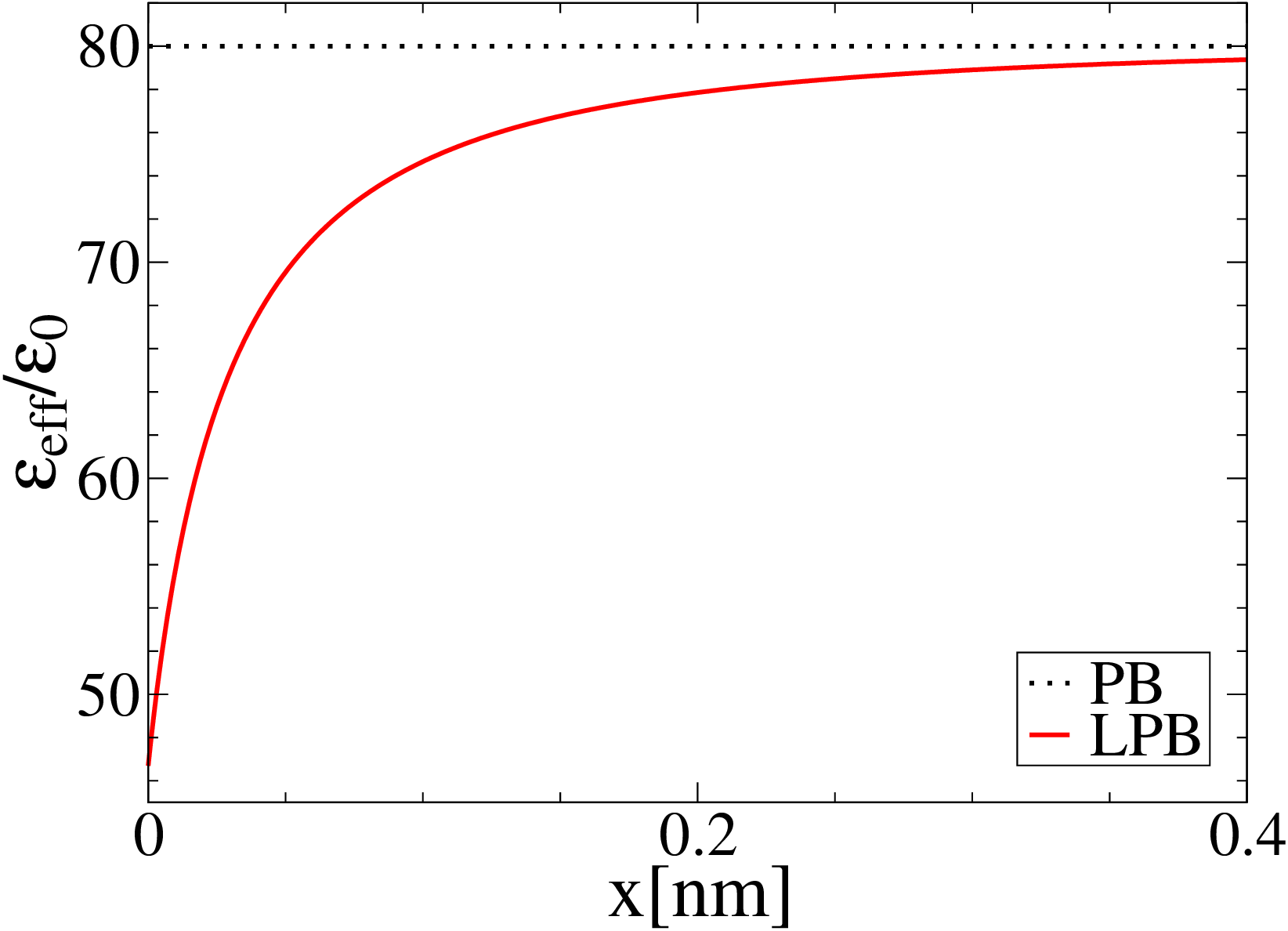}
  \includegraphics[width=0.5\textwidth]{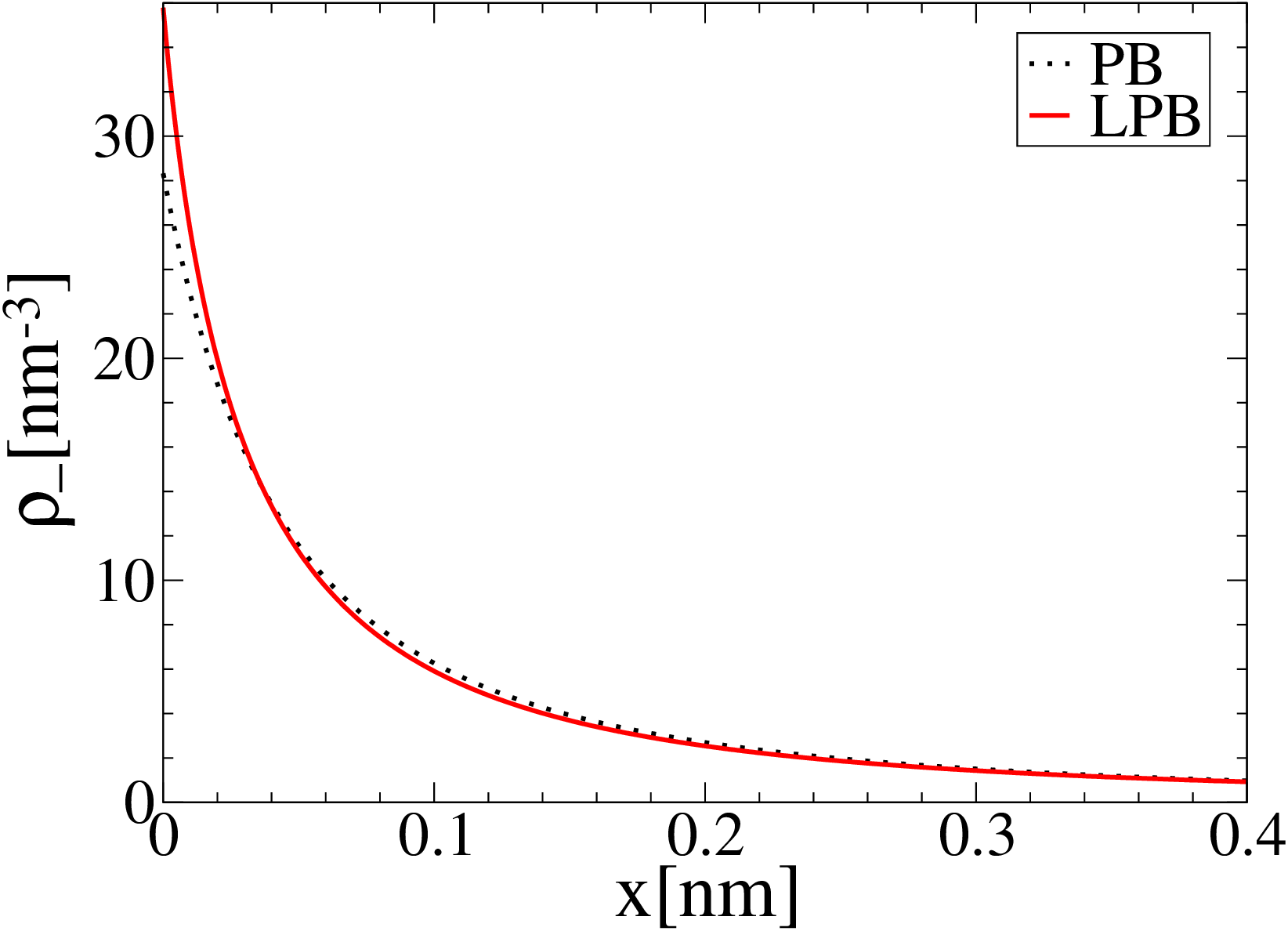}\\
\end{tabular}
 \end{center}
\caption{The effective dielectric constant, 
$\epsilon_{\rm eff}=\epsilon_0+p_0c_d{\mathcal L}/\psi'$, and the counterion density, 
$\rho_-$ (the density of coins is denoted as $\rho_+$) for a wall model and the Langevin 
PB equation.  The same parameters as in Fig. (\ref{fig:DPB}), except now the concentration 
of polar solvent is uniform and fixed at $c_d=55\,{\rm M}$.}
\label{fig:LPB}
\end{figure}

The results of the Langevin PB equation are shown in Fig. (\ref{fig:LPB}).  The dielectric 
constant near a wall decreases as the alignment of dipoles with field lines saturates, 
${\mathcal L}\to 1$. This is an opposite trend to that found in the dipolar PB equation, which
shows dielectric increment, see Fig. (\ref{fig:DPB}).  The dielectric decrement of the Langevin 
model generates stronger electrostatic interactions so that counterions stick more tightly to a 
charged wall.
\graphicspath{{figures/}}


\subsection{Point-dipoles with charge}
For the sake of illustration and as a way of transition to polarizable point-charges, 
we consider point-charges with a dipole moment.     
The mean-potential that an ion of a species $i$ feels involves two parts,
\begin{equation}
w_i({\bf r},\theta) = q_i\psi - p_i |\bnabla\psi |\cos\theta.  
\label{eq:w_CD}
\end{equation}
The corresponding mean-field distribution is
\begin{equation}
\varrho_i({\bf r},\theta_i) \sim c_i e^{-\beta (q_i\psi - p_i |\bnabla\psi|\cos\theta_i)}.
\end{equation}
After integrating out the orientational degrees of freedom we arrive at the usual number density,
\begin{eqnarray}
\rho_i({\bf r}) 
&=& \frac{c_ie^{-\beta q_i\psi}\epsilon_{\rm sol}\sinh\beta p_i|\bnabla\psi|}{\beta p_i|\bnabla\psi|}.
\end{eqnarray}
Before considering the polarization density we note that in the present model polarization 
density is associated with the density of ions.  In the Langevin model ions and dipoles were
separate species.  Following Eq. (\ref{eq:P}) we get
\begin{eqnarray}
{\bf P}&=& \bigg(\frac{\bnabla\psi}{|\bnabla\psi|}\bigg)\sum_{i=1}^K
\frac{p_ic_ie^{-\beta q_i\psi}\sinh\beta p_i |\bnabla\psi|}
{\beta p_i |\bnabla\psi|}{\mathcal L}(\beta p_i|\bnabla\psi|),
\end{eqnarray}
and the mean-field Poisson equation becomes
\begin{eqnarray}
\bnabla\cdot\Bigg[\Bigg(\epsilon &+&  
\sum_{i=1}^K\frac{p_ic_ie^{-\beta q_i\psi}\sinh\beta p_i|\bnabla\psi|}
{\beta p_i|\bnabla\psi|^2}{\mathcal L}(\beta p_i |\bnabla\psi|)\Bigg)\bnabla\psi\Bigg]
\nonumber\\
&=& 
-\sum_{i=1}^K\frac{c_iq_ie^{-\beta q_i\psi}\sinh\beta p_i|\bnabla\psi|}
{\beta p_i|\bnabla\psi|}.
\end{eqnarray}
Here ions themselves contribute to the dielectric response of the medium on account 
of their inherent dipole moment.  This leads to dielectric increment of a
solution medium.  Ions with permanent dipole moment are not common.  The typical ions 
such as Cl$^{-}$ and Na$^+$ have spherical distributions.  Ions with permanent dipole
are encountered in ionic liquids, where ions tend to be larger molecular structures.

\subsection{Polarizable Poisson-Boltzmann equation}
A dipole moment of a polarizable ion is not permanent but is induced by an external electrostatic 
field  
according to the linear relation
\begin{equation}
{\bf p} = \alpha{\bf E},
\label{eq:induced_dipole}
\end{equation}
where $\alpha$ is the ion specific polarizability.  Polarizability measures elasticity of an 
electron cloud of a molecule.  The larger the electron cloud, the more deformable is the cloud.  
This behavior is manifest in the sequence for halide ions: ${\rm F}^-<{\rm Cl}^-<{\rm Br}^-<{\rm I}^-$.

Polarizability is in fact a more general concept and measures response of an electron cloud
to a time dependent field leading to a frequency dependent polarizability.  
Our strict concern is with static, or zero-frequency polarizability as variations of an 
electric field induced by thermal fluctuations of an electrolyte operate at timescales much larger 
than timescales of inner dynamics of an electron cloud.  
Frequency dependent polarizability can lead to other effects, such as the London forces 
\cite{London37} when fluctuations in electron cloud of two nearby molecules become
synchronized.  
These, however, are less important  than the ion-dipole interactions generated by static 
polarizability \cite{Netz04a,Netz04b}.  

Polarizability can be treated by recourse to a harmonic oscillator model, where opposite 
charges can be displaced relative to one another under the action of an applied electric field,
and the resorting force is proportional to a displacement and the stiffness parameter $k$.  
The mean-potential for a species $i$ is then written as
\begin{equation}
w_i({\bf r}) = q_i\psi - p_i|\bnabla\psi| + \frac{k_i}{2}d_i^2,
\label{eq:w_i_PPB}
\end{equation}
where the last two terms characterize the energy of an induced dipole.  The electrostatic 
energy of a dipole is always in alignment with a field, and there is no orientational degree of 
freedom as for the case of a permanent dipole.  The last term is the energy of a harmonic 
oscillator.  
To relate the stiffness parameter $k$ to the polarizability $\alpha$ we start with the 
Hooke's law, ${\bf F} = k{\bf d}$, where the stretching force is electrostatic, ${\bf F}={q\bf E}$, 
and a dipole moment is related to a displacement, ${\bf p}=q{\bf d}$ ($q$ is the charge of 
the two charges being pulled apart).  Substituting these definitions into the Hooke's law we get
\begin{equation}
{\bf p} = \bigg(\frac{q^2}{k}\bigg){\bf E},
\end{equation}
so that from Eq. (\ref{eq:induced_dipole}) we get
\begin{equation}
k = \frac{q^2}{\alpha}.
\end{equation} 
The mean-potential in Eq. (\ref{eq:w_i_PPB}) can now be written as,
\begin{equation}
w_i({\bf r}) = q_i\psi  - \frac{\alpha_i}{2}|\bnabla\psi|^2, 
\end{equation}
and the mean-field density becomes
\begin{equation}
\rho_i({\bf r}) = c_ie^{-\beta (q_i\psi -\alpha_i|\bnabla\psi|^2/2)},
\end{equation}
leading to the following polarization density,
\begin{equation}
{\bf P} = -\bnabla\psi\Bigg(\sum_{i=1}^K\alpha_ic_i
e^{-\beta(q_i\psi-\alpha_i|\bnabla\psi|^2/2)}\Bigg).
\end{equation}
Polarization no longer depends on the Langevin function as it did for ions with a permanent
dipole moment.  All nonlinearity of the expression is linked to the local ion density.  
The polarizable PB equation that results is \cite{Outhwaite89,Frydel11b}
\begin{equation}
\bnabla\cdot\Bigg[\Bigg(\epsilon + \sum_{i=1}^K
\alpha_ic_ie^{-\beta (q_i\psi -\alpha_i|\bnabla\psi|^2/2)}\Bigg)\bnabla\psi\Bigg]
= \sum_{i=1}^Kq_ic_ie^{-\beta (q_i\psi -\alpha_i|\bnabla\psi|^2/2)}
\end{equation}

For a wall model and a symmetric $1:1$ electrolyte, where all ions have the same 
polarizability $\alpha$, the polarizable PB equation becomes
\begin{equation}
\epsilon\psi'' + 
\frac{\partial}{\partial x}\Big[2\alpha c_s\psi'\cosh(\beta e\psi)e^{\beta\alpha\psi'^2/2}\Big]
= 2ec_s\sinh(\beta e\psi)e^{\beta\alpha\psi'^2/2}.
\end{equation}
The boundary conditions at the wall are
\begin{equation}
-\epsilon\psi'_w = \sigma_c + \sigma_p, 
\label{eq:BC_PPB}
\end{equation}
where the polarization surface charge is
\begin{equation}
\sigma_p 
= 2\alpha c_s\psi'_w\cosh(\beta e\psi_w)e^{\beta\alpha\psi'^2_w/2}.
\label{eq:sigmap_PPB}
\end{equation}
Finally, the contact value theorem for the present model is
\begin{equation}
\rho_w = \rho_b + \frac{\beta}{2\epsilon}\Big(\sigma_c^2-\sigma_p^2\Big), 
\label{eq:CVT_PPB}
\end{equation}
obtained according to the procedure in Eq. (\ref{eq:CVT1}).  Equations (\ref{eq:BC_PPB}), 
(\ref{eq:sigmap_PPB}), and (\ref{eq:CVT_PPB}) can  be combined to yield a single equation
for either $\rho_w$ or $\sigma_p$.  Below we write down the equation for the ratio
$\sigma_p/\sigma_c$, which can be considered as a measure of polarizability,
\begin{equation}
\bigg(\frac{\sigma_p}{\sigma_c}\bigg)^3
+\bigg(\frac{\sigma_p}{\sigma_c}\bigg)^2
-\Bigg(1 + \frac{2\epsilon\rho_b}{\beta\sigma_c^2}
+\frac{2\epsilon^2}{\beta\alpha\sigma_c^2}\Bigg)\bigg(\frac{\sigma_p}{\sigma_c}\bigg)
-\Bigg(1+\frac{2\epsilon\rho_b}{\beta\sigma_c^2}\Bigg)=0.
\end{equation}
$\sigma_p/\sigma_c$ spans the range $[0,-1)$ as $\alpha$ increases.
$\sigma_p/\sigma_c=-1$ indicates that $\sigma_p$ cancels out the bare surface charge.   
From the cubic equation above we obtain the dimensionless polarizability parameter,
$$
\alpha^*=\bigg(\frac{\beta\sigma_c^2\alpha}{2\epsilon^2}\bigg),
$$
that controls the polarizability contributions at a charged surface.  The other parameter,
$\frac{2\epsilon\rho_b}{\beta\sigma_c^2}$, depends on a salt concentration.  

If we take an electrolyte at room temperature, the surface charge $\sigma_c=0.4\,{\rm Cm^{-2}}$, 
and the polarizability $\alpha/(4\pi\epsilon_0)=10\,{\rm \AA}^{3}$, then the dimensionless 
polarizability parameter is $\alpha^*\approx 0.04$.  Polarizability 
$\alpha/(4\pi\epsilon_0)=10\,{\rm \AA}^{3}$ corresponds roughly with the polarizability of
iodide ion I$^-$ and is already rather high.  We conclude then that polarizability of typical
salts has small effect on electrolytes.  

\begin{figure}[h] 
 \begin{center}%
 \begin{tabular}{rr}
  \includegraphics[width=0.5\textwidth]{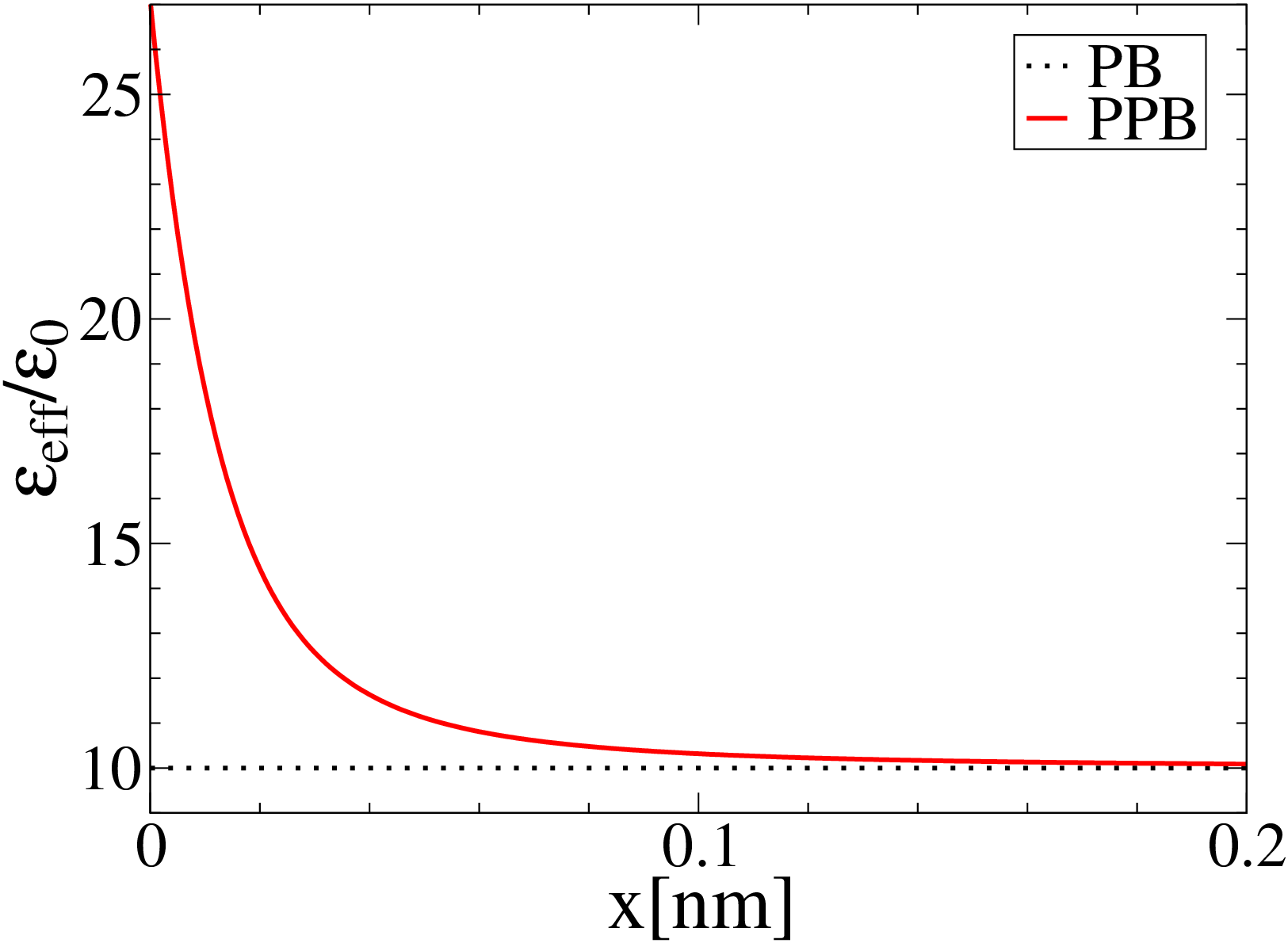}
  \includegraphics[width=0.5\textwidth]{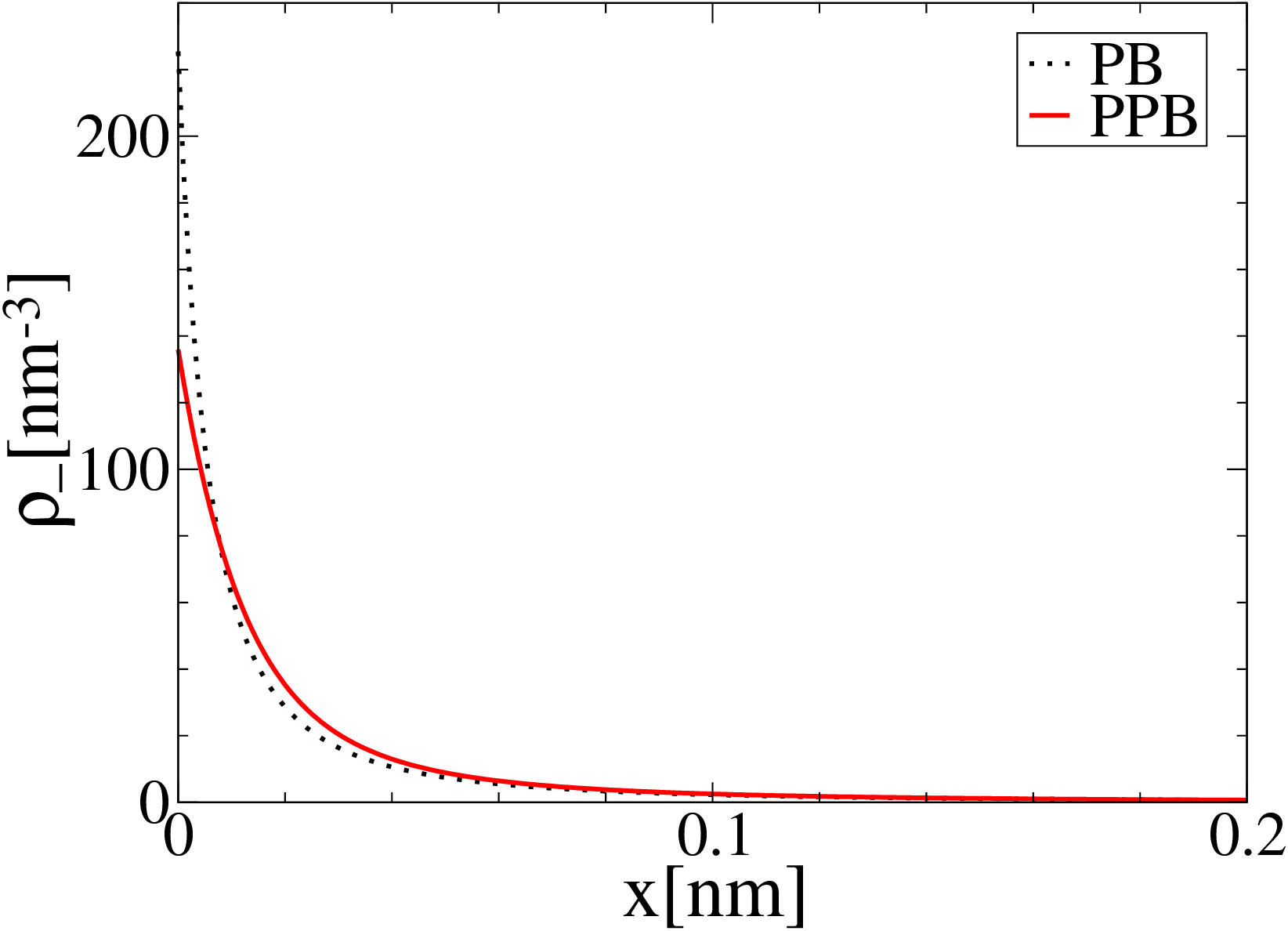}\\
\end{tabular}
 \end{center}
\caption{Effective dielectric constant $\epsilon_{\rm eff}=\epsilon+2\alpha c_s\cosh\beta e\psi$, 
and the counterion density profile for reduced dielectric constant, $\epsilon/\epsilon_0=10$ 
(in water $\epsilon/\epsilon_0=80$).  The relevant system parameters are:  
$\sigma_c=0.4\,{\rm Cm^{-2}}$, $\lambda_B=5.76\,{\rm nm}$, 
$c_s=0.1\,{\rm M}$, and $\alpha/(4\pi\epsilon_0)=10{\rm \AA}^{3}$.}
\label{fig:PPB_eps_low}
\end{figure}

Polarizability contributions can be increased for dielectric media with low dielectric 
constant.  Such a situation is realized in ionic liquids, where the absence of a polar solvent 
permits unscreened electrostatic interactions, as ionic liquids are melted salts.  
In Fig. (\ref{fig:PPB_eps_low}) we consider an electrolyte with reduced dielectric 
constant, $\epsilon/\epsilon_0=10$ (which yields a larger Bjerrum length, 
$\lambda_B=5.76\,{\rm nm}$).   
\graphicspath{{figures/}}
The increased dielectric constant near a wall region, which reflects a counterion profile,
$\epsilon_{\rm eff}=\epsilon+2\alpha c_s\cosh\beta e\psi$, generates a weaker attraction 
to a surface charge, so that counterions become more spread out.  

\begin{figure}[h] 
 \begin{center}%
 \begin{tabular}{rr}
  \includegraphics[width=0.5\textwidth]{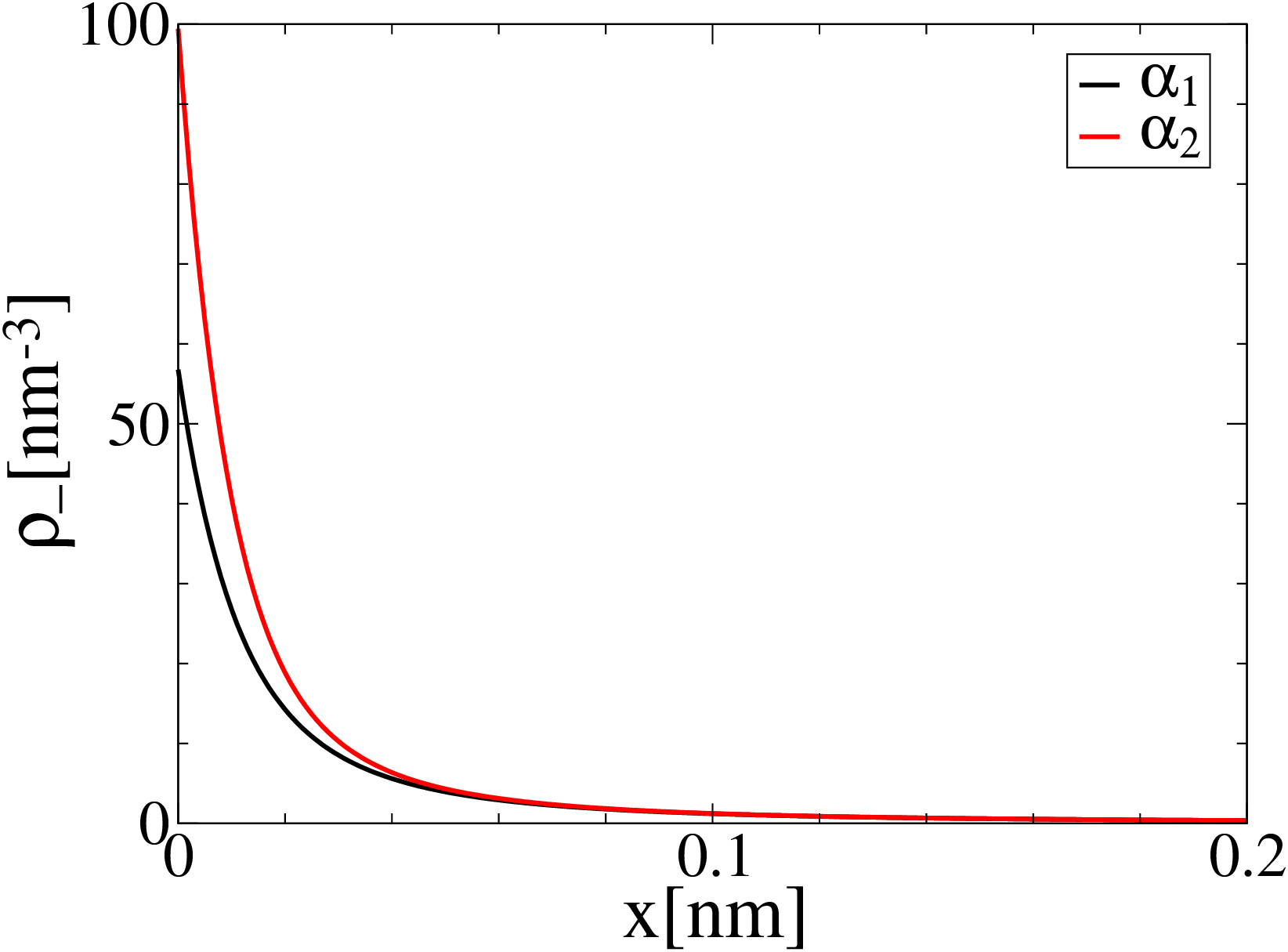}
\end{tabular}
 \end{center}
\caption{The counterion profiles for polarizable and non-polarizable ions.  The same 
parameters as in Fig. (\ref{fig:PPB_eps_low}) but now only half of the ions are polarizable:
$\alpha_1=0{\rm \AA}^{3}$ and $\alpha_2=10{\rm \AA}^{3}$.}
\label{fig:PPB2}
\end{figure}

The present model can by applied to study of ion specificity by assigning different 
polarizabilities to ion species.  In Fig. (\ref{fig:PPB2}) we show density profiles for counterions
with the same charge but different polarizability, $\alpha_1/(4\pi\epsilon_0)=0{\rm \AA}^{3}$
and $\alpha_2/(4\pi\epsilon_0)=10{\rm \AA}^{3}$.  
Polarizable counterions being a better screening agent are preferred near a wall.  
\graphicspath{{figures/}}

\subsubsection{Negative Excess polarizability}


The present mean-field framework developed for polarizable ions has been used to capture 
the physics of dielectric decrement caused by the restructuring of water as hydration shells 
form around dissolved 
ions \cite{Hatlo12,Rudi12}.  Since polarizable ions in general cause dielectric increment, dielectric
decrement can easily be realized when using negative values of polarizability.  Negative 
polarizability occurs in quantum mechanics for molecules in excited state or for non-static 
polarizabilities, but in soft-matter it is an effective phenomena.  Negative polarizability means 
that induced dipole acts opposite to the local field. 
This conceptually captures the fact that the water dipoles in a hydration shell do not respond 
to electrostatic field.  

According to Eq. (\ref{eq:eps_cs}), the dielectric increment/decrement depends linearly on the 
salt concentration. The same linear dependence is found for the polarizable PB model, 
\begin{equation}
\epsilon_{\rm eff} - \epsilon_{} = \alpha_+\rho_+ + \alpha_-\rho_-,
\end{equation}
where, after comparing with Eq. (\ref{eq:eps_cs}), the negative excess polarizability can be 
approximated as
\begin{equation}
\alpha_{\pm} = -M_{\pm}\bigg(\frac{\epsilon-\epsilon_0}{c_d}\bigg).
\end{equation}
Even for a modest value of a solvation number, $M_{\pm}=4$, the excess polarizability is
already significant, $\alpha/(4\pi\epsilon_0)\approx -300{\rm\AA}^3$, where we assume
$M_+=M_-=M$ and $\alpha_+=\alpha_-=\alpha$.  

\begin{figure}[h] 
 \begin{center}%
 \begin{tabular}{rr}
  \includegraphics[width=0.5\textwidth]{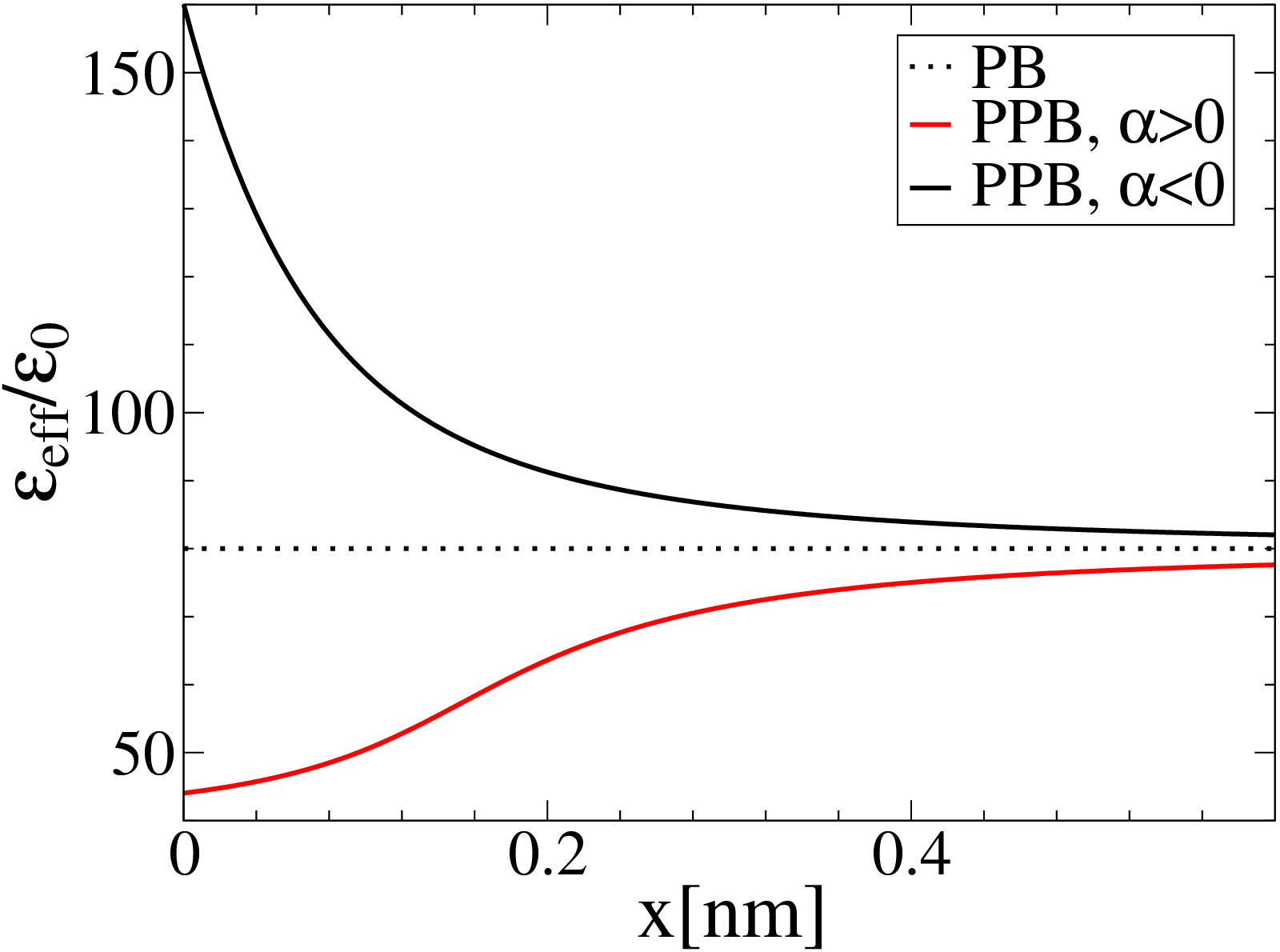}
  \includegraphics[width=0.5\textwidth]{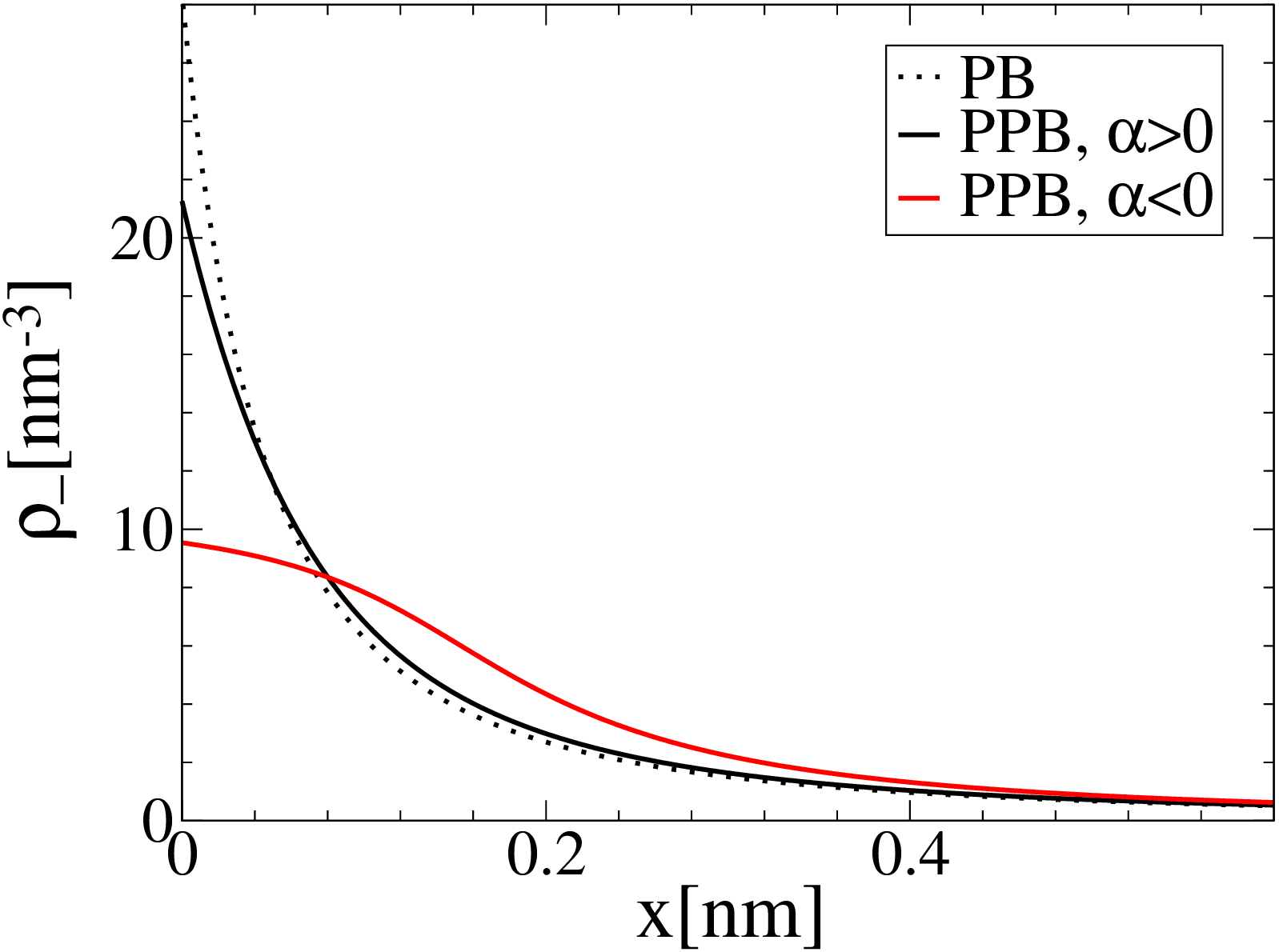}\\
\end{tabular}
 \end{center}
\caption{Effective dielectric constant $\epsilon_{\rm eff}=\epsilon+2\alpha c_s\cosh\beta e\psi$, 
and the counterion density profile for negative excess polarizability 
$\alpha/(4\pi\epsilon_0)= -300{\rm\AA}^3$.  The dielectric constant of a solvent background 
is that of water, $\epsilon/\epsilon_0=80$.  
The remaining parameters are:  
$\sigma_c=0.4\,{\rm Cm^{-2}}$, $\lambda_B=0.72\,{\rm nm}$, $c_s=0.1\,{\rm M}$. 
}
\label{fig:PPB_neg}
\end{figure}

In Fig. (\ref{fig:PPB_neg}) we show the results for $\alpha/(4\pi\epsilon_0)= -300{\rm\AA}^3$.  
We compare the plots with positive polarizability of the same magnitude, 
$\alpha/(4\pi\epsilon_0)=300{\rm \AA}^3$.  The negative polarizability, as expected, lowers the 
dielectric constant near a wall.  Less intuitive is the fact that this leads to depletion of 
counterions from the wall region.  The depletion is, furthermore, more significant than that for 
positive polarizability.  If the electrostatic screening is reduced in the wall region than 
counterions should hug to the wall more tightly.  This is, at least, what we see in 
Fig. (\ref{fig:LPB}) for the Langevin PB model.  So why does the Langevin PB model satisfies
our intuitions and the polarizable PB model for negative polarizabilities does not?  
Formal answer to this puzzle can be found by examining the contact value theorem in 
Eq. (\ref{eq:CVT_PPB}).  The sign of the polarization surface charge, $\sigma_p$, does not 
matter, and any polarizability lowers the contact density, $\rho_w$.   

\graphicspath{{figures/}}

The two models, the Langevin and the polarizable PB equation with $\alpha<0$
are designed to represent the same phenomena, the lowering of a dielectric 
constant as the hydration structures form around dissolved ions.  The results,
however, are not precisely comparable.  Decrement of a dielectric constant near a wall
are captured by both models, but density profiles are not comparable, even qualitatively.
Counterion profiles of the Langevin model are more concentrated, while those of the 
polarizable model are more dilute.  Without exact simulation results, it is hard to  know which
model is accurate.  In recent work by Ma {\sl et al.} \cite{Zhenli14} the Langevin PB equation 
with correlations has been solved and it yields a non-monotonic density profile with 
a valey at a wall followed by a peak further away from a wall.  The depletion is, therefore, 
captured but immediately at a wall and not for the entire profile.  
A possible weak point of the negative polarizability model is the linearity assumption, 
$\epsilon_{\rm eff}-\epsilon \sim -c_s$.  For homogenous solutions linearity breaks down 
for higher concentrations \cite{David13}, as hydration shells begin to overlap.  This suggests
that near a wall, where concentrations are high, this too could have its effect.  


\section[Finite-Spread Poisson-Boltzmann Equation]{Finite-Spread Poisson-Boltzmann Equation}

An alternative approach to introduce structure of a charged particle is to smear its net charge 
within a finite volume according to a desired distribution $\omega({\bf r}-{\bf r}_0)$ such that 
\begin{equation}
q=\int d{\bf r}\,\omega({\bf r}-{\bf r}_0).
\end{equation}
An arbitrary distribution is expected to depend on, in addition to the position ${\bf r}_0$, the
orientation characterized by three angles.  An arbitrary distribution embodies dipole, 
\begin{equation}
{\bf p} = \int d{\bf r}\,({\bf r}-{\bf r}_0)\omega({\bf r}-{\bf r}_0),
\end{equation}
and higher order multipoles.  If the two distributions 
at ${\bf r}_i$ and ${\bf r}_j$, 
do not overlap, there is no difference between the finite-spread and point-ion representation.
The difference occurs for overlapping separations and the resulting potential,
\begin{equation}
U({\bf r}_i-{\bf r}_j) 
= \int d{\bf r}\int d{\bf r}'\,
\frac{\omega({\bf r}-{\bf r}_i)\omega({\bf r}'-{\bf r}_j)}{4\pi\epsilon|{\bf r}-{\bf r}'|},
\label{eq:U_omega}
\end{equation}
is no longer described as a truncated series of multipoles.  

The finite-spread model does not want to provide a detailed electronic structure of an ion.  
This is beyond the scope of classical physics.  But there are particles whose charge distribution 
is better described as extended in space, rather than as a sequence of multipoles.  Among 
examples are charged rods, dumbbell shaped particles 
\cite{Pincus08,Bohinc08,Bohinc11,Bohinc12}, or macromolecules whose non-electrostatic 
interactions are "ultrasoft", allowing interpenetration, and the distribution of charge in space is a  
sensible representation \cite{Likos01}.  A perfect example is a polyelectrolyte in a good solvent 
whose charges along a polymer chain appear on average as a smeared-out cloud due to 
quickly alternating configurations.  Uncharged, two chain polymers interact via a Gaussian
potential representing steric interactions of two self-avoiding polymer chains \cite{Hansen00a}.  
Dendrimers offer another example of a soft, flexible macromolecule \cite{Likos01a}.

There is also a more fundamental aspect of smeared-out charges: a smear-out  
point-charge is rid of divergence.  For same-charged ions this eliminates effective excluded
volume effects of a Coulomb potential and permits interpenetration of two or more charges.  
For opposite-charged ions it leads to a new type of a Bjerrum pair where two ions collapse 
into a neutral but polarizable entity \cite{Hansen11a,Hansen11b,Hansen12,Warren13}.  
The usual Bjerrum pair, formed between ions with hard-core interactions, is represented as
a permanent dipole \cite{Yan93}.  

Ultrasoft repulsive interactions (without the long-range Coulomb part) have been extensively 
studied, both for its theoretical aspects and as a description of a soft matter system.  Studies 
reveal two distinct behaviors.  Some ultrasoft potentials supports "stacked" configurations, where 
two or more particles collapse, even though no true attractive interactions come into play
\cite{Lowen01,Likos06}.  This behavior leads to a peak in a correlation function around $r=0$.  
To this class of potentials belongs the penetrable sphere model \cite{Witten89}.  The Gaussian 
core model \cite{Stillinger76}, on the other hand, represents the class of sort particles unable
to support stacked configurations.

As a note of interest, we mention that the interest in ultrasoft interactions is not confined to 
soft-matter.  The soft-core boson model with interactions $U(r)\sim (R^{6}+r^{6})^{-1}$, where $R$ 
is the soft-core radius, has been studied in \cite{Pohl14} in connection to superfluidity.  
By removing the singularity from the potential, boson particles cluster and form crystal with multiple
particles occupying the same lattice sites.

\subsection{Spherical Distribution $\omega(|{\bf r}-{\bf r}_0|)$}
In this section we consider ion species with spherically symmetric distribution 
$\omega_i(|{\bf r}-{\bf r}_0|)$.  As in previous mean-field constructions, we start with 
the mean-potential that an ion of the species $i$ feels, 
\begin{equation}
w_i({\bf r}) = \int d{\bf r}'\,\omega_i(|{\bf r}-{\bf r}'|)\psi({\bf r}').
\end{equation}
The non-locality of the expression reflects the finite distribution of an ion charge in space and 
the fact that every part of this distribution interacts with an electrostatic field. 
The number density that follows is
\begin{equation}
\rho_i({\bf r}) = c_i e^{-\beta \int d{\bf r}'\,\omega_i(|{\bf r}-{\bf r}'|)\psi({\bf r}')}.
\end{equation}
To obtain the appropriate PB equation we need an expression for the charge density,
which is given as the convolution of the number density,
\begin{equation}
\rho_c({\bf r}) = \sum_{i=1}^K\int d{\bf r}'\,\omega_i(|{\bf r}-{\bf r}'|)\rho_i({\bf r}').
\label{eq:rhoc_fspb}
\end{equation}
Convolution is, again, a result of the finite extension of a charge.  
Using an explicit expression for $\rho_i$ we get
\begin{equation}
\rho_c({\bf r}) = \sum_{i=1}^Kc_i\int d{\bf r}'\,\omega_i({\bf r}-{\bf r}')
e^{-\beta \int d{\bf r}''\,\omega_i({\bf r}'-{\bf r}'')\psi({\bf r}'')}, 
\end{equation}
and the finite-spread PB equation for smeared-out ions is \cite{Frydel13}
\begin{equation}
\epsilon\nabla^2\psi = -\sum_{i=1}^Kc_i\int d{\bf r}'\,\omega_i(|{\bf r}-{\bf r}'|)
e^{-\beta \int d{\bf r}''\,\omega_i(|{\bf r}'-{\bf r}''|)\psi({\bf r}'')}.  
\label{eq:FSPB}
\end{equation}

\begin{figure}[h] 
 \begin{center}%
\includegraphics[width=0.5\textwidth]{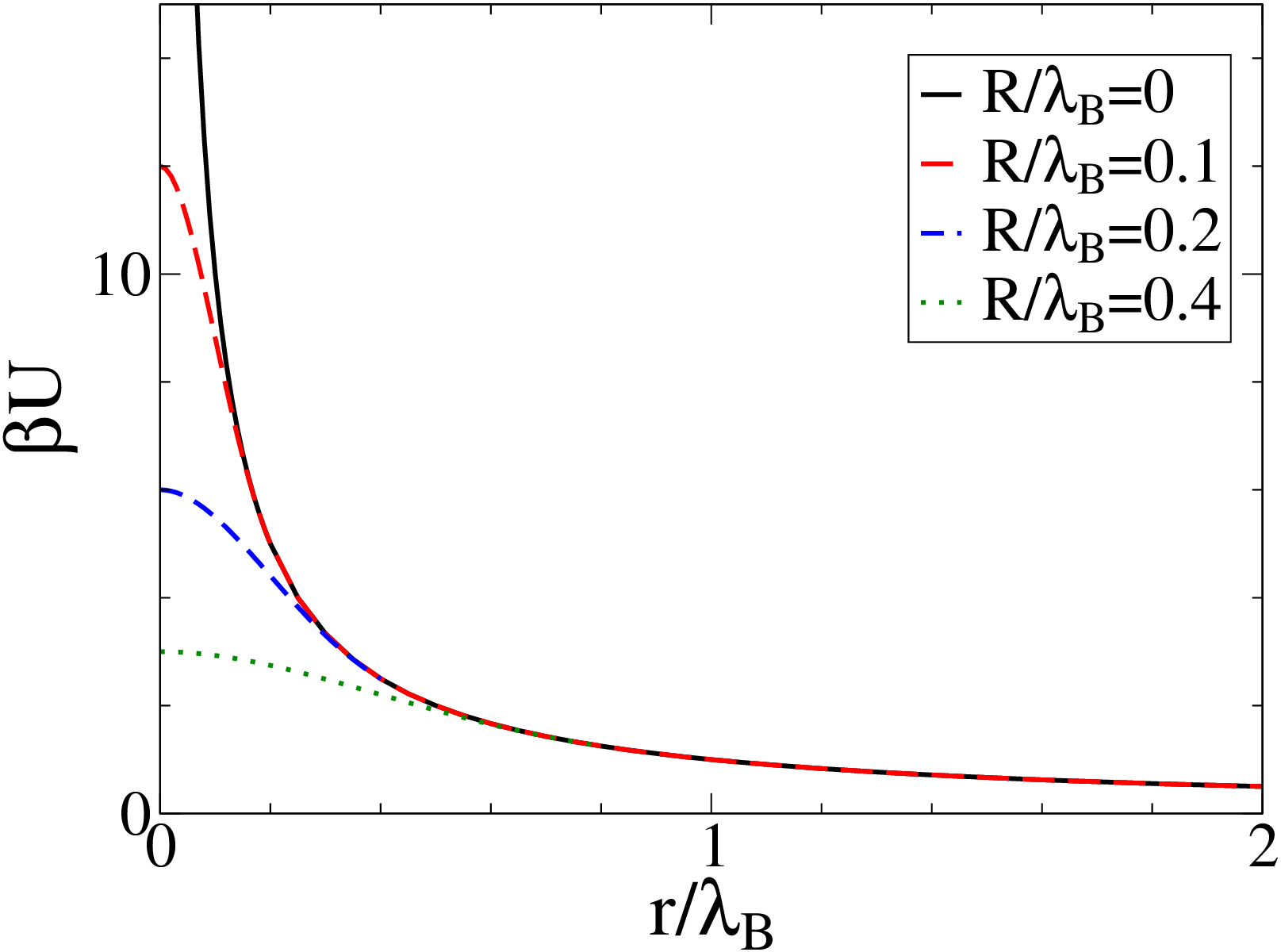}
\end{center}
\caption{Pair potential between two charge distributions in Eq. (\ref{eq:omega}) for different
$R$.  At overlapping separations the functional form of a pair potential is that in 
Eq. (\ref{eq:U_sphere}). }
\label{fig:UR}
\end{figure}

To complete the model, we still need to choose a specific spherical distribution.   
We model ions as uniformly distributed charges within a spherical volume,
\begin{equation}
\omega_i(|{\bf r}-{\bf r}'|) = \frac{3q_i}{4\pi R_i^3}\theta(R_i-|{\bf r}-{\bf r}'|),
\label{eq:omega}
\end{equation}
where $q_i$ and $R_i$ are the charge and the radius of an ion species $i$, respectively.  
The pair interaction between two ions with charge $q$ and size $R$, when the two ions 
overlap, is
\begin{equation}
U(r\le 2R) = \frac{q^2}{4\pi\epsilon R}\Bigg[\frac{6}{5}-\frac{1}{2}\Big(\frac{r}{R}\Big)^2
+\frac{3}{16}\Big(\frac{r}{R}\Big)^3-\frac{1}{160}\Big(\frac{r}{R}\Big)^5\Bigg],
\label{eq:U_sphere}
\end{equation}
when not overlapping the usual Coulomb potential is recovered, 
\begin{equation}
U(r> 2R) = \frac{q^2}{4\pi\epsilon R}\Big(\frac{r}{R}\Big)^{-1}.
\end{equation}
If overlap is complete the pair interaction remains finite,
\begin{equation}
U(0) = \frac{6}{5}\frac{q^2}{4\pi\epsilon R},
\end{equation}
and is said to be bounded.  
In Fig. (\ref{fig:UR}) we plot various realizations of the pair potential $U(r)$ for different ion
size $R$.  The degree of penetration clearly increases with increasing $R$.  
\graphicspath{{figures/}}

The mean-field Poisson equation for a symmetric $1:1$ electrolyte, with ion distributions 
in Eq. (\ref{eq:omega}), is
\begin{eqnarray}
\epsilon\nabla^2\psi &=& \frac{6ec_s}{4\pi R^3}\int d{\bf r}'\,\theta(R-|{\bf r}-{\bf r}'|)\nonumber\\
&\times&\sinh\bigg[{-\frac{3\beta e}{4\pi R^3}\int d{\bf r}''\,\theta(R-|{\bf r}'-{\bf r}''|)\psi({\bf r}'')}\bigg].\nonumber\\
\label{eq:FSPB11}
\end{eqnarray}
For the wall model we 
the integral terms simplify, 
\begin{equation}
\int d{\bf r}'\,\theta(R-|{\bf r}-{\bf r}'|)f(z) = \pi\int_{-R}^{R}dz'\,f(z+z')(R^2-z'^2),
\end{equation}
where for $f(z)=1$ we recover $4\pi R^3/3$, a volume of a sphere.  And the finite-spread PB 
equation becomes
\begin{equation}
\epsilon\psi'' = \frac{6ec_s}{4 R^3}\int_{-R}^{R}dz'\,(R^2-z'^2)
\sinh\bigg[{-\frac{3\beta e}{4 R^3}\int_{-R}^R dz''\,\psi(z+z'+z'')(R^2-z''^2)}\bigg].
\end{equation}

For the wall model particle centers are confined to the half-space $x>0$ but a charge
density starts from $x>-R$ due to finite size of an ionic charge as half of a sphere sticks out.  
The boundary conditions, therefore, are not determined at the wall, $x=0$, but at $x=-R$,
\begin{equation}
-\epsilon\psi'(-R) = \sigma_c.
\end{equation}
This implies that the surface charge is at $x=-R$.
The contact value theorem, however, is not effected and is the same as for the standard
PB equation,
\begin{equation}
\rho_w = \rho_b + \frac{\beta \sigma_c^2}{2\epsilon},
\end{equation}
where $\rho_w=\rho(0)$.

\begin{figure}[h] 
 \begin{center}%
\begin{tabular}{rr}
\includegraphics[width=0.48\textwidth]{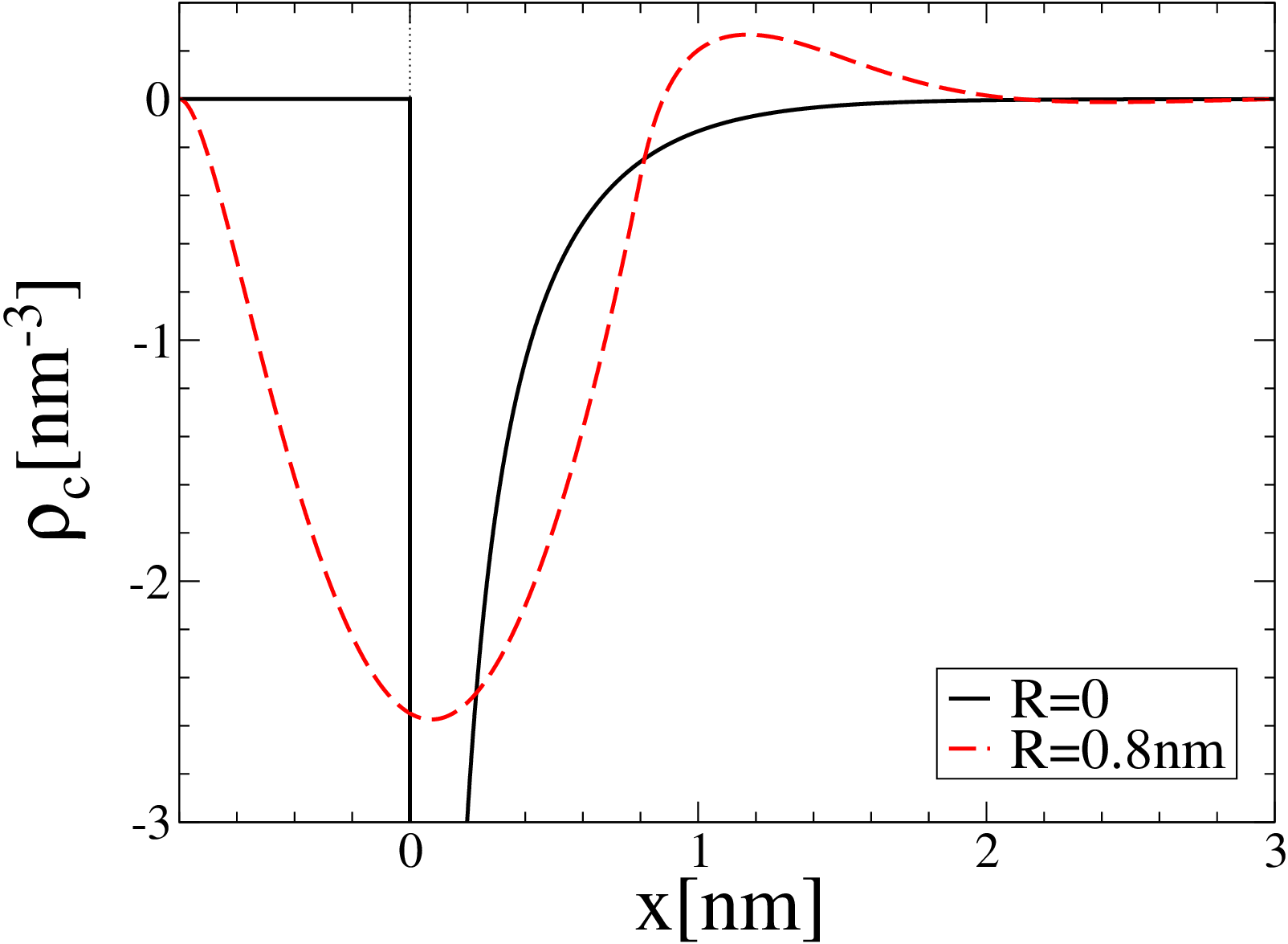}&
\includegraphics[width=0.48\textwidth]{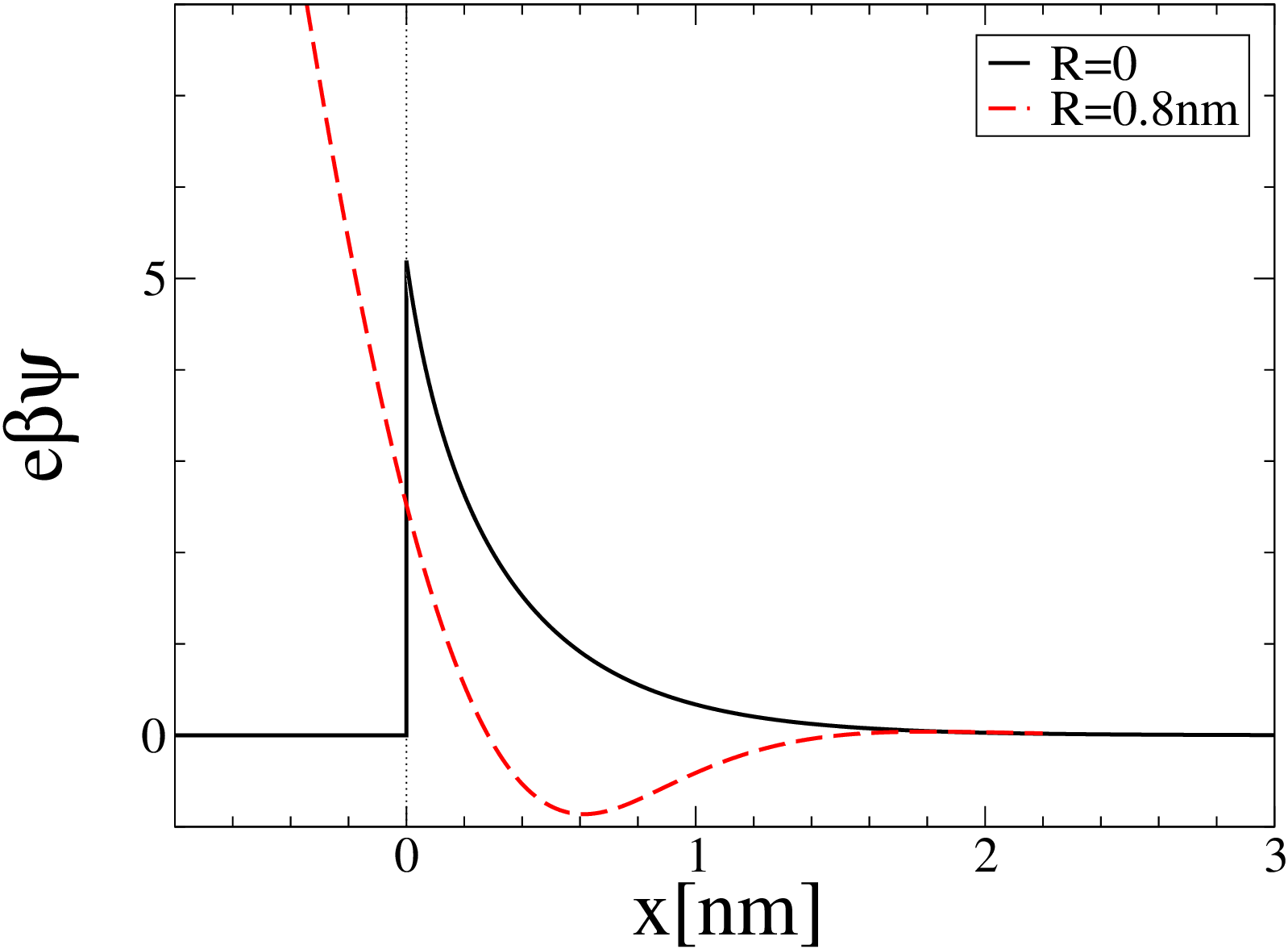}\\
\end{tabular}
\end{center}
\caption{The charge density and electrostatic potential for penetrable ions with charge 
distribution in Eq. (\ref{eq:omega}) with $R=0.8\,{\rm nm}$.   The ion centers are confined to 
the half-space $x>0$ and the vertical line at $x=0$ marks the half-space available to ion centers.  
The results for $R=0$ correspond to those for the standard PB equation. 
The system parameters are $\sigma_c=0.4\,{\rm C/m^2}$, $\lambda_B=0.72\,{\rm nm}$, 
and $c_s=1\,{\rm M}$.  }
\label{fig:FSPB_rhoc}
\end{figure}

In Fig. (\ref{fig:FSPB_rhoc}) we plot electrostatic quantities of penetrable ions: a charge 
density and an electrostatic potential.  Unlike the number density, these quantities are not 
confined to the region $x>0$, and extend to $x=-R$ as the charge of an ion sticks out.  Note 
how the sharp peak in the charge density for the standard PB model is smoothed-out in the 
finite-spread model.  
\graphicspath{{figures/}}

\begin{figure}[h] 
 \begin{center}%
 \begin{tabular}{rr}
  \includegraphics[width=0.49\textwidth]{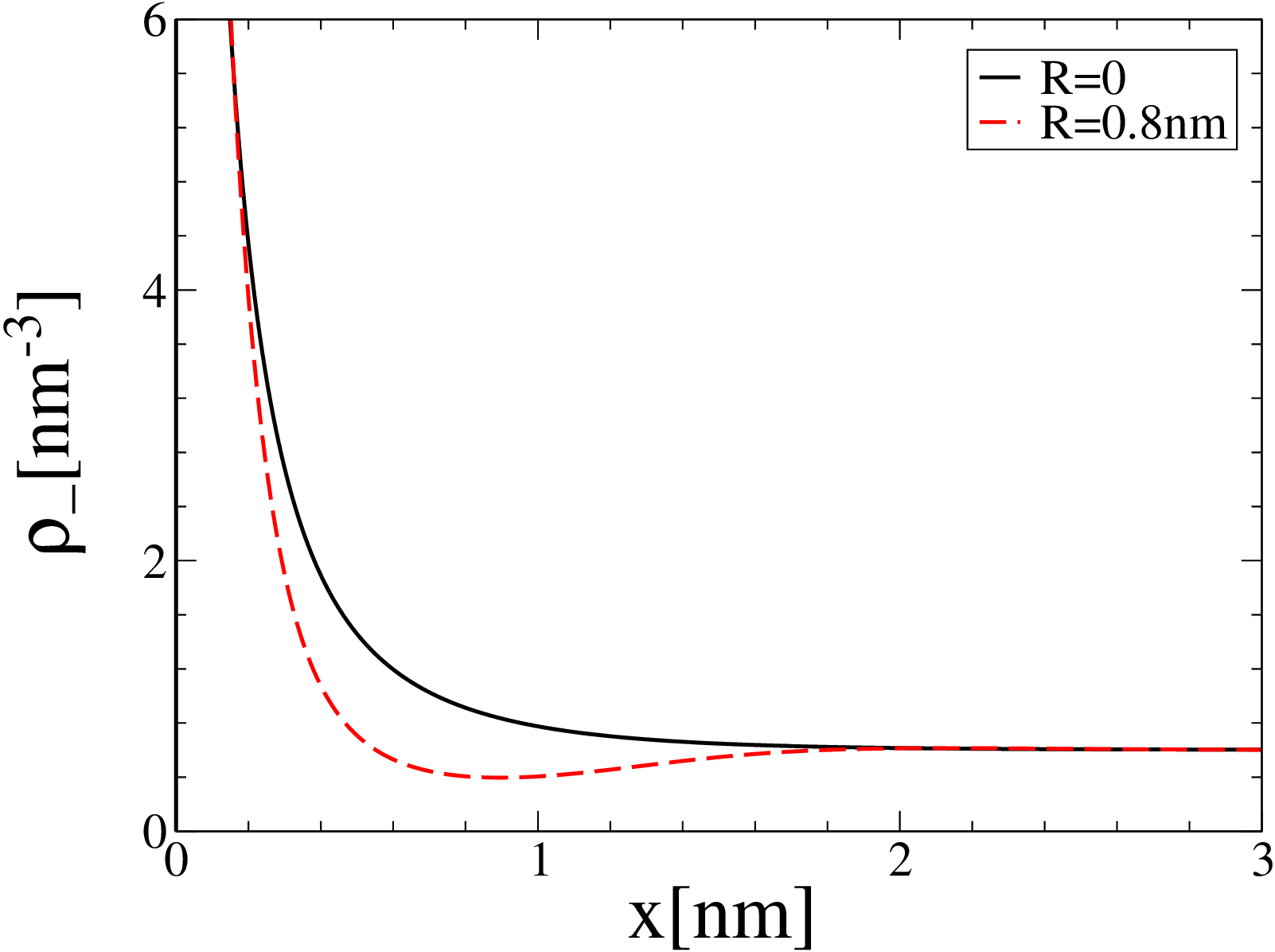}
  \includegraphics[width=0.49\textwidth]{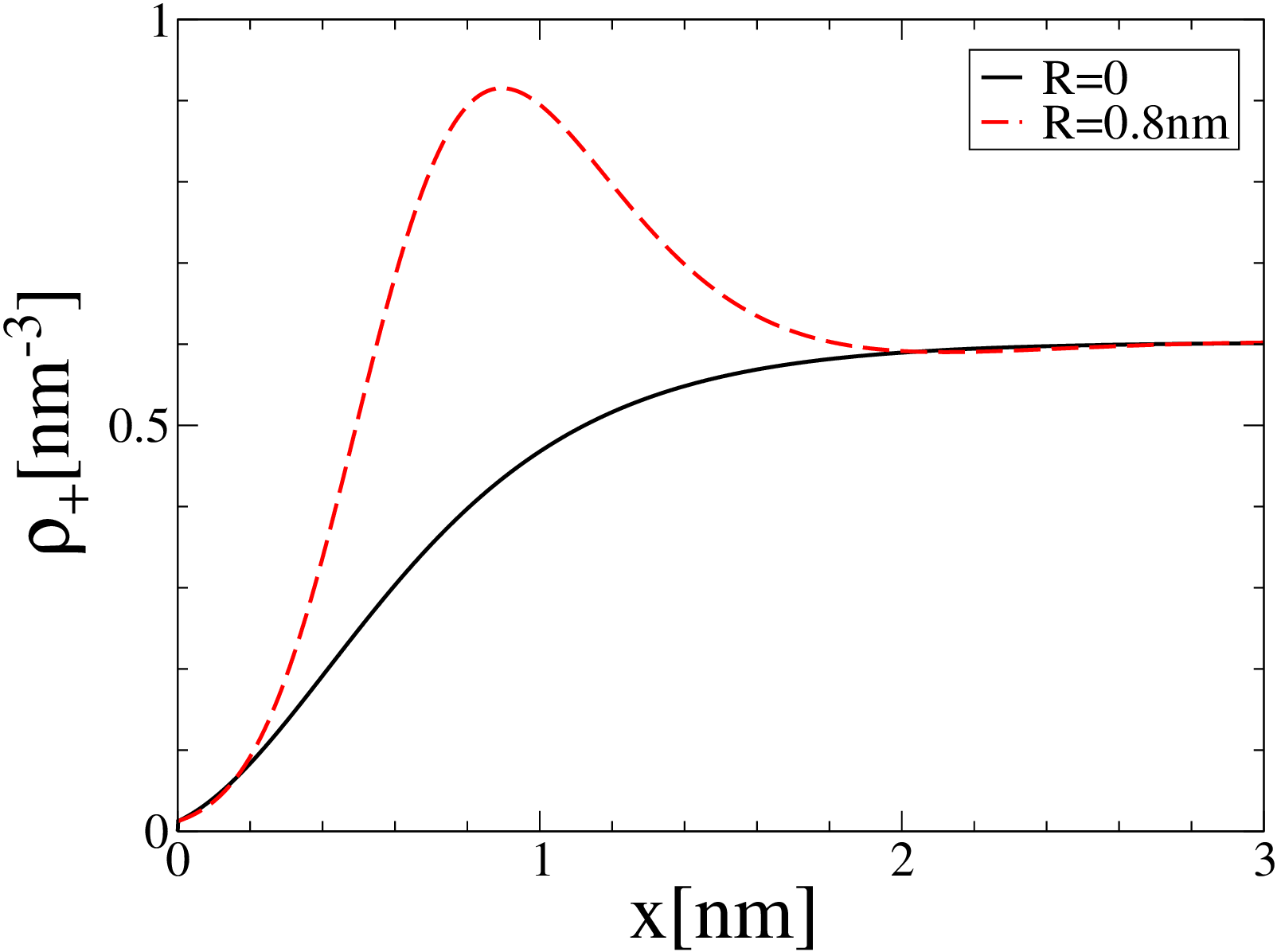}\\
\end{tabular}
 \end{center}
\caption{The number density profiles for counter- and co-ions for penetrable ions near a charged 
wall.  The same parameters as in Fig. (\ref{fig:FSPB_rhoc}).
}
\label{fig:FSPB_1}
\end{figure}

Fig. (\ref{fig:FSPB_1}) shows number density profiles for penetrable ions.  
The first striking feature is that profiles are non-monotonic.  More surprising still is the fact that
a surface charged is overcharged:  more counterions accumulate at a wall than needed for 
neutralizing it.   Overcharging is signaled by a peak in coion density and indicates attraction of
coions to a same-charged surface.  
Attraction between same-charged plates was found for dumbbell shaped counterions in 
\cite{Pincus08}, suggesting that charge inversion is a common feature of charges extended 
in space.
\graphicspath{{figures/}}

\begin{figure}[h] 
 \begin{center}%
\includegraphics[width=0.5\textwidth]{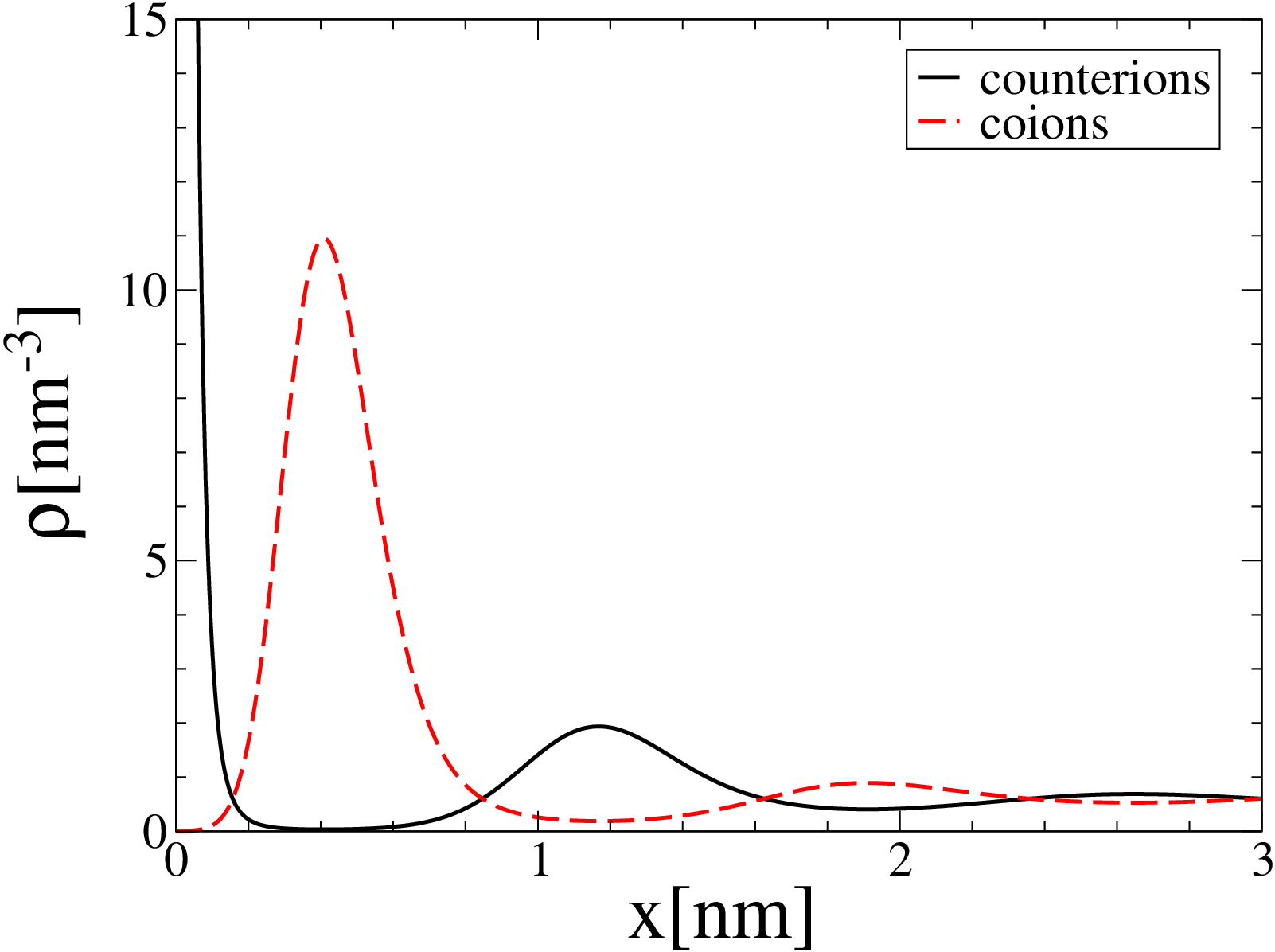}
\end{center}
\caption{Number density profiles for counterions and coions with valance number $3$.  
The increased electrostatic interactions magnify the features in Fig.  (\ref{fig:FSPB_1}).
Otherwise the same parameters as those in Fig. (\ref{fig:FSPB_rhoc}). }
\label{fig:rho_3}
\end{figure}

A closer look into plots reveals that overcharging and consequent charge inversion is a more
complex phenomenon.  To magnify these features we plot in Fig. (\ref{fig:rho_3}) the number 
densities for a symmetric $3:3$ electrolyte.  What we see is not a simple charge inversion 
but rather an alternating layers of counterions and coions leading to oscillations in density
profiles.   This behavior is reminiscent of polyelectrolyte layer-by-layer 
adsorption onto a charged substrate \cite{Borukhov98,Polylayer05,Polylayer10}.  
\graphicspath{{figures/}}

\begin{figure}[h] 
 \begin{center}%
\begin{tabular}{rr}
\includegraphics[width=0.48\textwidth]{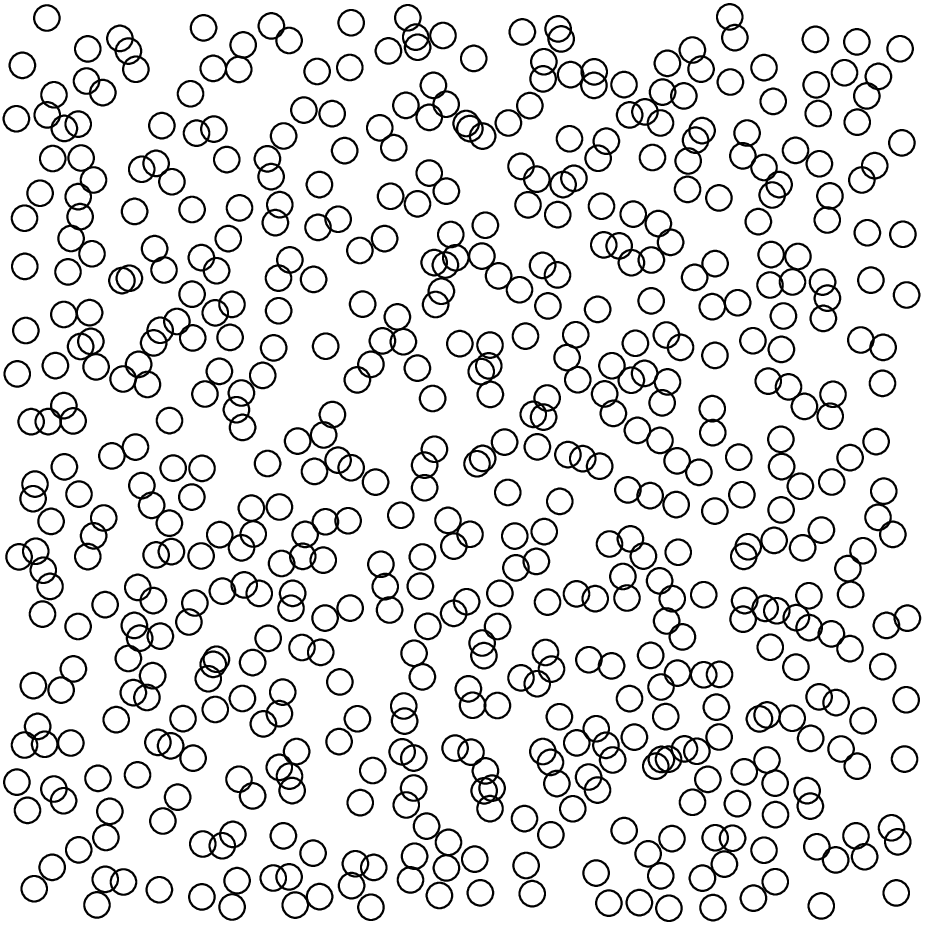}&
\includegraphics[width=0.48\textwidth]{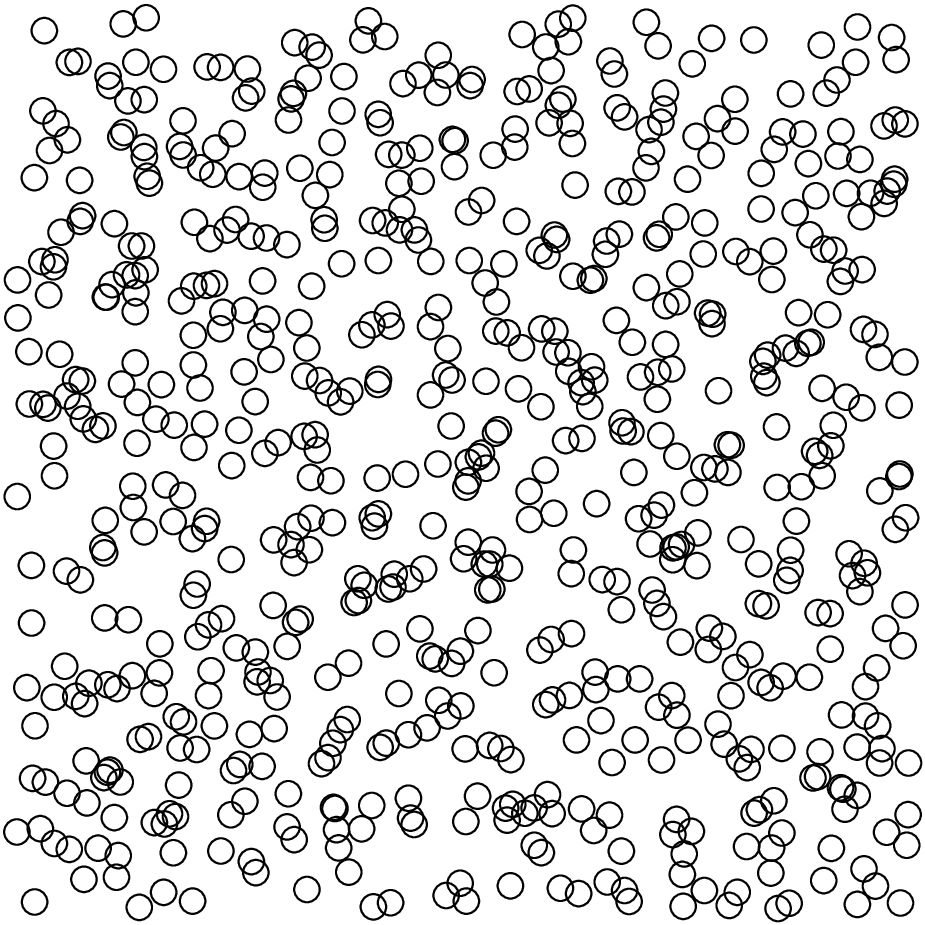}\\
\end{tabular}
\end{center}
\caption{Monte Carlo configuration snapshots for counterions adsorbed onto a charged wall 
(within the slice $0<x<0.35\,{\rm nm}$).  The conditions are the same as for results in Fig. (\ref{fig:FSPB_sim}).  The circles representing particles have diameter $\sigma=0.5\,{\rm nm}$ 
and are selected arbitrarily for visualization.  
The first snapshot is for $R=0.1\,{\rm nm}$, essential non-penetrable ions, and
the second snapshot is for $R=0.8\,{\rm nm}$, the fully penetrable ions.  
The 2D densities of each snapshot are 
$\rho_{2d}=2.34{\rm nm^{-2}}$ and $\rho_{2d}=2.68{\rm nm^{-2}}$, respectively.  For 
comparison, the surface charge density is $\sigma_c/e=2.50\,{\rm nm^{-2}}$, indicating 
overcharging for $R=0.8\,{\rm nm}$ counterions.  }
\label{fig:snapshot}
\end{figure}

\begin{figure}[h] 
 \begin{center}%
\includegraphics[width=0.5\textwidth]{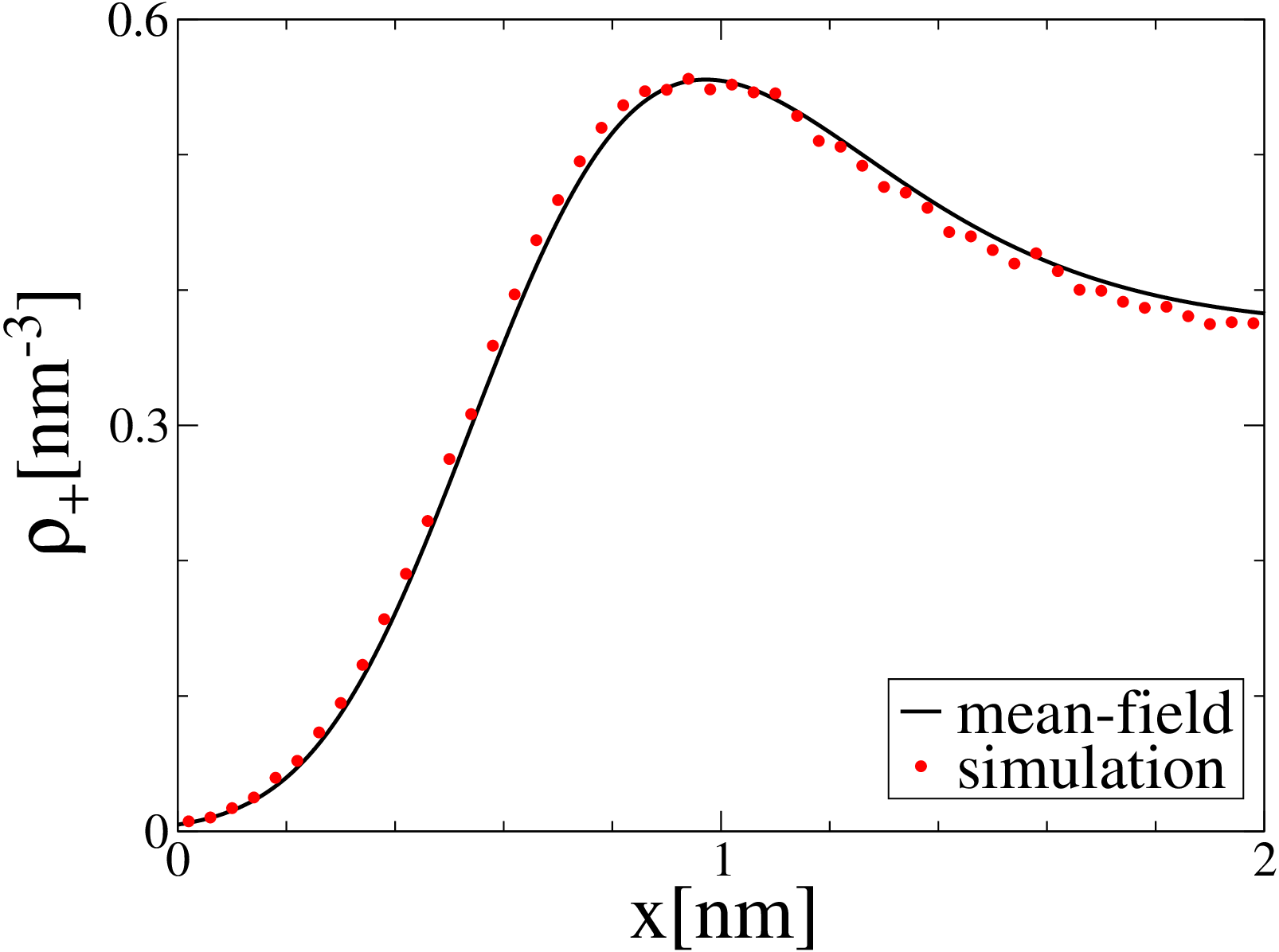}
\end{center}
\caption{The coion density profile near a charged wall.  The mean-field theory very accurately reproduces
the exact results of the Monte Carlo simulation.  The number of particles in the simulation box is 
$N_+=N_-=1200$, and the box size is $L_y=L_z=16{\rm nm}$ and $L_x=12{\rm nm}$.  The 
periodic boundary conditions are in the lateral $(y,z)$-directions.  The other parameters are:  
$\lambda_B=0.72\,{\rm nm}$, $R=0.8\,{\rm nm}$, and $\sigma_c=0.4\,{\rm Cm^{-2}}$.  }
\label{fig:FSPB_sim}
\end{figure}

In Fig. (\ref{fig:snapshot}) we show Monte Carlo snapshots for counterions adsorbed onto a 
charged wall.   The figure compares two systems: counterions that approximate a point-charges 
with size $R=0.1\,{\rm nm}$, and counterions that are fully penetrable with size 
$R=0.8\,{\rm nm}$.  A configuration for smaller ions appears more or less evenly distributed, 
indicating the strong presence of correlations, although the structure is still far from the Wigner 
crystal \cite{Trizac11}.  
There is no overcharging observed for this system, although deviations from the mean-field
are sufficiently significant to yield counterion density profiles different than those of 
the standard PB equation by being shifted closer to a wall.  
On the other hand, the configuration for $R=0.8\,{\rm nm}$ is more arbitrary and there are 
numerous overlaps.  In this system counterions are in excess and overcharge the surface charge.  
Furthermore, the correlations have no part in the overcharging mechanism as the Monte
Carlo and the finite-spread PB equation yield identical profiles, see Fig. (\ref{fig:FSPB_sim}).
\graphicspath{{figures/}}
\graphicspath{{figures/}}

The agreement between the mean-field profiles and those from the Monte Carlo simulation
rules out correlations as being behind the observed charge inversion.  The mechanism 
must then come from other quarters.  According to an orthodox charge inversion mechanism, 
driven by correlations, adsorbed onto a charged wall counterions come close to form a Wigner 
crystal, at least locally if not globally \cite{Yan02,Shklovskii02}.  High degree of ordering
leads to 
irregularities in the potential landscape of a now neutralized surface at distances smaller
or comparable to the lattice size.  These irregularities have attractive spots which can 
accommodate an additional counterion.  
For penetrable ions the absence of correlations makes this picture obsolete.  What drives
overcharging is the low energy cost for overlapping configurations.  This energy 
cost in inversely proportional to the radius $R$, see Eq. (\ref{eq:U_sphere}).   By taking
the limit $R\to\infty$ we recover the ideal gas limit.  
A divergence in the pair interactions for point-ions acts as an effective hard-shpere interaction
whose radius depends on the Bjerrum length as well as a surface charge.  This leads to the
effective excluded volume interactions.  By removing a divergence and permitting  
interpenetration, the effective excluded volume interactions are eliminated.  This completely 
changes the lateral structure of adsorbed counterions.

\subsubsection{stacked configurations}
A configuration snapshot for penetrable ions
in Fig. (\ref{fig:snapshot}) gives impression that counterions form stacked configurations, 
detected from the correlation function by the presence of a peak around $r=0$.  The presence 
of stacked formations is, furthermore, linked to instability of the Kirkwood analysis
\cite{Klein77b,Klein77a,Klein80,Frisch69,Nagai74}.  The Kirkwood analysis is the mean-field 
type of an analysis.  
As already said in section \ref{sec:MF}, even though the mean-field approximation does not 
incorporate explicit correlations in its free energy formulation, the correlations can be 
extracted using exact thermodynamic relations.  This is possible because the mean-field is 
not a self-consistent approximation.  Thus, the exact relation in Eq. (\ref{eq:delta_rho})
yields Eq. (\ref{eq:delta_rho2}) from the mean-field density, which then can be put in the form 
of the Ornstein-Zernike equation, as done in Eq. (\ref{eq:h_mf}), which, when Fourier 
transformed, becomes,
\begin{equation}
h_{\rm }(k) = -\frac{\beta u(k)}{1 + \rho\beta u(k)},
\end{equation}
and the corresponding structure factor, defined as $S(k) = 1 + \rho h(k)$, is
\begin{equation}
S_{\rm }(k) = \frac{1}{1 + \rho\beta u(k)}.
\end{equation}
All is good as long as $u(k)$ is a non-negative function.  But if for some modes $k$ the
Fourier transformed pair potential is negative, $S(k)$ becomes divergent for some wave 
number $k_0$, indicating divergent fluctuations.  In addition, the onset of this so-called 
Kirkwood instability coincides with an onset of the long-range correlation order and  
with a bifurcation point where a constant density no longer 
yields minimum free energy and another periodic solution takes precedence 
\cite{Klein77b}.  The instability was later linked to the spinodal of the supercooled liquid.  

Simulations of the penetrable sphere model (exhibiting the Kirkwood instability) showed 
the existence of stacked configurations (referred to as "clumps" in that work) at temperatures 
below instability \cite{Klein94}. Individual stacks arranged into crystal structure 
(corresponding to the global minimum) or amorphous glassy structures (corresponding to a 
local minimum).  Stacked structures gave rise to a peak around $r=0$ in the correlation function.  
A more thorough analysis supported by simulations linked the instability to crystals with 
multiply occupied lattice sites \cite{Lowen98,Schmidt99,Lowen01}.  Potentials whose Fourier 
transformed potential $u(k)$ is positive remain always stable and are not found to 
form stacked crystal formations.  Instead their solid phase exhibits reentrant melting 
upon squeezing, a behavior seen in water \cite{Stillinger97}.

In the present work we are interested in the liquid structure before the onset of instability.  
In particular, we want to know if the penetrable ions exhibit Kirkwood instability linked 
to stacked formations, and if yes, what role they play in a charge inversion mechanism.  

It is enough to consider the one component plasma of penetrable ions.  
The Fourier transformed pair potential depends on the distribution $\omega$ and is obtained 
from Eq. (\ref{eq:U_omega}),  
\begin{equation}
U(k) = \frac{\omega^2(k)}{\epsilon k^2}.
\end{equation}
For point-ions $\omega(k)=q$, and for penetrable ions $\lim_{k\to 0}\omega(k)=q$, since
at large separations the usual Coulomb interactions are recovered.  The difference between point 
and spread-out ions is seen for large $k$, where $S(k)$ reflects behavior for small separations.  
Regardless of the distribution $\omega(k)$, $U(k)\ge 0$ for any $k$ and the Kirkwood instability
does not occur.   Stacked formations, therefore, can be excluded as playing any part in a 
mechanism for charge inversion.

It is not clear, however, whether the conclusion holds for all Coulomb potentials with 
soft-core, or if it is specific to smeared-out ions.  To address this concern, we consider
the following soft-core Coulomb potential, 
\begin{equation}
 u(r) =
  \begin{cases}
   \frac{q^2}{4\pi\epsilon r} & \text{if } r \geq \sigma \\
   \frac{q^2}{4\pi\epsilon\sigma}       & \text{if } r < \sigma,
  \end{cases}
  \label{eq:U_sc}
\end{equation}
whose Fourier transform is 
\begin{equation}
u(k) = \frac{q^2}{\epsilon k^2}\frac{\sin k\sigma}{k\sigma}.
\end{equation}
$U(k)$ is no longer a non-negative function and yields instability in the mean-field structure 
factor.  The mean-field correlation function, furthermore, shows a peak at $r=0$,
\begin{equation}
h_{\rm }(0) = 
-\frac{1}{2\pi^2}\int_0^{\infty} dk\,\frac{\beta U(k)k^2}{1+\rho\beta u(k)},
\end{equation}
under certain conditions.  
We conclude that while the smearing-out procedure cannot lead to Kirkwood instability, 
this behavior is not general to all Coulomb potentials with soft-core.

.

\subsection{Needle-ions:  the case for non-spherical $\omega({\bf r}-{\bf r}')$}

The modified PB equation for non-spherical distributions is more complicated as there are
three additional degrees of freedom for a particle orientation to cope with.  Levy {\sl et al.} 
\cite{David13} derived the modified PB equation for a general distribution using the 
field-theory methodology.  

In this section we consider, as an example of non-spherical distribution, needle-ions.
We perform our construction as before, by writing down a mean-potential from which we obtain a 
number and charge densities.  A needle-ion consists of a charge $q$ uniformly distributed along 
a line of length $d$. The charge distribution of a needle-ion is 
\begin{equation}
\omega({\bf r}-{\bf r}_0,{\bf n}) = \frac{q}{d}\int_{-d/2}^{d/2} ds\,\delta({\bf r}_0+s{\bf n}-{\bf r}),
\label{eq:omega_needle}
\end{equation}
where ${\bf r}_0$ is a midpoint and ${\bf n}$ is a unit vector that designates orientation. 
The mean-potential that a needle-ion of a species $i$ feels when its center is at ${\bf r}$ is
\begin{equation}
w_i({\bf r},{\bf n}) = \frac{q_i}{d_i}\int d{\bf r}'\,\psi({\bf r}')\int_{-d_i/2}^{d_i/2} ds\,\delta({\bf r}+s{\bf n}-{\bf r}'),
\end{equation}
or, suppressing the delta function, we may alternatively write 
\begin{equation}
w_i({\bf r},{\bf n}) = \frac{q_i}{d_i}\int_{-d_i/2}^{d_i/2} ds\,\psi({\bf r}+s{\bf n}).  
\end{equation}
A nonlocal contribution comes from particle's finite extension in space.  The mean-field 
expression for the number density in space and orientation is
\begin{equation}
\varrho_i({\bf r},{\bf n}) \sim 
c_i\exp\Bigg[-\frac{\beta q_i}{d_i}\int_{-d_i/2}^{d_i/2} \!\!\!ds\,\psi({\bf r}+s{\bf n})\Bigg].
\label{eq:varrho_i}
\end{equation}
We still need an expression for a charge density to complete the construction.  
A charge density at location ${\bf r}$ has nonlocal contributions from neighboring 
ions that lie within a spherical region of radius $d/2$.  This region is described by  
the Heaviside step function $\theta(d/2-|{\bf r}-{\bf r}'|)$.  
However, being located within this region is not sufficient condition for contributing to the
charge density at ${\bf r}$.  There is additional condition of orientation: only ions with 
orientation
\begin{equation}
{\bf n} = \frac{{\bf r}'-{\bf r}}{|{\bf r}'-{\bf r}|},
\end{equation}
contribute to the charge density at ${\bf r}$.  
Each species' contribution to the charge density is
\begin{equation}
\rho_c^{i}({\bf r}) \sim {q_i}\int d{\bf r}'\,\theta\bigg(\frac{d_i}{2}-|{\bf r}'-{\bf r}|\bigg) 
\varrho_i\bigg({\bf r}',\frac{{\bf r}'-{\bf r}}{|{\bf r}'-{\bf r}|}\bigg).
\end{equation}
Using Eq. (\ref{eq:varrho_i}) to substitute for $\varrho_i$, the total charge density becomes
\begin{eqnarray}
\rho_c({\bf r})&=&
\sum_{i=1}^N{q_ic_i}\Bigg(\frac{6}{\pi d_i^3}\Bigg)\int d{\bf r}'\,\theta\bigg(\frac{d_i}{2}-|{\bf r}'-{\bf r}|\bigg)
\nonumber\\&\times&
\exp\Bigg[-\frac{\beta q_i}{d_i}
\int_{-d_i/2}^{d_i/2} \!\!\!ds\,\psi\bigg({\bf r}'+s\frac{{\bf r}'-{\bf r}}{|{\bf r}'-{\bf r}|}\bigg)\Bigg],\nonumber\\
\end{eqnarray}
where the coefficient $6/(\pi d_i^3)$ comes from the limit $\psi\to 0$, where all orientations are 
equally probable and the charge density properly recovers its bulk value,
$\rho_c\to\sum_{i=1}^Kc_iq_i$.  For a symmetric $1:1$ electrolyte and ions of the same 
length we get
\begin{eqnarray}
\rho_c({\bf r})&=&\frac{12ec_s}{\pi d^3}\int d{\bf r}'\,\theta\bigg(\frac{d}{2}-|{\bf r}'-{\bf r}|\bigg)
\nonumber\\&\times&
\sinh\Bigg[-\frac{\beta e}{d}
\int_{-d/2}^{d/2} \!\!\!ds\,\psi\bigg({\bf r}'+s\frac{{\bf r}'-{\bf r}}{|{\bf r}'-{\bf r}|}\bigg)\Bigg].\nonumber\\
\end{eqnarray}
By inserting this result into the Poisson equation, $\epsilon\nabla^2=-\rho_c$, we obtain the 
desired modified PB equation for needle-ions,
\begin{eqnarray}
-\epsilon\nabla^2\psi &=&
\frac{12ec_s}{\pi d^3}\int d{\bf r}'\,\theta\bigg(\frac{d}{2}-|{\bf r}'-{\bf r}|\bigg)\nonumber\\&\times&
\sinh\Bigg[-\frac{\beta e}{d}
\int_{-d/2}^{d/2} \!\!\!ds\,\psi\bigg({\bf r}'+s\frac{{\bf r}'-{\bf r}}{|{\bf r}'-{\bf r}|}\bigg)\Bigg].\nonumber\\
\end{eqnarray}

\subsection{Dumbbell ions}
There are cases when multivalent organic ions, such as certain DNA condensing agents or short 
stiff polyeletrolyes, have a rod-like structure wherein charges are spatially separated from each
other \cite{Pincus08,Bohinc04,Bohinc08,Bohinc11,Bohinc12}.  
These separations are not small and are comparable to typical screening lengths, 
$\sim 1{\rm nm}$.  The simplest representation of such ions is a dumbbell, a structure  
made of two point charges at fixed separation $d$.   
The distribution of a dumbbell ion located at ${\bf r}_0$ is
\begin{equation}
\omega({\bf r}-{\bf r}_0,{\bf n}) = q\delta({\bf r}-{\bf r}_0) + q\delta({\bf r}-{\bf r}_0-d{\bf n}).  
\label{eq:omega_dumbbell}
\end{equation}
These dumbbell counterions were found to give rise to attraction between two same-charged 
plates by bridging two surfaces, thereby 
providing a finite equilibrium distance between surfaces.  

The mean-field construction for dumbbell ions follows the usual route.  
The mean-potential for an ion species $i$ is
\begin{equation}
w_i({\bf r}) = q_i\psi({\bf r}) + q_i\psi({\bf r}+d_i{\bf n}),
\end{equation}
which then leads to the following distribution,
\begin{equation}
\varrho_i({\bf r},{\bf n}) \sim c_ie^{-\beta q_i\psi({\bf r})-\beta q_i\psi({\bf r}+d_i{\bf n})}.
\end{equation}
A properly normalized charge density then is written as
\begin{equation}
\rho_c({\bf r}) = \sum_{i=1}^K2q_ic_ie^{-\beta q_i\psi({\bf r})}
\int d{\bf r}'\,\frac{\delta(d_i-|{\bf r}-{\bf r}'|)}{4\pi d_i^2}e^{-\beta q_i\psi({\bf r}')},
\end{equation}
and it remains now to put this into the Poisson equation.  
The mean-field Poisson equation for a $1:1$ dumbbell electrolyte with equal sized particles 
becomes
\begin{equation}
\epsilon\nabla^2\psi({\bf r}) = 4ec_s
\int d{\bf r}'\,\frac{\delta(d-|{\bf r}-{\bf r}'|)}{4\pi d^2}
\sinh\bigg[\beta e\psi({\bf r})+\beta e\psi({\bf r}')\bigg].
\end{equation}
For the wall model this becomes
\begin{equation}
\epsilon\psi'' = \frac{2ec_s}{d} \int_{-d}^{d}ds\,\sinh\big[\beta e\psi(x)+\beta e\psi(x+s)\big]
\theta(x)\theta(x+s),
\end{equation}
where the Heaviside step functions ensure that dumbbell ions do not go through the wall.

\section{Short-range non-electrostatic interactions}

So far we have considered only interactions due to electrostatic structure of an ion.  
In addition to these there are also non-electrostatic interactions, generally short-ranged
and repulsive, the most obvious of which are the excluded volume interactions whose
source is traced to 
the Pauli exclusion principle which prohibits two electrons from occupying the same quantum 
state \cite{Dyson}.

For ions in aqueous solution the excluded volume interactions are enhanced due to formation
of a hydration shell.  
Excluded interactions can, furthermore, lead to effective, softer type of interactions.  For example, 
the effective interactions between two linear polymers in a good solvent can be represented
with the Gaussian functional form and result from a self-avoiding walk between dissolved 
polymer chains \cite{Hansen00a}.  There are many other types of exotic interactions in 
soft-matter systems, and some further examples include effective interactions between star 
polymers, dendrimers, etc. \cite{Likos01}.  In this section we consider different schemes for 
incorporating short-range interactions of non-electrostatic origin.

\subsection{the mean-field implementation}
The simplest way to implement short-range interactions is to use the mean-field 
framework.  Considering point-charges, the mean-potential for such 
an implementation is
\begin{equation}
\beta w_i({\bf r}) = \beta q_i\psi({\bf r}) + \beta\sum_{j=1}^K\int d{\bf r}' \rho_j({\bf r}') u_{ij}({\bf r}-{\bf r}'),
\end{equation}
where $u_{ij}$ designates the non-electrostatic interactions between particles of the species $i$ 
and $j$.  The corresponding mean-field density is
\begin{equation}
\rho_i = c_ie^{-\beta q_i\psi}
e^{-\beta\sum_{j=1}^K\int d{\bf r}' (\rho_j({\bf r}')-c_j) u_{ij}({\bf r}-{\bf r}')},
\label{eq:rho_i}
\end{equation}
which recovers bulk density in the limit $\psi\to 0$.  
The charge density is $\rho_c=\sum_{i=1}^Kq_i\rho_i$ and the resulting mean-field Poisson 
equation is
\begin{equation}
-\varepsilon\nabla^2\psi  = 
\sum_{i=1}^Kc_iq_ie^{-\beta q_i\psi}
e^{-\beta\sum_{j}\int \!d{\bf r}' (\rho_j({\bf r}')-c_j) u_{ij}({\bf r}-{\bf r}')}.
\label{eq:PB+MF}
\end{equation}
The approximation consists of two coupled equations, Eq. (\ref{eq:rho_i}) and (\ref{eq:PB+MF}).  
Note that the implementation of non-electrostatic interactions leads to nonlocal approximation
where the density is convoluted with the pair interaction $u_{ij}({\bf r}-{\bf r}')$.  

If the electrolyte is symmetric, $1:1$, and there is only one type of short-range interactions 
for each particles, Eq. (\ref{eq:PB+MF}) reduces to a more familiar form,
\begin{equation}
\varepsilon\nabla^2\psi({\bf r}) = 2c_s\sinh(\beta e\psi)\,e^{-\beta\int \!d{\bf r}' (\rho({\bf r}')-2c_s) 
u({\bf r}-{\bf r}')},
\label{eq:PB+MF1}
\end{equation}
where the total number density, $\rho=\rho_++\rho_-$, is given by
\begin{equation}
\rho = 2c_s\cosh(\beta e\psi)e^{-\beta\int d{\bf r}' (\rho({\bf r}')-2c_s) u({\bf r}-{\bf r}')}.
\label{eq:rho_i1}
\end{equation}

As a specific example, we consider penetrable sphere ions (PSM) whose short-range 
repulsive interaction is 
\begin{equation}
\beta u(|{\bf r}-{\bf r}'|) = \varepsilon\,\theta(\sigma-|{\bf r}-{\bf r}'|),
\end{equation}
where $\sigma$ is the diameter of a penetrable sphere, and $\varepsilon$ is the strength.  
In the limit $\varepsilon\to\infty$ the hard-core interactions are recovered.  

\begin{figure}[h] 
 \begin{center}%
\includegraphics[width=0.5\textwidth]{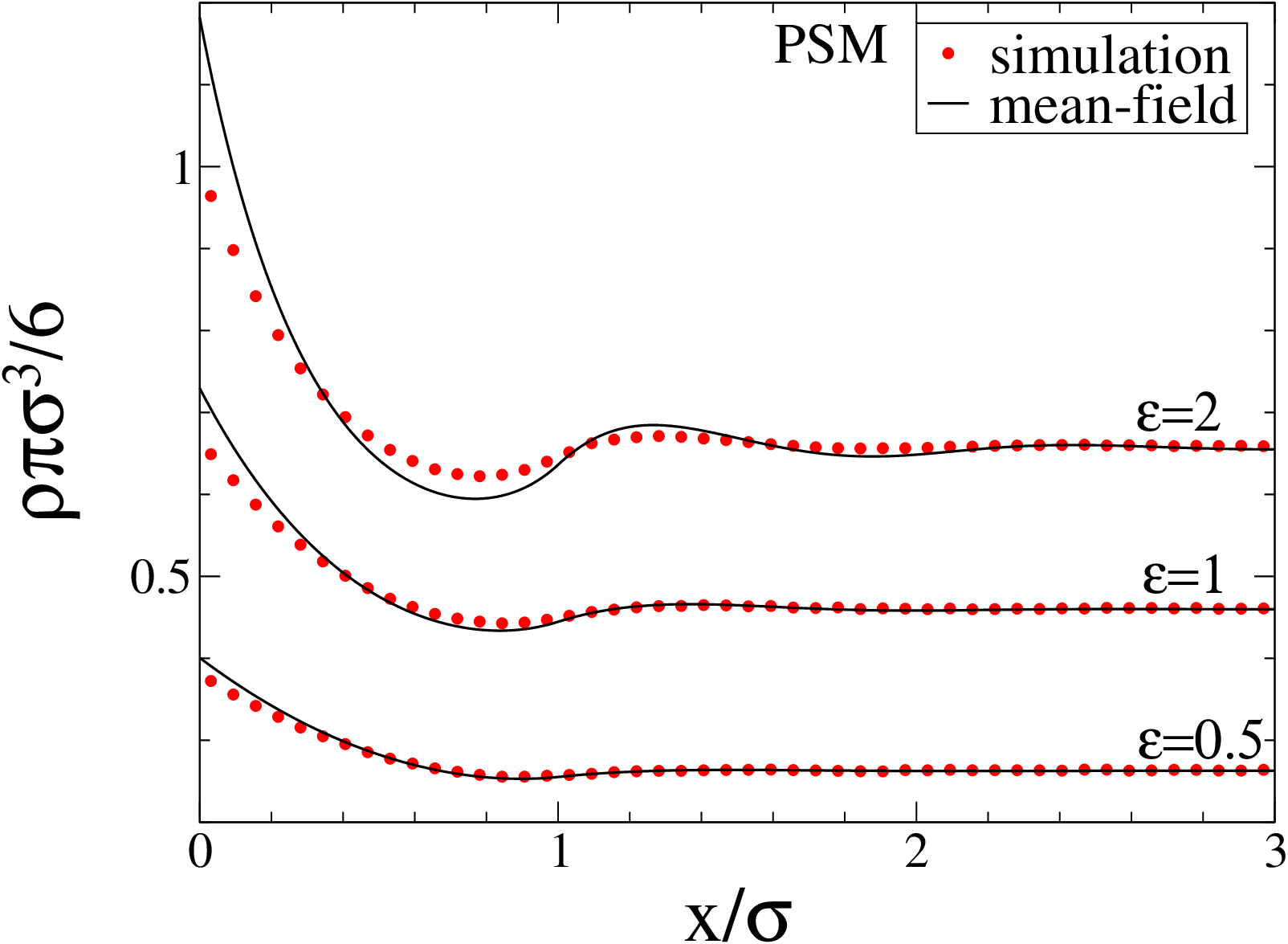}
\end{center}
\caption{Density profiles of uncharged penetrable spheres near a planar wall.  
Particle centers are confined in the $x$-axis, $x\in[0,12.5\sigma]$.  The number of particles is fixed,
$\int dx\, \rho(x)=N=1000$.  For Monte Carlo simulation the dimensions of the simulation 
box are $12.5\times 12.5\times 12.5\sigma$.  The box encloses $N=1000$ particles.  
In the $y$ and $z$ directions periodic boundary conditions are used.  The influence of the 
second wall is minor, and we refer to this system as the wall model.  }
\label{fig:rho_PSM}
\end{figure}

At first we consider uncharged penetrable spheres and compare results with those from 
simulation.  
Results for a wall model are plotted 
in Fig. (\ref{fig:rho_PSM}) which shows density profiles of penetrable spheres near a planar wall.  
\graphicspath{{figures/}}
The mean-field becomes less accurate as $\varepsilon$ increases where it overestimates  
the contact values, $\rho_w$, which are related to the bulk pressure via the contact value 
theorem,
\begin{equation}
\rho_w = \beta P.
\end{equation}
Overestimated contact density values imply that the mean-field pressure is larger than
the true one, $P_{\rm mf}>P$.  
To obtain the mean-field pressure we use the virial equation \cite{Hansen86},
\begin{equation}
\beta P = \rho_b-\frac{2\pi}{3}\rho_b^2\int_0^{\infty} dr\,r^3g(r)
\frac{\partial\beta u(r)}{\partial r},
\end{equation}
from which we discard correlations, $g(r)=1$, according to the mean-field procedure, and we get
 \begin{equation}
\beta P_{\rm mf} = \rho_b + \frac{\varepsilon\rho_b^2}{2}\Big(\frac{4\pi\sigma^3}{3}\Big),
\end{equation}
where we used $\frac{\partial \beta u(r)}{\partial r}=-\varepsilon\delta(r-\sigma)$.  
The mean-field contact value theorem for penetrable spheres, therefore, is
\begin{equation}
\rho_w = \rho_b + \frac{\varepsilon\rho_b^2}{2}\Big(\frac{4\pi\sigma^3}{3}\Big).
\end{equation}
The lower contact density for a true system implies a neglect of correlations in the 
mean-field approximation.  

Having in mind hard-spheres as a model system for excluded volume interactions, we can 
make contact with it by setting $\varepsilon=1$, where the resulting mean-field pressure, 
\begin{equation}
\beta P_{\rm mf} = \rho_b(1 + 4\eta),
\end{equation}
agrees to the second virial term with the the pressure for hard-spheres, where 
$\eta=\pi\rho\sigma^3/6$ is the packing fraction.

\begin{figure}[h] 
 \begin{center}%
\includegraphics[width=0.5\textwidth]{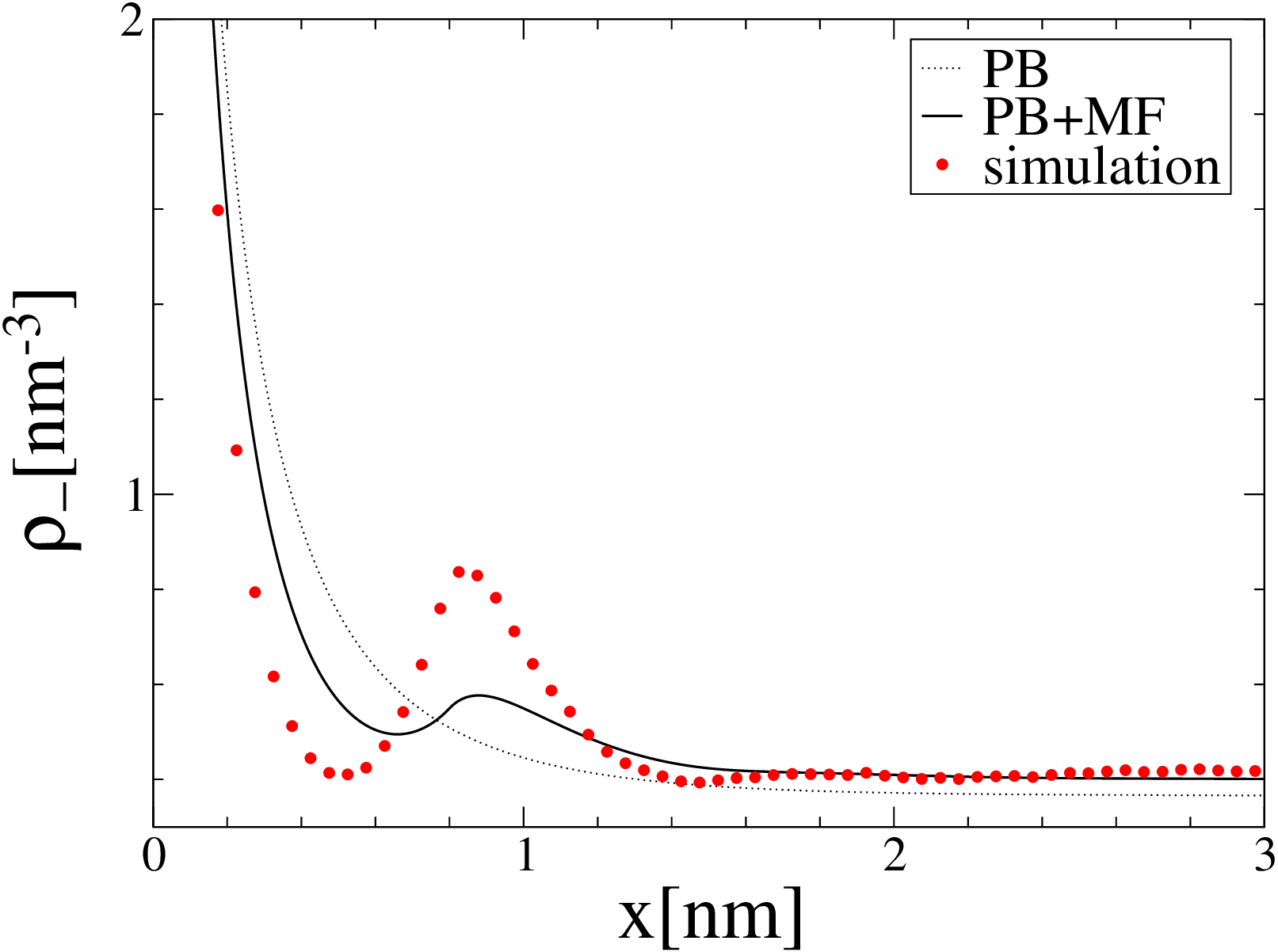}
\end{center}
\caption{The counterion density near a charged wall.  
The system is confined between two charged walls at $x=0$ and $x=6\,{\rm nm}$ and the 
surface charge is $\sigma_c=-0.2\,{\rm Cm^{-2}}$.  The other parameters are: the Bjerrum 
length $\lambda_B=0.72\,{\rm nm}$, and the diameter of penetrable spheres 
$\sigma=0.8{\rm nm}$.  There is no dialectric discontinuity across an interface.  The number of 
cations is $\int dx\, \rho_i(x)=N_i=300$ and $N_+=N_-$.  The penetrability parameter is set to 
$\varepsilon=1$.  For simulation we used the hard-sphere limit, $\varepsilon\to\infty$.  }
\label{fig:PB+MF}
\end{figure}

We next consider charged penetrable spheres with $\varepsilon=1$ and solve 
Eq. (\ref{eq:PB+MF1}) and Eq. (\ref{eq:rho_i1}) for symmetric $1:1$ electrolyte.  For the wall 
model the boundary conditions are the same as for the standard PB equation, and the contact 
value theorem is
\begin{equation}
\rho_w = \rho_b(1+4\eta) + \frac{\beta\sigma_c^2}{2\epsilon}.  
\end{equation}
In Fig. (\ref{fig:PB+MF}) 
we plot density profiles for counterions near a charged wall.  In comparison with the standard 
PB equation, the penetrable sphere ions generate a non-monotonic structure of a 
double-layer, where we see the emergence of a secondary peak.  
The structure is a result of overcrowding, where counterions coming to neutralize the 
surface charge cannot be packed too closely together.  
\graphicspath{{figures/}}
The simulation results for hard-sphere particles with the same diameter yield a profile
with stronger structure.  The penetrable sphere model captures only qualitatively 
these features.  

\begin{figure}[h] 
 \begin{center}%
 \begin{tabular}{rr}
  \includegraphics[width=0.5\textwidth]{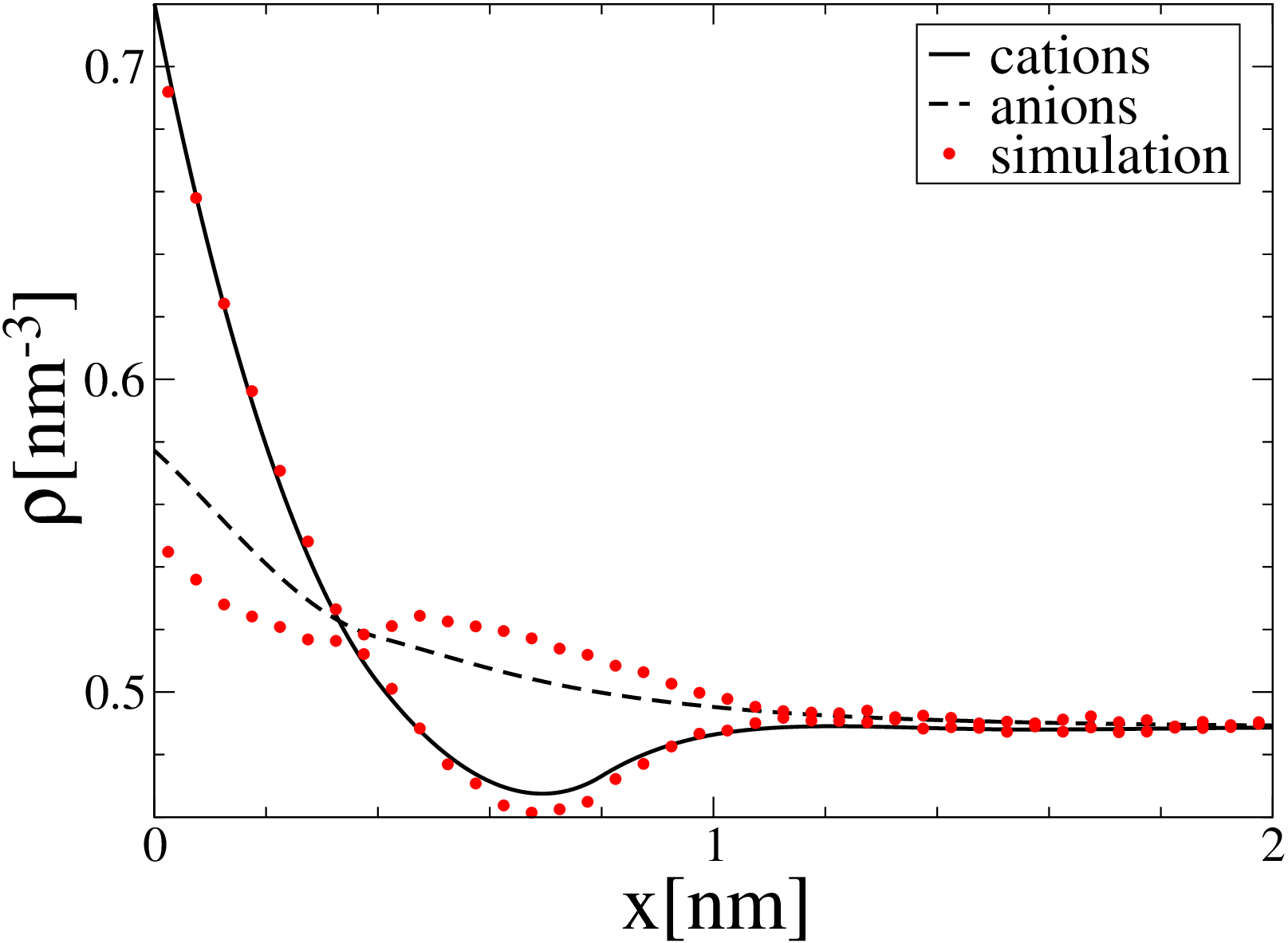}
  \includegraphics[width=0.5\textwidth]{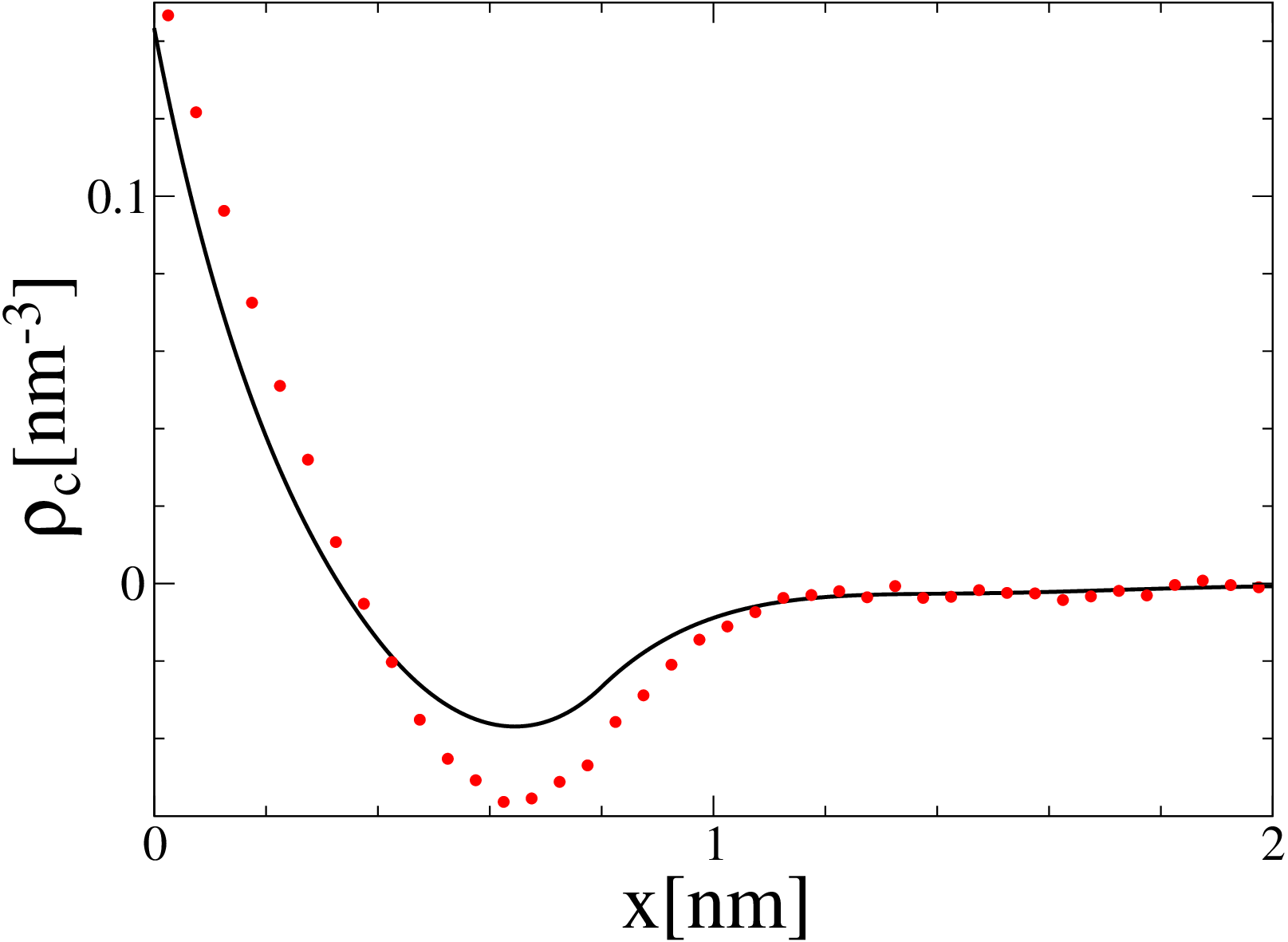}\\
\end{tabular}
 \end{center}
\caption{Density profiles of cations, anions, and of total charge near an uncharged wall.  Two 
parallel uncharged plates at $x=0$ and $x=6\,{\rm nm}$ confine all particle centers.  The dielectric 
constant is the same across an interface.  The Bjerrum lenght is $\lambda_B=0.72\,{\rm nm}$, 
the ion sizes are $\sigma_{++}=0.8\,{\rm nm}$, 
$\sigma_{--}=0\,{\rm nm}$, and $\sigma_{+-}=0.4\,{\rm nm}$, and 
$\int dx \rho_i(x)=N_i$ where $N_+=N_-=300$.  The simulation was done for the same system 
but for hard-spheres, $\varepsilon\to\infty$, while in the numerical model we used $\varepsilon=1$.
}
\label{fig:PB+MF0}
\end{figure}

The present model can be used to study ion specific effects.  For neutral surfaces, 
size asymmetry can lead to different density profiles of ions with the same valance number.  
This leads to charge build-up across an interface.   In Fig. (\ref{fig:PB+MF0}) we show 
density profiles near a neutral wall for a $1:1$ electrolyte with size asymmetry.
The larger cations exhibit greater structure and are squeezed against the wall which, 
in turn, leads to a charge build-up that pulls anions.  
Simulation results for hard-spheres with the same diameters show similar profiles, however,
anions have greater structure as they are depleted from the immediate wall vicinity.  
\graphicspath{{figures/}}

\subsection{short-range interactions beyond the mean-field}
In this section we develop a more accurate implementation of short-range interactions, 
while keeping electrostatics at the mean-field level.  Such a procedure introduces asymmetric 
treatment of different parts of a pair potential.  To formally set up and 
justify this asymmetry of methods, we consider the scaled pair potential,
\begin{equation}
u_{ij}^{\lambda}=u_{ij}^{\rm hs}+\lambda \Bigg(\frac{q_iq_j}{4\pi\epsilon|{\bf r}-{\bf r}'|}\Bigg), 
\end{equation}
where $u^{\rm hs}_{ij}$ is the hard-sphere potential, and $\lambda$ is the scaling parameter.  
For $\lambda=0$ a hard-sphere system is recovered.   The density of all ions is independent
of $\lambda$ and is kept fixed by the external electrostatic potential, $\psi_{\rm ext}^{\lambda}$.
The partition function for this system is
\begin{equation}
Z_{\lambda} = \frac{1}{\prod_{j=1}^K (N_j!\Lambda^{3N_j})}
\int\prod_{i=1}^{N} d{\bf r}_i
e^{-\beta \sum_{i}^N Q_i\psi_{\rm ext}^{\lambda}({\bf r}_i)}e^{-\frac{\beta}{2}\sum_{i,j}^N
\big[\lambda \frac{Q_iQ_j}{4\pi\epsilon|{\bf r}_i-{\bf r}_j|}+u_{ij}^{\rm hs}({\bf r}_i,{\bf r}_j)\big]},
\end{equation}
where $N=\sum_{j=1}^KN_j$ is the total number of particles, $N_j$ is the number of particles 
of a species $i$, and the charges $\{Q_i\}$ have the following values
$Q_{N_{j-1}+1}=Q_{N_{j-1}+2}=\dots=\dots=Q_{N_{j-1}+N_j}=q_j$, where $q_i$ is the charge
of a species $i$.  
The exact functional form of $\psi_{\rm ext}^{\lambda}({\bf r})$ is not needed and 
it will not appear in the final result.  It is sufficient to know that it keeps densities fixed
at their physical shape for any value $\lambda$, and $\psi_{\rm ext}^{\lambda=1}$ recovers the 
true external potential, $\psi_{\rm ext}$.

The free energy, $\beta F_{}=-\log Z_{}$, is obtained from thermodynamic integration, 
\begin{eqnarray}
F[\{\rho_i\}] &=& 
F_{\lambda=0}[\{\rho_i\}] + \int_0^1 d\lambda\,\frac{\partial F_{\lambda}}{\partial\lambda}\nonumber\\
&=&F_{\rm id}[\{\rho_i\}] + F^{\rm hs}_{\rm ex}[\{\rho_i\}] + 
\int d{\bf r}\,\rho_c({\bf r})\psi_{\rm ext}^{\lambda=0}({\bf r}) + 
\int_0^1 d\lambda\,\frac{\partial F_{\lambda}}{\partial\lambda},\nonumber\\
\label{eq:F_lambda}
\end{eqnarray}
where $F_{\rm ex}^{\rm hs}[\rho]$ is the excess free energy due to hard-sphere interactions and 
is a functional of density only.  The integrand of the last term after evaluation is
\begin{equation}
\frac{\partial F_{\lambda}}{\partial\lambda} = 
\int d{\bf r}\,\rho_c({\bf r})\frac{\partial \psi_{\rm ext}^{\lambda}({\bf r})}{\partial\lambda}
+\frac{1}{2}\sum_{i,j}^K
\int d{\bf r}\int d{\bf r}'\,\frac{q_i\rho_i({\bf r})q_j\rho_j({\bf r}')}{4\pi\epsilon|{\bf r}-{\bf r}'|}
g^{\lambda}_{ij}({\bf r},{\bf r}'),
\end{equation}
where $g_{ij}=1+h_{ij}$.  Inserting this into Eq. (\ref{eq:F_lambda}) we get
\begin{eqnarray}
F[\{\rho_i\}] &=& 
F_{\rm id}[\{\rho_i\}] + F^{\rm hs}_{\rm ex}[\{\rho_i\}] + \int \!\!d{\bf r}\,\rho_c({\bf r})\psi_{\rm ext}({\bf r})
+\frac{1}{2}\int\!\!d{\bf r}\!\!\int \!\!d{\bf r}'\,
\frac{\rho_c({\bf r})\rho_c({\bf r}')}{4\pi\epsilon|{\bf r}-{\bf r}'|}\nonumber\\
&+&\frac{1}{2}\sum_{i,j}^K\int \!\!d{\bf r}\!\!\int \!\!d{\bf r}'\,
\frac{q_i\rho_i({\bf r})q_j\rho_j({\bf r}')}{4\pi\epsilon|{\bf r}-{\bf r}'|}
\int_0^1 d\lambda \,
h_{ij}^{\lambda}({\bf r},{\bf r}').
\nonumber\\
\label{eq:F_lambda2}
\end{eqnarray}
By setting correlations to zero, $h_{ij}^{\lambda}=0$,
\begin{equation}
F[\{\rho_i\}] \approx F_{\rm id}[\{\rho_i\}] + F^{\rm hs}_{\rm ex}[\{\rho_i\}] + 
\int d{\bf r}\,\rho_c({\bf r})\psi_{\rm ext}({\bf r})
+\frac{1}{2}\int d{\bf r}\int d{\bf r}'\,\frac{\rho_c({\bf r})\rho_c({\bf r}')}{4\pi\epsilon|{\bf r}-{\bf r}'|},
\label{eq:F_MF+HS}
\end{equation}
we have an approximation that completely neglects terms coupling the electrostatic
and hard-core interactions, and that consists of the free energy for hard-spheres plus the 
mean-field electrostatic correction.

Densities are obtained from the minimum condition,   
$\frac{\delta F}{\delta \rho_i({\bf r})} = 0$,
and
\begin{equation}
\rho_i({\bf r}) = 
c_i\exp\Big[-\beta q_i\psi({\bf r}) -
\frac{\delta \beta F^{\rm hs}_{\rm ex}}{\delta \rho_i({\bf r})} + \beta\mu_{\rm ex}\Big],  
\label{eq:rhoi_bmf}
\end{equation}
where $\mu_{\rm ex}=\frac{\partial F_{\rm ex}^{\rm hs}}{\partial\rho}|_{\rho=\rho_b}$ is the excess 
chemical potential over the ideal contribution, $\mu_{\rm id}=\log\rho_b\Lambda^3$, 
and $\psi$ is the total electrostatic potential.  
Implementing this into the Poisson equation we get
\begin{equation}
-\epsilon\nabla^2\psi = \sum_{i=1}^K c_iq_i 
\exp\Big[-\beta ez_i\psi-\frac{\delta \beta F^{\rm hs}_{\rm ex}}{\delta \rho_i({\bf r})}
+\beta\mu_{\rm ex}\Big].  
\label{eq:PB+HS}
\end{equation}
Then for $1:1$ electrolyte with all ions having the same size,
\begin{equation}
\epsilon\nabla^2\psi = 2c_s e\sinh(\beta e\psi)
\exp\Big[-\frac{\beta\delta F_{\rm ex}^{\rm hs}}{\delta\rho} 
+ \beta\mu_{\rm ex}\Big]
\end{equation}
where $\rho = \rho_++\rho_-$ is
\begin{equation}
\rho = 2c_s \cosh(\beta e\psi)\exp\Big[-\frac{\beta\delta F_{\rm ex}^{\rm hs}}{\delta\rho} 
+ \beta\mu_{\rm ex}\Big].
\end{equation}

\subsubsection{perturbative expansion and the dilute limit}	
To complete the approximation it remains to find an expression for the excess free energy.
The first two terms of the virial expansion for $F^{\rm hs}_{\rm ex}$ are \cite{Evans79}
\begin{eqnarray}
F^{\rm}_{\rm ex} &=& 
\frac{1}{2}\int \!\!d{\bf r}_1\!\!\int\!\! d{\bf r}_2\,\rho({\bf r}_1)\rho({\bf r}_2)\bar f(r_{12})
\nonumber\\&+&
\frac{1}{6}\int \!\!d{\bf r}_1\!\!\int \!\!d{\bf r}_2\!\!\int \!\!d{\bf r}_3\,
\rho({\bf r}_1)\rho({\bf r}_2)\rho({\bf r}_3)\bar f(r_{12})\bar f(r_{23})\bar f(r_{31})
\nonumber\\&+&\dots
\label{eq:Fex_vir}
\end{eqnarray}
where 
$\bar f_{}(r)=1-e^{-\beta u_{}(r)}$ is the negative Mayer $f$-function, and for 
hard-spheres is given by the Heaviside function, $\bar f_{}(r)=\theta(\sigma_{}-r)$.  
In the dilute limit the first term dominates and constitutes an accurate approximation,
\begin{equation}
\lim_{\rho\to 0}F^{\rm hs}_{\rm ex} =
\frac{1}{2}\int d{\bf r}\int d{\bf r}'\,
\rho_{}({\bf r}')\rho_{}({\bf r}')\theta(\sigma_{} - |{\bf r}-{\bf r}'|),
\label{eq:F_ex_dilute}
\end{equation}
which yields
\begin{equation}
\frac{\delta F^{\rm hs}_{\rm ex}}{\delta\rho({\bf r})} = 
\int d{\bf r}'\,\rho_{}({\bf r}')\theta(\sigma_{} - |{\bf r}-{\bf r}'|),
\end{equation}
and the number density becomes
\begin{equation}
\rho_{\pm}({\bf r}) = 
c_be^{\mp\beta e\psi({\bf r}) - \int d{\bf r}'\,(\rho({\bf r}')-2c_b)\theta(\sigma_{}-|{\bf r}-{\bf r}'|)}. 
\end{equation}
Incidentally, the dilute limit approximation is the same as the mean-field implementation of 
the penetrable sphere interactions with $\varepsilon=1$, as both approximations are designed 
to give the lowest order term of the virial expansion for $F_{\rm ex}$ (see Fig. (\ref{fig:PB+MF})
and Fig. (\ref{fig:PB+MF0}) for performance of the dilute limit approximation).

\subsubsection{nonperturbative approach}
Further expansion of the excess free energy does not constitute an efficient scheme. 
Already the second lowest term involves the three-body overlap contributions that 
numerically is difficult to deal with.  A more powerful approach is a nonperturbative scheme.  
A nonperturbative construction keeps numerical complexity of the dilute limit approximation 
but incorporates additional terms (generally an infinite set of terms) that lead to accurate 
behavior for some limiting condition. 

One example is the weighted density approximation 
of Ref.\cite{Tarazona84,Tarazona84b}.  This approximation is constructed in terms of 
weighted density which constitutes a building block of the theory and is suggested from the 
lowest order term of the virial series for $F_{\rm ex}^{\rm hs}$,
\begin{equation}
\bar\rho({\bf r}) = \frac{1}{8}\int d{\bf r}'\,\rho({\bf r}')\theta(\sigma-|{\bf r}-{\bf r}'|).
\label{eq:rhow}
\end{equation}
$\bar\rho$ is dimensionless and normalized 
to recover the packing fraction in a bulk, $\bar\rho_b=\eta=\pi\sigma^3\rho_b/6$.  
The approximation assumes that $F^{\rm hs}_{\rm ex}$ has a general form
\begin{equation}
F_{\rm ex}^{\rm hs} = \int d{\bf r}\,\rho({\bf r})\phi_{\rm ex}(\bar\rho({\bf r})),
\end{equation}
where $\phi_{\rm ex}$ denotes an excess free energy per particle and is a function of 
$\bar\rho({\bf r})$.  
Tarazona suggested a generalized Carnahan-Starling approach, 
where for $\phi_{\rm ex}$ he used the quasi-exact Carnahan-Starling equation 
\cite{Tarazona84}, 
\begin{equation}
\phi_{\rm ex}^{\rm cs}(\bar\rho({\bf r})) = 
\frac{\bar\rho({\bf r})(4-3\bar\rho({\bf r}))}{(1-\bar\rho({\bf r}))^2},
\end{equation}
but defined as a function of a weighted density.  Now, in addition to recovering the 
dilute limit exactly, the approximation recovers the homogenous limit.  
Furthermore, the construction satisfies 
the contact value theorem, $\rho_w=P_{\rm cs}$, where $P_{\rm cs}$ is the 
Carnahan-Starling expression for hard-sphere pressure.   
The excess chemical potential of this generalized Carnahan-Starling approach is
\begin{equation}
\frac{\delta F^{\rm hs}_{\rm ex}}{\delta\rho({\bf r})} = f_{\rm ex}^{\rm cs}({\bar\rho({\bf r})})
+ \frac{1}{8}\int d{\bf r}'\,
\rho({\bf r}')\frac{\partial f_{\rm ex}^{\rm cs}}{\partial\bar\rho({\bf r}')}\theta(\sigma-|{\bf r}-{\bf r}'|),
\end{equation}
and the densities of ionic species are obtained from Eq. (\ref{eq:rhoi_bmf}).  

A nonperturbative construction can be further improved by increasing the number of  
weighted densities as building blocks of the theory.  Some improvements where implemented
as a result of careful studies of the direct correlation function, which suggested
a density dependent weight function \cite{Evans92}.  The breakthrough approach, however, 
came with the Rosenfeld's fundamental measure theory \cite{Rosenfeld89}.  Motivated 
(at least in part) by desire to construct a theory that recovers the 1D limit behavior (a property 
later referred to as the dimensional crossover), Rosenfeld obtained a new set of 
weight functions by decomposing the Heaviside step function,
\begin{eqnarray}
\theta(\sigma_{ij}-|{\bf r}_i-{\bf r}_j|) &=& \omega_{3}^i\!\otimes\!\omega_{0}^j
+ \omega_{0}^i\!\otimes\!\omega_{3}^j
+\omega_{2}^i\!\otimes\!\omega_{1}^j + \omega_{1}^i\!\otimes\!\omega_{2}^j\nonumber\\
&-&{\boldsymbol\omega}_{2}^i\!\otimes\!{\boldsymbol\omega}_{1}^j
-{\boldsymbol\omega}_{2}^i\!\otimes\!{\boldsymbol\omega}_{1}^j,
\end{eqnarray}
where 
\begin{equation}
\omega_{\alpha}\otimes\omega_{\beta} = 
\int d{\bf r}'\,\omega_{\alpha}^i({\bf r}'-{\bf r}_i)\omega_{\beta}^j({\bf r}'-{\bf r}_j),
\end{equation}
and the relevant weight functions are 
$$
\omega_3^i({\bf r}) = \theta(R_i-r),
$$
$$
\omega_2^i({\bf r}) = \delta(R_i-r),
$$
$$
{\boldsymbol \omega}_2^i({\bf r}) = \frac{\bf r}{r}\delta(R_i-r),
$$
and $\omega_1^i({\bf r})=\omega_2^i({\bf r})/(4\pi R_i)$, 
$\omega_0^i({\bf r})=\omega_2^i({\bf r})/(4\pi R_i^2)$, and
${\boldsymbol \omega}_1^i({\bf r})={\boldsymbol \omega}_2^i({\bf r})/(4\pi R_i)$.  
The six weighted densities that result are,
\begin{equation}
n_{\alpha} = \sum_{i=1}^K\int d{\bf r}'\,\rho_i({\bf r}')\omega^i_{\alpha}({\bf r}-{\bf r}'),
\end{equation}
and a general formula for the excess free energy is
\begin{equation}
\beta F_{\rm ex}^{\rm hs} = \int d{\bf r}\,\Phi^{\rm RF}(\{n_{\alpha}({\bf r})\}).  
\end{equation}
Based on the scaled particle theory results \cite{Lebowitz59,Corti04,Stillinger06}, 
Rosenfeld came up with the following functional form 
\cite{Rosenfeld89,Tarazona08,Evans09,Roth10},
\begin{eqnarray}
\Phi^{\rm RF} &=& -n_0\log(1-n_3)+\frac{n_1n_2-{\bf n}_1\cdot{\bf n}_2}{1-n_3}
+\frac{n_2^3-3n_2({\bf n}_2\cdot{\bf n}_2)}{24\pi(1-n_3)^2}.\nonumber\\
\label{eq:Phi_RF}
\end{eqnarray}
The construction also recovers the PY direct correlation function for homogenous liquids.  

\begin{figure}[h] 
 \begin{center}%
\includegraphics[width=0.5\textwidth]{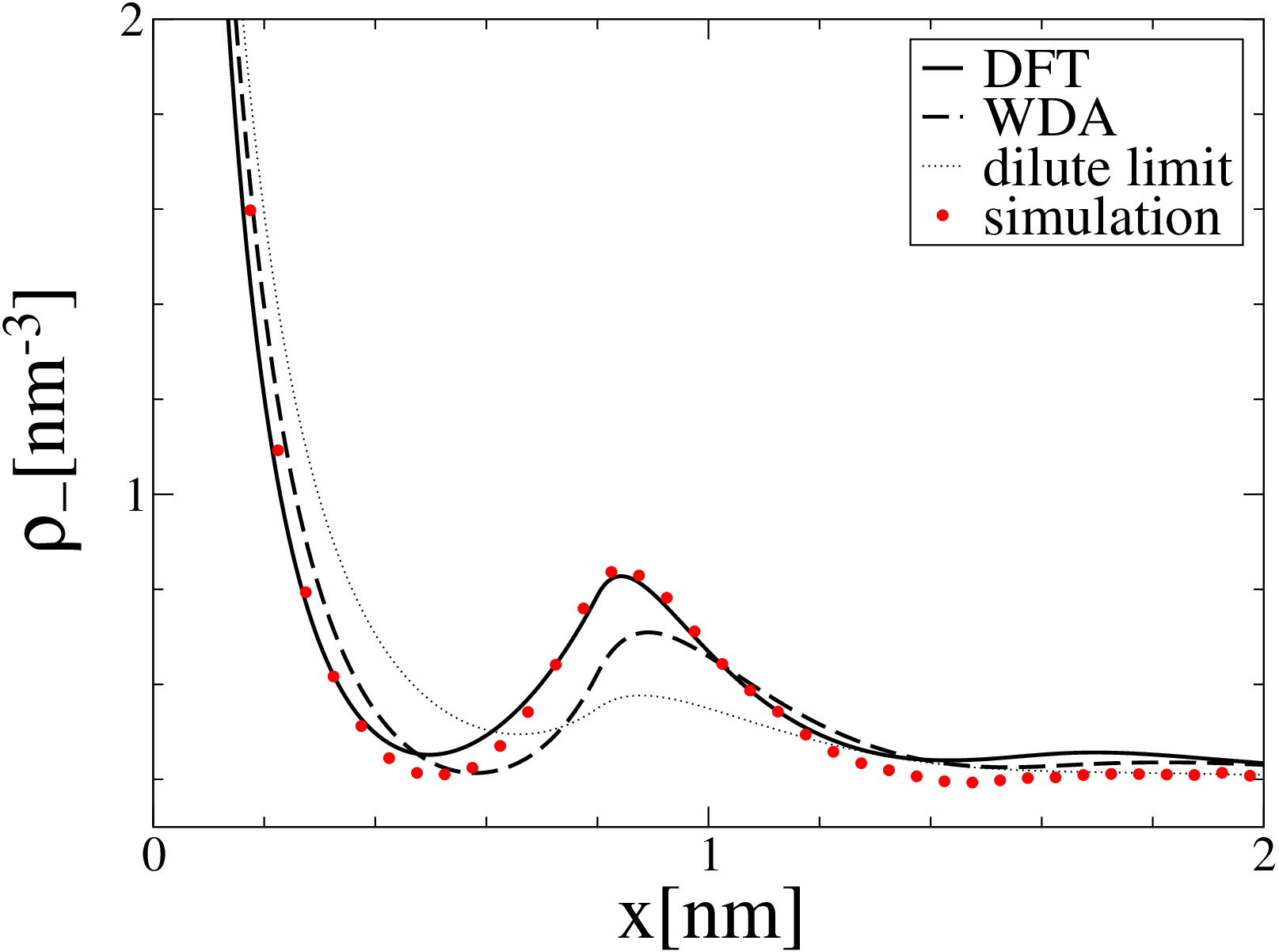}
\end{center}
\caption{The counterion density near a charged wall for the DFT scheme.  Conditions as in Fig. (\ref{fig:PB+MF}).  DFT denotes the density functional theory based on the fundamental 
measure theory \cite{Rosenfeld89}, and WDA denotes the weighted density approximation
based on the generalized Carnahan-Starling equation \cite{Tarazona84}.  
}
\label{fig:rho_DFT}
\end{figure}

\begin{figure}[h] 
 \begin{center}%
 \begin{tabular}{rr}
\includegraphics[width=0.5\textwidth]{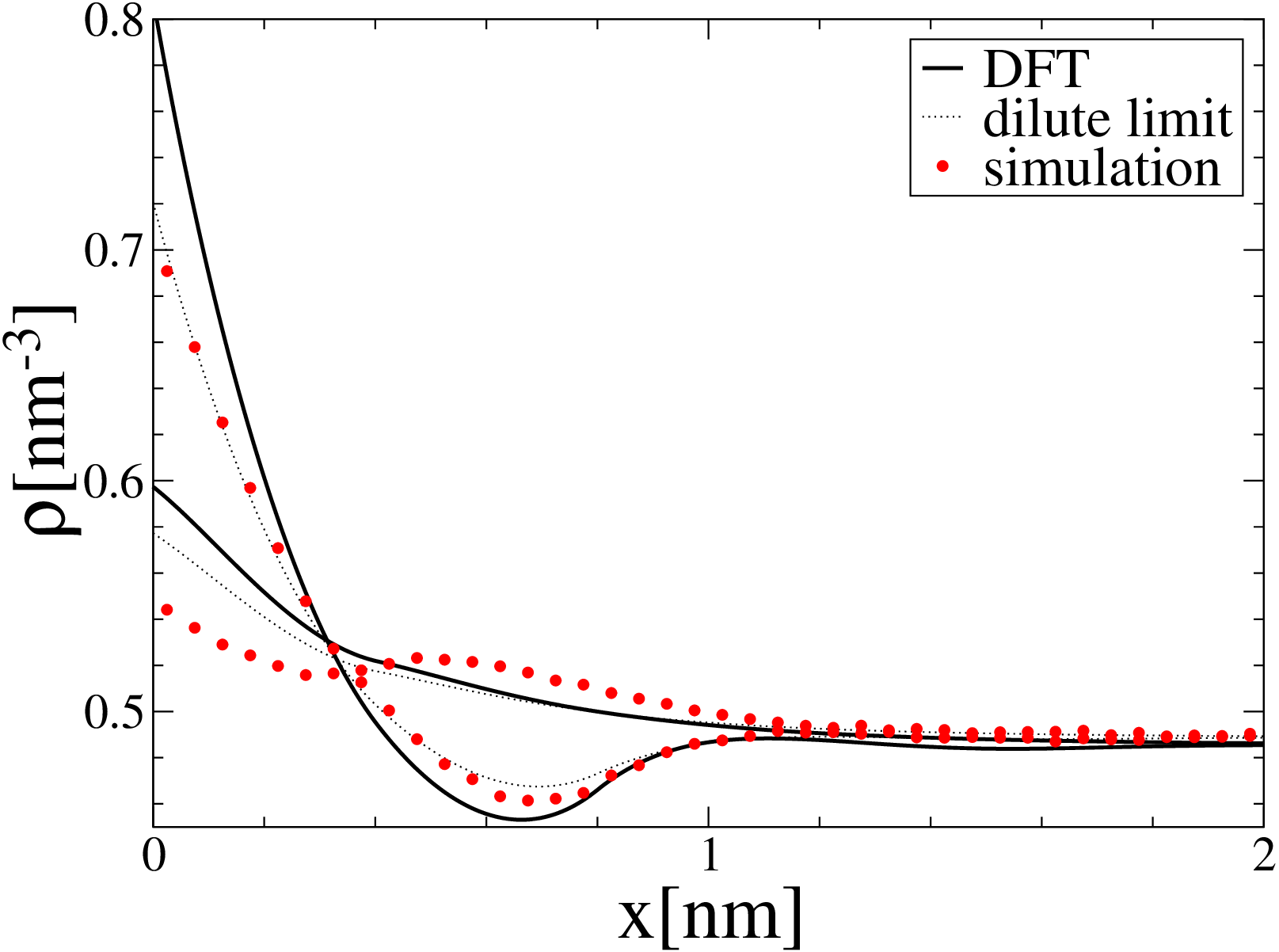}&
\includegraphics[width=0.5\textwidth]{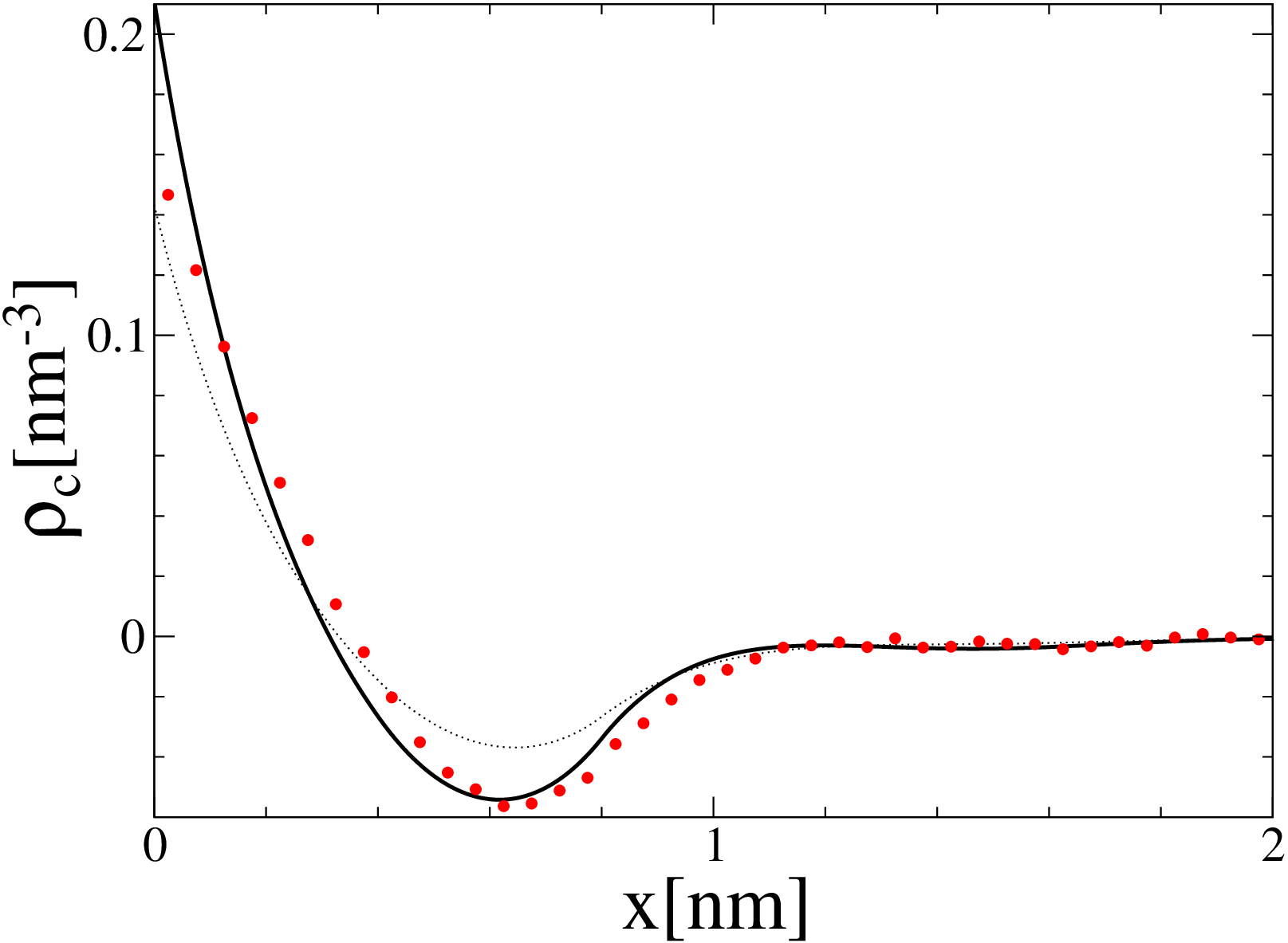}\\
\end{tabular}
\end{center}
\caption{Density profiles near an uncharged wall for various approximations.  
Conditions as in Fig. (\ref{fig:PB+MF0}).   }
\label{fig:rho_neutral_DFT}
\end{figure}

In Fig. (\ref{fig:rho_DFT}) we plot the counterion density profiles for a symmetrical 
electrolyte $1:1$ confined between two parallel hard walls.  The conditions are the same as in 
Fig. (\ref{fig:PB+MF}).  The WDA gives improvements over the dilute limit approximation, but
the DFT fundamental measure theory results agree most closely with the simulation.  
\graphicspath{{figures/}}
In Fig. (\ref{fig:rho_neutral_DFT}) we plot density profiles for a $1:1$ electrolyte with 
size asymmetry confined by uncharged walls.  The conditions are the same as in 
Fig. (\ref{fig:PB+MF0}).  The DFT here is less accurate, although the charge density
profile quite well agrees with simulation.  The disparity can be traced to the lack of correlations
in the mean-field treatment of electrostatics, which become important for 
an uncharged wall system.  The presence of correlations is best seen in the contact density,
related to the pressure via the contact value theorem, which is lower in the simulation results,
and which indicates negative correlational contributions due to formation of Bjerrum 
pairs \cite{Yan93}.  
\graphicspath{{figures/}}

\subsubsection{Correlations}
The results in Fig. (\ref{fig:rho_neutral_DFT}) for neutral confinement 
indicate that despite of highly accurate expression 
for hard-core interactions, $F_{\rm ex}^{\rm hs}$, the mean-field treatment of electrostatics is 
not sufficient and the correlations play a dominant role.  The neglected correlational contribution 
to the free energy, taken out of the complete expression in Eq. (\ref{eq:F_lambda2}), is
\begin{equation}
F_{\rm c} = \frac{1}{2}\sum_{i,j}^K\int \!\!d{\bf r}\!\!\int \!\!d{\bf r}'\,
\frac{q_i\rho_i({\bf r})q_j\rho_j({\bf r}')}{4\pi\epsilon|{\bf r}-{\bf r}'|}
\int_0^1 d\lambda \,
h_{ij}^{\lambda}({\bf r},{\bf r}').
\label{eq:F_c}
\end{equation}
As $\lambda\to 0$, $h^{\lambda}_{ij}$ does not vanish, but instead 
$h^{\lambda}_{ij}\to h^{\rm hs}_{ij}$.  

Applied to homogenous electrolytes, the formula in Eq. (\ref{eq:F_c}) yields the charging
process formula,
\begin{eqnarray}
f_{\rm c} &=& \frac{1}{2}\sum_{i=1}^Kq_ic_i\int_0^1 d\lambda\,
\Bigg[4\pi\int_0^{\infty} dr\,r^2\sum_{j=1}^{K}
\frac{q_jc_jh^{\lambda}_{ij}(r)}{4\pi\epsilon r}\Bigg],
\end{eqnarray}
where $f_c=F_c/V$.  By identifying the term
\begin{equation}
\rho_{c,i}^{\lambda}(r)=\sum_{j=1}^Kq_jc_jh^{\lambda}_{ij}(r)
\end{equation}
as the charge distribution around an ion of the species $i$ fixed at the origin (constituting
the charge correlation hole), the term in brackets becomes an electrostatic potential that a test 
ion of the species $i$ feels due to surrounding ions in the system,
\begin{equation}
f_{\rm c} = \frac{1}{2}\sum_{i=1}^Kq_ic_i\int_0^1 d\lambda\,\psi_i^{\lambda},
\end{equation}
where the superscript $\lambda$ indicates that the interactions are scaled.  The formula
is the expression of the "charging process", a common route for obtaining the correlational 
free energy.  Substituting for $\psi^{\lambda}$ the linear Debye-H\"uckel solution leads to the 
Debye charging process, which for $1:1$ electrolyte of ions of the same size becomes 
\cite{Yan02}, 
\begin{equation}
q_i\psi_i^{\lambda} = -\frac{e}{4\pi\epsilon}
\Bigg[\frac{\kappa\sqrt{\lambda}}{1+\kappa\sigma\sqrt{\lambda}}\Bigg],
\end{equation}
and
\begin{eqnarray}
\beta f_c 
&=&-\frac{1}{4\pi\sigma^3}\Bigg[\log(\kappa\sigma+1)-\kappa\sigma+\frac{(\kappa\sigma)^2}{2}\Bigg].
\end{eqnarray}
This correlation term constitutes a weak-coupling correction and it does not capture the 
formation of Bjerrum pairs \cite{Yan93}.  It gives, however, some estimate of what the 
contributions of neglected correlations are.  

Application of the charging process to inhomogeneous electrolytes is not easy
as it requires the precise functional form for $\psi_{\rm ext}^{\lambda}({\bf r})$ which ensures 
that densities remain constant throughout charging.  The implementation of coupled
contributions of hard-core and electrostatic contributions remains a challenge.  
There are some perturbative extensions to the DFT theory based on the reference fluid density 
\cite{Rosenfeld93,Eisenberg03} and which address this issue.  



\subsection{local schemes}
After reviewing nonlocal approximations for hard-sphere interactions, it may seem a 
regression to discuss next local approximations. Nonlocal construction based on weighted 
densities captures discrete structure of a fluid and is found to satisfy the contact value 
theorem sum rule.  A local construction, on the other hand, is expressed in terms of local 
density (or a weighted density with a delta weight function), and as such, it does not posses 
discrete structure of a liquid that is implicit in weight functions, and fails to satisfy the contact 
value theorem \cite{Frydel12}.  In other words, density profiles produced by the local type of 
an approximation are unphysical.  

Then why even bother with local approximations?  The first answer is simplicity.  But this
does not justify a model as a description of the world.  A more reasonable justification may
sound like this.  For true electrolytes as they are found in laboratories the exact nature of the 
excluded volume interactions is not known with precision and there are many different and 
complex contributions.  The hard-sphere model is itself an idealization.  The local approximation, 
in spite of its shortcomings, can offer a first glance and an estimate of excluded volume effects.  
One, however, has to know how to interpret such a local approximation.  The structureless 
density profile cannot be read as physical.  The saturation effect of a local density triggered 
by overcrowding is an artifact of the model.  But although density is unphysical, it does not mean 
that every other quantity that follows is equally so.  For example, the incorrect contact density does 
not imply an incorrect contact potential.  In fact, the contact potential values were found to be 
reasonably well estimated by a local scheme \cite{Frydel12}.  The local saturation of a density
profile, its flattening near a charged surface, captures qualitatively the fact that a double-layer
is elongated due to excluded volume effects, and this in turn reproduces, at least qualitatively,
the increase in electrostatic potential that is less efficiently screened.  

Within local approximation the excess free energy is 
\begin{equation}
F_{\rm ex}^{\rm hs} = \int d{\bf r}\,f_{\rm ex}(\rho({\bf r})) ,
\end{equation}
where the excess free energy density $f_{\rm ex}$ is a function of a local density,
and a functional derivative becomes classical derivative, 
\begin{equation}
\frac{\delta F_{\rm ex}^{\rm hs}}{\delta\rho({\bf r})} 
= \frac{\partial f_{\rm ex}}{\partial \rho({\bf r})}
= \mu_{\rm ex}(\rho({\bf r})),
\end{equation}
which yields the following density 
\begin{equation}
\rho_i({\bf r}) = c_ie^{-\beta\big[q_i\psi({\bf r}) + 
\mu_{\rm ex}(\rho({\bf r}))-\mu_{\rm ex}(\rho_b)\big]},
\end{equation}
The mean-field Poisson equation becomes,
\begin{equation}
\epsilon\nabla^2\psi({\bf r}) = -\sum_{i=1}^K
q_ic_ie^{-\beta\big[q_i\psi({\bf r}) + \mu_{\rm ex}(\rho({\bf r}))-\mu_{\rm ex}(\rho_b)\big]}.  
\end{equation}
We assume that all diameters are the same.  

To complete the model, it remains to choose expression for $\mu_{\rm ex}$.  
There are several equations of state we can choose from.  
The hard-sphere model (or the quasi-exact Carnahan-Starling equation) is
\begin{equation}
\frac{\beta P}{\rho} = \frac{1+\eta+\eta^2-\eta^3}{(1-\eta)^3}
~~~\longrightarrow~~~\beta\mu_{\rm ex} = \frac{8\eta-9\eta^2+3\eta^3}{(1-\eta)^3}.
\end{equation}
A cruder van der Waals model for excluded volume interactions is
\begin{equation}
\frac{\beta P}{\rho} = \frac{1}{1-\nu\rho}
~~~\longrightarrow~~~\beta\mu_{\rm ex} = -\log(1-\nu\rho) + \frac{\nu\rho}{1-\nu\rho},
\end{equation}
where $\nu$ denotes the excluded volume.  Finally, the lattice-gas model is \cite{David97}
\begin{equation}
\frac{\beta P}{\rho} = -\frac{\log(1-\eta)}{\eta}
~~~\longrightarrow~~~\beta\mu_{\rm ex} = -\log(1-\eta).
\label{eq:LG}
\end{equation}

\begin{figure}[tbh]
\vspace{0.6cm}
\centerline{\resizebox{0.55\textwidth}{!}
{\includegraphics{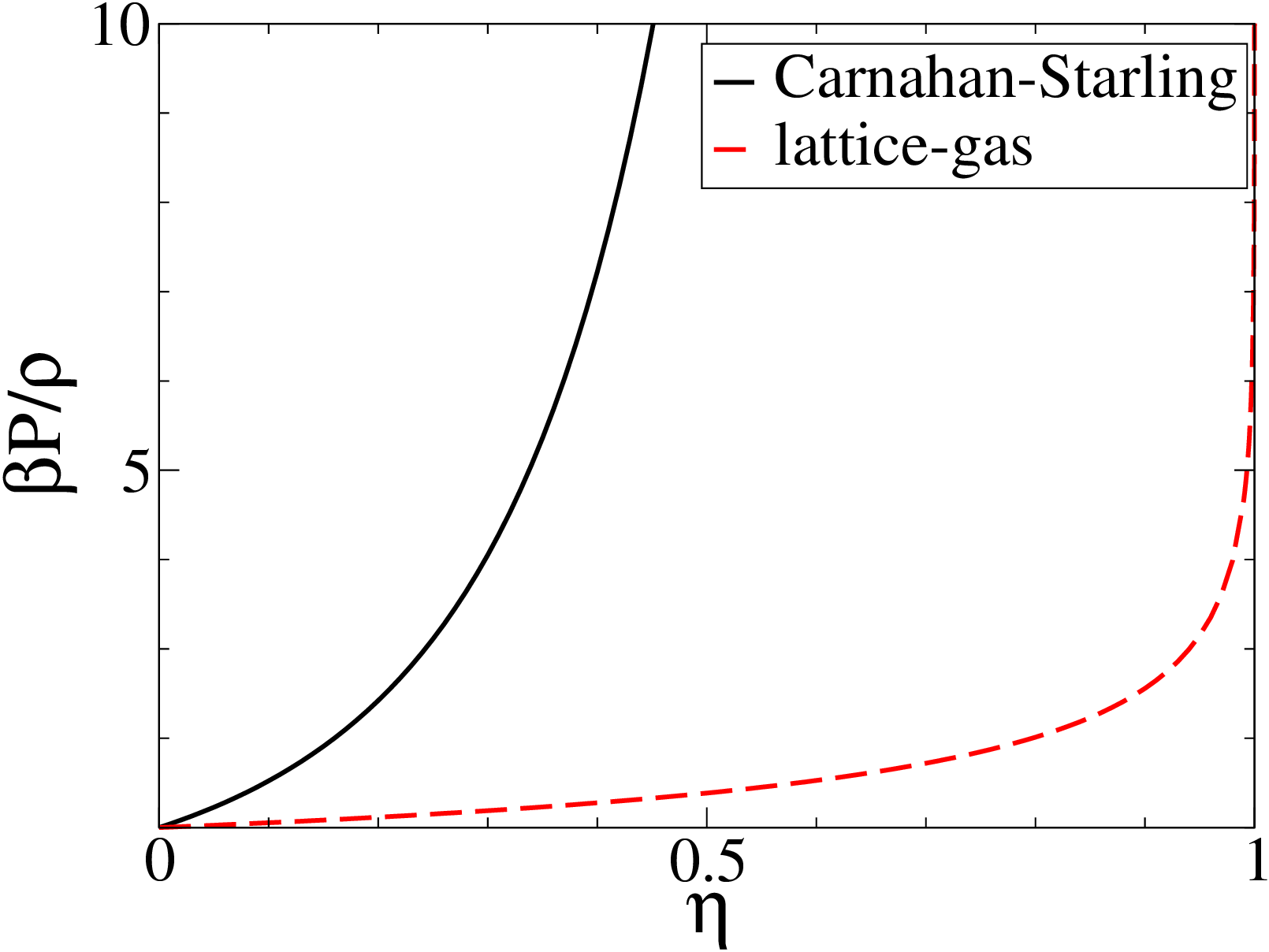}}}
\caption{The equation of state $\beta P/\rho$ as a function of a packing fraction 
$\eta=\pi\sigma^3/6$.  
}
\label{fig:Z}
\end{figure}

In Fig. (\ref{fig:Z}) we compare the lattice-gas equation of state with that for hard-spheres.  The 
two models are completely different.  There is no agreement in any limit.  The lattice-gas 
curve is relatively flat and then exhibits a sharp rise as $\eta\to 1$.   

The lattice-gas model has advantage in its simple analytical form.
The probability for successful insertion of a particle into a hard-sphere fluid is 
$e^{-\beta\mu_{\rm ex}}=1-\eta$.  The result is intuitive and expresses the fraction of an
available volume not taken up by other particles. 
This simple result leads to the following density 
\begin{equation}
\rho_i = c_ie^{-\beta q_i\psi}\bigg(\frac{1-\nu\sum_{i=1}^K\rho_i}{1-\nu \sum_{i=1}^Kc_i}\bigg),
\end{equation}
where $\nu=\pi\sigma^3/6$ is the sphere volume.  After some algebraic manipulation we get 
\begin{equation}
\rho_i = \frac{c_ie^{-\beta q_i\psi}}{1+\nu \sum_{i=1}^Kc_i(e^{-\beta q_i\psi}-1)}.
\end{equation}
In the limit $\psi\to 0$, the standard Poisson-Boltzmann equation is recovered, 
$\rho_i\to c_ie^{-\beta q_i\psi}$.  But if potential becomes large a density cannot 
increase indefinitely as it is bounded from above, $\rho\le\nu^{-1}$.  
The modified Poisson-Boltzmann equation that results is 
\begin{equation}
\epsilon\nabla^2\psi = 
-\frac{\sum_{i=1}^Kq_ic_ie^{-\beta q_i\psi}}{1+\nu \sum_{i=1}^K c_i(e^{-\beta q_i\psi}-1)}.
\end{equation}
Specializing to the $1:1$ electrolyte we get \cite{David97}
\begin{equation}
\epsilon\nabla^2\psi = 
-\frac{2e c_s\sinh\beta e\psi}{1 + 2\nu c_s(\cosh\beta e\psi-1)}
\end{equation}

This is the modified Poisson-Boltzmann equation as derived in \cite{David97}.  It yields
the same boundary condition as the standard PB equation.
Also, the model does not lead to the true contact value theorem, 
$\rho_w=P+\frac{\beta\sigma_c^2}{2\epsilon}$, where in place of $P$ we use the 
lattice-gas pressure in Eq. (\ref{eq:LG}).  Instead, it obeys another contact value relation,  
\begin{equation}
\rho_w = \frac{1}{\nu}\bigg[1 - (1-\eta_b)e^{-\beta\sigma_c^2\nu/2\epsilon}\bigg],
\end{equation}
since, as was said before, the density is not physical.  
The model introduces a new length scale, $\nu\sigma_c$, that corresponds to the width 
of counterion layer that would form if all counterions were allowed to come to a charged
surface and the exceeded volume effect was the only interaction.  
Now, even for the vanishing screening length, $\kappa^{-1}\to 0$, a double-layer
will have thickness $\nu\sigma_c$ (where $\kappa = \sqrt{8\pi c_s\lambda_B}$ is the 
Debye screening parameter).  Based on these two competing length scales
it is possible to estimate the importance of the excluded volume effects.  
If $\nu\sigma_c > \kappa^{-1}$, we should expect the excluded volume effects to play
a significant role.

\begin{figure}[h] 
 \begin{center}%
 \begin{tabular}{rr}
\includegraphics[width=0.48\textwidth]{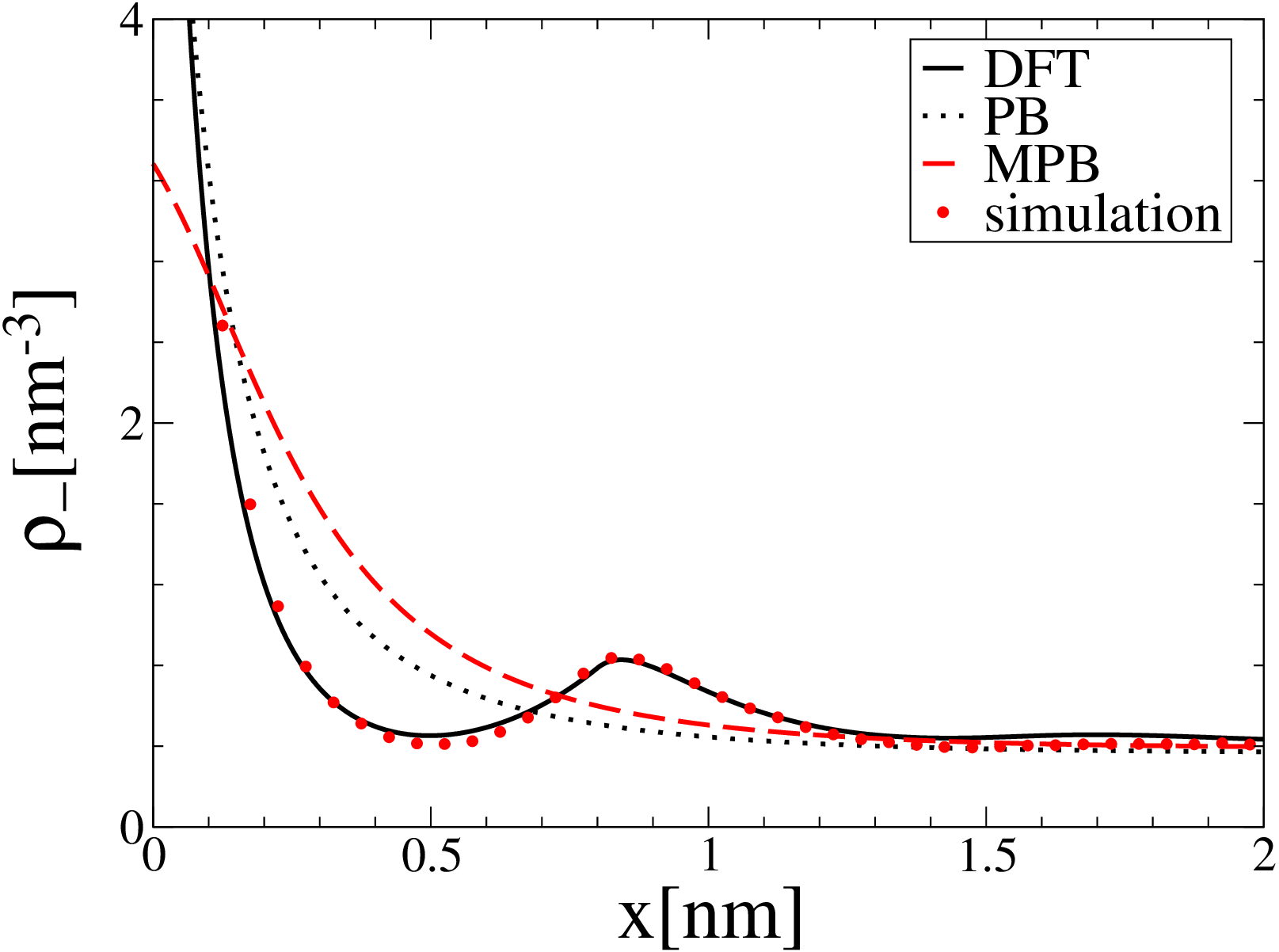}&
\includegraphics[width=0.48\textwidth]{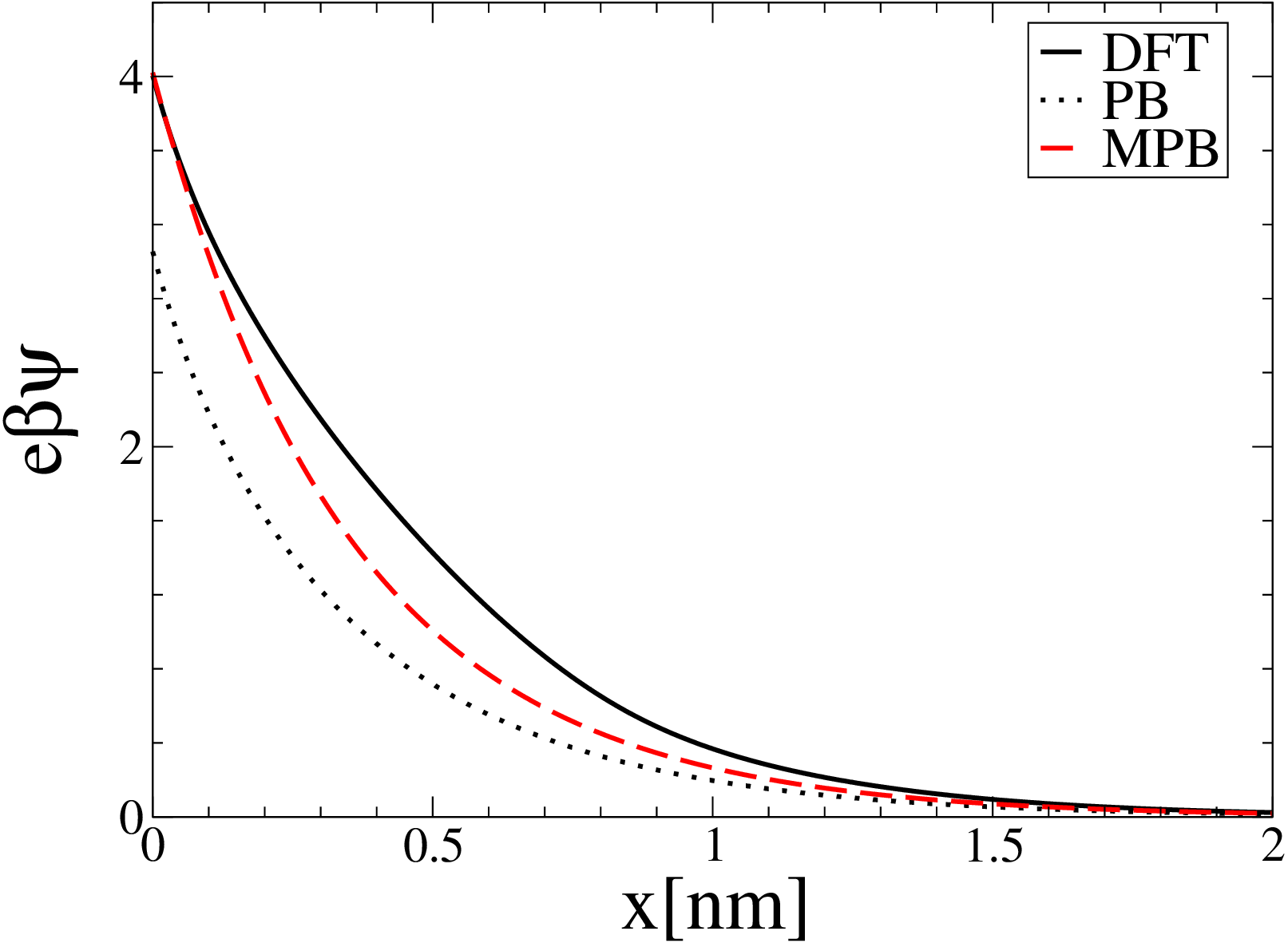}\\
\end{tabular}
\end{center}
\caption{Counterion density and potential profiles for the conditions as in Fig. (\ref{fig:PB+MF})
and Fig. (\ref{fig:PB+MF}).   }
\label{fig:MPB}
\end{figure}

\begin{figure}[h] 
 \begin{center}%
\includegraphics[width=0.5\textwidth]{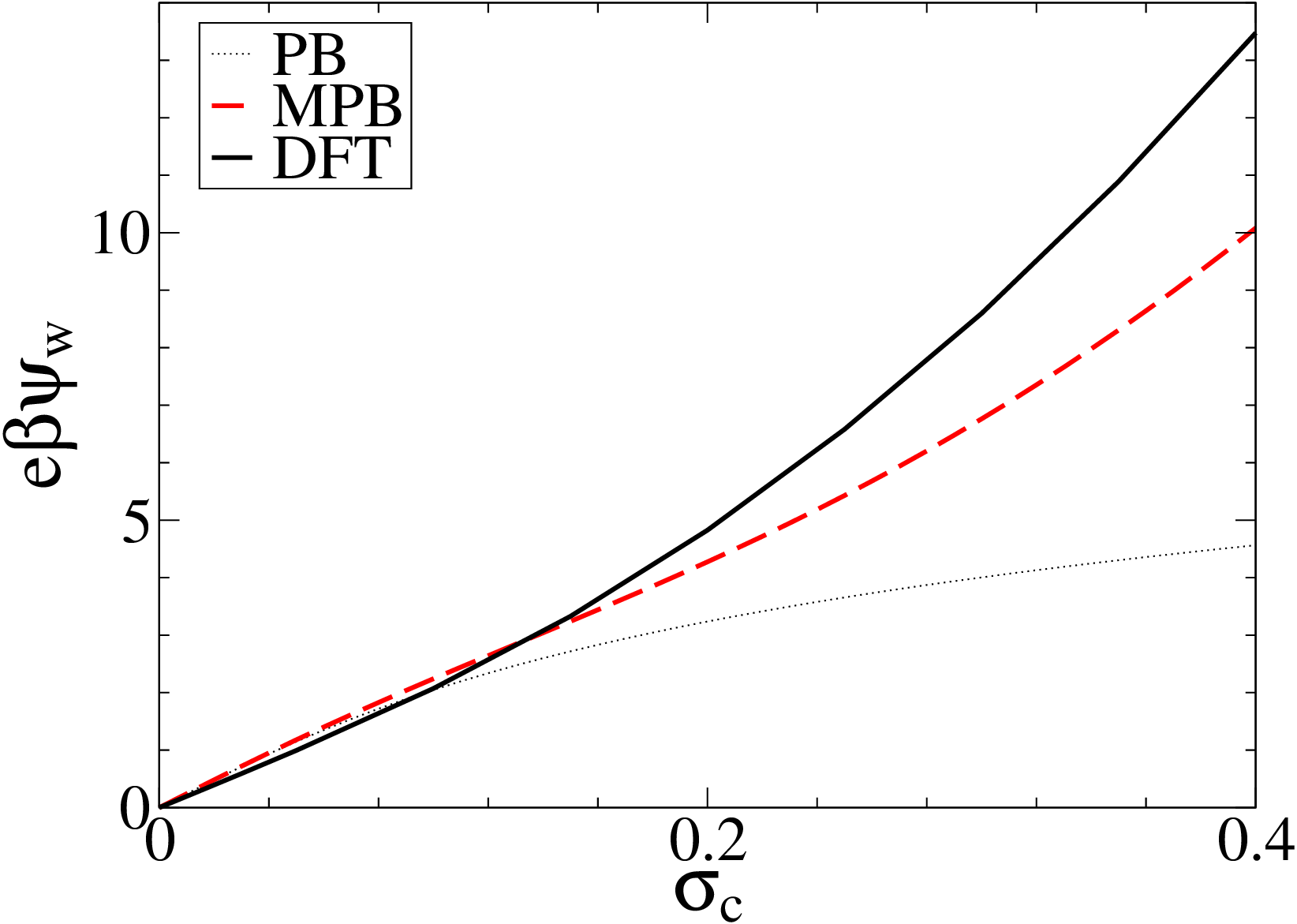}\\
\end{center}
\caption{Electrostatic potential at a wall contact as a function of a surface charge.
The system parameters are :  $\lambda_B=0.72\,{\rm nm}$, $\sigma=0.8\,{\rm nm}$, 
$c_s=0.3\,{\rm M}$.   }
\label{fig:psi_w}
\end{figure}

In Fig. (\ref{fig:MPB}) we compare the results of the modified PB equation with other approximations.
The density profile of the modified PB equation shows unphysical saturation of a local density.  
The standard PB equation, in fact, yields better agreement with the DFT near a wall including 
a contact density.  The modified PB equation, however, yields better results for electrostatic
potential which comes close to the DFT approximation at a wall contact.  
\graphicspath{{figures/}}
The modified PB equation captures the fact that the surface charge is less efficient screening 
when the excluded volume interactions  are involved.  
In Fig. (\ref{fig:psi_w}) we plot contact potential as a function of the surface charge.  
The modified PB equation captures the influence of the excluded volume effects
in relation to the standard PB equation.  As $\sigma_c$ becomes large, the agreement
with the DFT is less perfect.  
\graphicspath{{figures/}}

The question still remains how using the Carnahan-Starling equation of state for the local
approximation would change the results.  After all, when choosing the lattice-gas equation of 
state, the only criterion that was followed was simplicity.  It turns out that the equation of state for 
hard-spheres gives worse agreement with the DFT and it exaggerates overcrowding by yielding 
too high contact potentials.  It somewhat seems a stroke of luck that the lattice-gas
equation of state provides both simplicity and relative accuracy, suggesting that some 
cancellation of errors is being involved.




\section{Conclusion}

The present review provides a framework for constructing various mean-field models for 
ions with some sort of structure and provides a number of modified PB equations to which
this construction leads.  All possibilities, of course, 
cannot be exhausted, but with a large number of detailed constructions it should not be 
difficult to build a model that fits a given situation.  One possible direction to pursue further 
is to explore in more detail models for ions with finite charge distribution, all the way 
until making contact with polyelectrolytes, whose single configuration is represented with 
a Brownian walk type of a distribution.  
What was also left out from the review were models that combine several structures together. 
For example, a spherical charge distribution could be supplemented with repulsive Gaussian 
interactions representing a self-avoiding walk of two polymer chains.  This would render a
more realistic representation of polyelectrolytes.  But again, such combinations are not
difficult to infer from provided constructions.  It is, in fact, one of the goals of this review to 
motivate such new constructions.  

The review also leaves some suggestions for the future work.  As there is an ever increasing 
number of new macromolecules with interactions ranging from ultrasoft to hard-core,
particles whose shape is not fixed but flexible, there is an ever growing demand for their 
accurate representation.  One possible way to proceed is to explore the present mean-field 
framework and implement some sort of elastic behavior to allow charge distributions to deform 
into most optimal shape, so that near an interface particles and their mutual interactions are 
modified.  Similar elastic behavior could be supplemented to non-electrostatic type of 
interactions.

As far as the treatment of short-range non-electrostatic interactions is concerned, the 
mean-field is sufficient if interactions are soft.  The handling of hard-sphere 
interactions, however, is far more challenging.  The most efficient theory for hard-core 
interactions, the fundamental measure DFT, shows shortcomings even for the weak-coupling 
limit conditions.  To construct a more accurate theory it is, therefore, necessary to incorporate 
correlations, thus, to go beyond the conveniences of the mean-field.  But for charged hard-core 
particles correlations are difficult to implement as they couple hard-core and electrostatic 
interactions.  The treatment of these coupled, short- and long-range, interactions is, in fact, one 
of the most outstanding problems of soft-matter electrostatics.  

The idea of a modified PB equation, of representing more physics and more accurately through 
a model based on a single differential equation, has caught some momentum, and there are 
models that go beyond the mean-field and attempt to implement effects seen in the intermediate-
and strong-coupling limit.  An exemplary case is the work by Bazant {\sl et al.} 
\cite{Bazant11,Bazant12} where the authors suggested the modified PB equation with dielectric 
constant represented as a linear differential operator.  The model aims to represent ionic liquids 
were opposite ions strongly associate and coexist as a Bjerrum pair rather than 
as free ions.  

There is an additional motivation for pursuing various mean-field constructions.  
Simple models such as charged hard-spheres are easy to simulate, and for these systems
one could simply stick to simulations to cover the entire range of electrostatics, from weak- to
strong-coupling regime.  But there are systems that are not easy to simulate.  This is especially 
true for polarizable ions and for explicit treatment of water.  For these cases 
the mean-field construction provides a real alternative, sometimes the only choice.  

Finally, 
it should be reminded and stressed one more time what the boundaries of the mean-field
treatment are and what type of electrostatics it is capable of representing.  Not only it neglects 
correlational corrections already active in the weak-coupling regime, but it completely 
fails in the strong-coupling limit as a predictive tool.  The strong-coupling limit electrostatics
calls for different treatments.

 \section*{ACKNOWLEDGMENT}
 The author would like to thank Tony Maggs as well as the ESPCI lab, were bulk of this work 
 was being done, for helpful and friendly working atmosphere.  
This work was supported in part by the agence nationale de la recherche via the project FSCF.

{99}



 \bibliographystyle{plain}
 \nocite{*}  
 \bibliography{Sample}








% ...


\end{document}